\documentclass[11pt]{article}\usepackage[]{graphicx}\usepackage[]{color}
\makeatletter
\def\maxwidth{ %
  \ifdim\Gin@nat@width>\linewidth
    \linewidth
  \else
    \Gin@nat@width
  \fi
}
\makeatother

\definecolor{fgcolor}{rgb}{0.345, 0.345, 0.345}
\newcommand{\hlnum}[1]{\textcolor[rgb]{0.686,0.059,0.569}{#1}}%
\newcommand{\hlstr}[1]{\textcolor[rgb]{0.192,0.494,0.8}{#1}}%
\newcommand{\hlcom}[1]{\textcolor[rgb]{0.678,0.584,0.686}{\textit{#1}}}%
\newcommand{\hlopt}[1]{\textcolor[rgb]{0,0,0}{#1}}%
\newcommand{\hlstd}[1]{\textcolor[rgb]{0.345,0.345,0.345}{#1}}%
\newcommand{\hlkwa}[1]{\textcolor[rgb]{0.161,0.373,0.58}{\textbf{#1}}}%
\newcommand{\hlkwb}[1]{\textcolor[rgb]{0.69,0.353,0.396}{#1}}%
\newcommand{\hlkwc}[1]{\textcolor[rgb]{0.333,0.667,0.333}{#1}}%
\newcommand{\hlkwd}[1]{\textcolor[rgb]{0.737,0.353,0.396}{\textbf{#1}}}%

\usepackage{framed}
\makeatletter
\newenvironment{kframe}{%
 \def\at@end@of@kframe{}%
 \ifinner\ifhmode%
  \def\at@end@of@kframe{\end{minipage}}%
  \begin{minipage}{\columnwidth}%
 \fi\fi%
 \def\FrameCommand##1{\hskip\@totalleftmargin \hskip-\fboxsep
 \colorbox{shadecolor}{##1}\hskip-\fboxsep
     \hskip-\linewidth \hskip-\@totalleftmargin \hskip\columnwidth}%
 \MakeFramed {\advance\hsize-\width
   \@totalleftmargin\z@ \linewidth\hsize
   \@setminipage}}%
 {\par\unskip\endMakeFramed%
 \at@end@of@kframe}
\makeatother

\definecolor{shadecolor}{rgb}{.97, .97, .97}
\definecolor{messagecolor}{rgb}{0, 0, 0}
\definecolor{warningcolor}{rgb}{1, 0, 1}
\definecolor{errorcolor}{rgb}{1, 0, 0}
\newenvironment{knitrout}{}{} 

\usepackage{alltt}

\usepackage{amsthm}
\usepackage{amsmath}
\usepackage{amsfonts}
\usepackage{amscd}
\usepackage{amssymb}
\usepackage{natbib}
\usepackage{url}
\usepackage{times}
\usepackage{epsfig}
\usepackage[export]{splitbib}
\usepackage{mathtools}

\usepackage{algorithmic}
\usepackage{dsfont} 

\usepackage{geometry}
\newcommand{\R}{\mathbb{R}}
\newcommand{\calR}{\mathcal{R}}
\newcommand{\E}{\mathcal{E}}
\newcommand{\Prob}{\mathbb{P}}
\newcommand{\tr}{\text{tr}}
\newcommand{\Mhat}{\widehat{M}}
\newcommand{\Mstar}{\widehat{M}^{\textstyle{*}}}
\newcommand{\Uhat}{\widehat{U}}
\newcommand{\Ustar}{\widehat{U}^{\textstyle{*}}}
\newcommand{\That}{\widehat{\theta}}
\newcommand{\TFG}{\widehat{\theta}^{\text{FG}}}

\newcommand{\TD}{\widehat{\theta}^{\text{1D}}}
\newcommand{\Ttil}{\widetilde{\theta}}
\newcommand{\Tstar}{\widehat{\theta}^{\textstyle{*}}}

\newcommand{\TstarFG}{\widehat{\theta}^{{\text{FG}}^{\textstyle{*}}}}
\newcommand{\TstarD}{\widehat{\theta}^{{\text{1D}}^{\textstyle{*}}}}
\newcommand{\Ttilstar}{\widetilde{\theta}^{\textstyle{*}}}
\newcommand{\wstar}{w^{\textstyle{*}}}
\newcommand{\Ghat}{\widehat{G}}
\newcommand{\Gamhat}{\widehat{\Gamma}}
\newcommand{\Gamstar}{\widehat{\Gamma}^{\textstyle{*}}}
\newcommand{\GamstarT}{\widehat{\Gamma}^{\textstyle{*}^T}}

\newcommand{\Omhat}{\widehat{\Omega}}
\newcommand{\Sigmahat}{\widehat{\Sigma}}
\newcommand{\ghat}{\widehat{g}}
\newcommand{\rootn}{\sqrt{n}}

\newcommand{\EnvwFG}{\widehat{\theta}^{\text{FG}}_w}

\newcommand{\EnvwoneD}{\widehat{\theta}^{\text{1D}}_w}

\newcommand{\EnvuFGstar}{\widehat{\theta}^{\text{FG}^{\textstyle{*}}}_u}

\newcommand{\phistar}{\phi^{\textstyle{*}}}
\newcommand{\IoneD}{\mathcal{I}_n^{\text{1D}}}
\newcommand{\IFG}{\mathcal{I}_n^{\text{FG}}}
\newcommand{\uoneD}{\hat{u}_{\text{1D}}}
\newcommand{\In}{\mathcal{I}_n}
\newcommand{\uFG}{\hat{u}_{\text{FG}}}
\newcommand{\wFG}{w^{\text{FG}}}

\newcommand{\wstarFG}{w^{{\text{FG}}^{\textstyle{*}}}}

\newcommand{\woneD}{w^{\text{1D}}}

\newcommand{\X}{\mathbf{X}}
\newcommand{\A}{\mathcal{A}}
\newcommand{\Xstar}{\X^{\textstyle{*}}}

\newcommand{\gstar}{\hat{g}^{\textstyle{*}}}
\newcommand{\Gstar}{\widehat{G}^{\textstyle{*}}}
\newcommand{\GstarT}{\widehat{G}^{\textstyle{*}^T}}
\newcommand{\Pstar}{\widehat{P}^{\textstyle{*}}}
\newcommand{\vstar}{v^{\textstyle{*}}}

\newcommand{\Jstar}{J^{\textstyle{*}}}
\newcommand{\lstar}{l^{\textstyle{*}}}

\newtheorem{lem}{Lemma} 
\newtheorem{thm}{Theorem} 
\newtheorem{defn}{Definition}

\title{General model-free weighted envelope estimation}
\author{Daniel J. Eck}
\IfFileExists{upquote.sty}{\usepackage{upquote}}{}
\begin{document}

\maketitle

\begin{abstract}
Envelope methodology is succinctly pitched as a class of procedures for 
increasing efficiency in multivariate analyses without altering traditional 
objectives \citep[first sentence of page 1]{cook2018introduction}.  
This description is true with the additional caveat that the efficiency gains 
obtained by envelope methodology are mitigated by model selection volatility 
to an unknown degree.  The bulk of the current envelope methodology 
literature does not account for this added variance that arises from the 
uncertainty in model selection.  Recent strides to account for model selection 
volatility have been made on two fronts: 1) development of a 
weighted envelope estimator, in the context of multivariate regression, to 
account for this variability directly; 2) development of a model selection 
criterion that facilitates consistent estimation of the correct envelope model 
for more general settings.  In this paper, we unify these two directions and 
provide weighted envelope estimators that directly account for the 
variability associated with model selection and are appropriate for 
general multivariate estimation settings for vector valued parameters.  
Our weighted estimation technique provides practitioners with robust and 
useful variance reduction in finite samples.
Theoretical justification is given for our estimators and validity of a 
nonparametric bootstrap procedure for estimating their asymptotic variance are 
established. 
Simulation studies and a real data analysis support our claims and demonstrate 
the advantage of our weighted envelope estimator when model selection 
variability is present.
\\[1em]
\textbf{Keywords}: dimension reduction; model averaging; envelope methodology; 
  nonparametric bootstrap; bootstrap smoothing; model selection
\end{abstract}

\section{Introduction}

Let $\X_1$, $\ldots$, $\X_n$ be an independent sample where $\theta \in \R^p$ is a target parameter that we want to estimate. Suppose that $\Ttil = \Ttil(\X_1$, $\ldots$, $\X_n)$ is a $\rootn$-consistent and asymptotically normal estimator of $\theta$ with asymptotic variance $\Sigma > 0$ such that  
\begin{equation}
  \rootn\left(\Ttil - \theta\right) \overset{d}{\to} N(0, \Sigma).
\label{context}
\end{equation}
The idea of envelope methodology is to estimate $\theta$ with less asymptotic variability than $\Sigma$ through the exploitation of a parametric link between $\theta$ and $\Sigma$ \citep{cook2010, found, cook2018introduction}. Envelope methodology originated as a method to reduce the variability of a regression coefficient matrix $\beta$ in the multivariate linear regression model \citep{cook2010, su, su-inner, cook-scale, cook2018introduction}. The key insight behind using envelope methodology as a variance reduction tool was the observation that some linear combinations of the response vector may be invariant to changes in the predictors. Such linear combinations represent variability in the response vector that is not directly relevant to the estimation of $\beta$ and should be discarded. \cite{found} extended envelope methodology to the general setting where one only has a target parameter $\theta$, a $\rootn$ consistent and asymptotically normal estimator of $\theta$ as in \eqref{context}, and a $\rootn$ consistent estimator $\Sigmahat$ of $\Sigma$. 

In both the multivariate linear regression model and the general estimation framework in \eqref{context}, variance reduction obtained through envelope methodology arises from exploiting a subspace of the spectral structure of the variance matrix with dimension $u < p$ that contains span$(\theta)$. The dimension $u$ of the envelope space is unknown in practice. In many envelope modeling contexts one can estimate $u$ with information criteria, likelihood ratio tests, or cross-validation. Information criteria and likelihood ratio tests can be estimated with full Grassmannian optimization \citep{cook2010, algo, found, zhangmai}, the 1D algorithm \citep{algo, found}, or the fast envelope estimator \citep{fast-algo}. Recently, \cite{zhangmai} proposed new model-free information criteria that can estimate $u$ consistently. With $u$ estimated, the variability of the envelope estimator is assessed via the bootstrap. However, most bootstrap implementations are conditional on $u = \hat{u}$ where $\hat{u}$ is the estimated dimension of the envelope space. These procedures ignore the variability associated with model selection. \cite{eck2017weighted} provided a weighted envelope bootstrap to alleviate this problem in the context of the multivariate linear regression model. In this context the variability of the weighted envelope estimator was appreciably lower than that obtained by bootstrapping the multivariate linear regression model parameters as in \citet{eck2018bootstrapping}. \citet{eck2020aster} provided a double bootstrap procedure for envelope estimation of expected Darwinian fitness from an aster model \citep{geyer2007aster, shaw2008aster} which demonstrated useful variance reduction empirically. That being said, the theoretical motivations for each of these bootstrap procedures are not applicable for envelope estimation in general settings. The weights in \cite{eck2017weighted} are constructed from the Bayesian Information Criterion values of the multivariate linear regression model evaluated at all envelope estimators fit at dimension $u = 1$, $\ldots$, $p$. Model selection volatility is taken into account in \citet{eck2020aster} by criteria that also require a likelihood.

In this paper we provide weighted envelope estimators that are appropriate for envelope estimation in the general setting described by the setup in \eqref{context}. These settings do not require the existence of a likelihood function and avoid having to condition on an estimated envelope dimension. We then provide bootstrap procedures which estimate the variability of our weighted envelope estimators. More importantly, our bootstrap procedures estimate the variability of the envelope estimator at the true, unknown, dimension $u$. This is because as $n\to\infty$, the weight at $u$ converges to $1$ fast enough to not incorporate influences from other envelope dimensions. The envelope estimators developed in this paper are a powerful and practical unification of the methodology in \cite{eck2017weighted} and \cite{zhangmai}. Our approach greatly generalizes the appropriateness of envelope model averaging beyond the context of multivariate linear regression models while simultaneously maintaining robust variance reduction in finite-samples as well as comparable to the estimators in \cite{zhangmai}. In practical applications, our approach encapsulates the variability associated with model selection volatility. We now motivate envelope methodology and weighted estimation techniques.

\section{Envelope properties}

We first provide the definition of a reducing subspace and an envelope. 

\begin{defn}[Reducing subspace]
A subspace $\calR \subset \R^p$ is a reducing subspace of a matrix $M$ if $M\calR \subset \calR$ and $M\calR^{c} \subset \calR^{c}$ where $\calR^{c}$ is the orthogonal complement of $\calR$ relative to the usual inner product. 
\end{defn}

A reducing subspace $\calR$ of a matrix $M$ allows one to decompose $M$ as $M = P_{\calR}MP_{\calR} + Q_{\calR}MQ_{\calR}$ where $P_{\calR}$ is the projection into $\calR$ and $Q_{\calR} = I - P_{\calR}$. When the eigenvalues of $M$ are distinct, a reducing subspace is a direct sum of eigenspaces of $M$. 

\begin{defn}[Envelope]
The $M$ envelope of span($U$) is defined as the intersection of all reducing subspaces $\calR$ of $M$ which satisfies span$(U) \subseteq \calR$. The envelope subspace is denoted by $\E_M(U)$. 
\end{defn}

The subspace $\E_M(U)$ is a small targeted part of the spectral structure of $M$ which contains the span of $U$. Denote $u, 0 \leq u \leq p$, as the dimension of $\E_M(U)$. All things being equal, a smaller value of $u$ indicates that stronger inferences can be obtained by taking advantage of the envelope structure when $U$ and $M$ are a parameter of interest and a covariance matrix respectively. 

To see the benefit of envelope methodology, first consider the case when $\E_M(U)$ is known. Let $\Gamma \in \R^{p \times u}$, $u < p$, be a known semi-orthogonal basis matrix for $\E_M(U)$. Let $U = \theta\theta^T$ and  $M = \Sigma$ and suppose that we have $\rootn$-consistent $\Ttil$ as in the initial setup \eqref{context}. The envelope estimator of $\theta$ is then $P_{\Gamma}\Ttil$ where $P_{\Gamma} = \Gamma\Gamma^T$. Notice that, 
\begin{align*}
  &\rootn\left(P_{\Gamma}\Ttil - \theta\right) 
    = P_{\Gamma}\left\{\rootn\left(\Ttil - \theta\right)\right\} 
    \overset{d}{\to} N\left(0, P_{\Gamma}\Sigma P_{\Gamma}\right), \\
  &\rootn\left(Q_{\Gamma}\Ttil\right) 
    =Q_{\Gamma}\left\{\rootn\left(\Ttil - \theta\right)\right\}
    \overset{d}{\to} N\left(0, Q_{\Gamma}\Sigma Q_{\Gamma}\right),
\end{align*}
where $Q_{\Gamma} = I - P_{\Gamma}$ and $Q_{\Gamma}\theta = 0$ by definition. In this demonstration we see that $P_{\Gamma}\Ttil$ consistently estimates $\theta$ with less asymptotic variability than that of the original estimator. The remaining piece of $\Ttil$, given by $Q_{\Gamma}\Ttil$, is a $\rootn$ consistent estimator of $0$ with non-trivial asymptotic variability given by $Q_{\Gamma}\Sigma Q_{\Gamma}$. Therefore $P_{\Gamma}\Ttil$ is all that is needed to estimate $\theta$ while $Q_{\Gamma}\Ttil$ produces extra variability that is nonessential to the estimation of $\theta$. Let $\|\cdot\|$ denote the spectral matrix norm. Envelope methodology leads to massive efficiency gains in settings where $\|Q_{\Gamma}\Sigma Q_{\Gamma}\| \gg \|P_{\Gamma}\Sigma P_{\Gamma}\|$. When one has the true $u$ but needs to estimate all other quantities then the asymptotic variability of the envelope estimator $P_{\Gamma}\Sigma P_{\Gamma}$ incurs an additional cost resulting from said estimation, see Section 5.2 of \citet{cook2010} and Section 3.3 of \citet{found} for specific examples.

In practical settings, all envelope modeling quantities, including $u$, require estimation. In such general cases where the likelihood function need not be known, \citet{zhangmai} proposed to estimate a semi-orthogonal basis matrix $\Gamma \in \R^{p \times u}$ of $\E_M(U)$ by minimizing the generic moment-based objective function:
\begin{equation} \label{Jn}
  J_n(\Gamma) = \log\mid \Gamma^T\Mhat\Gamma \mid 
    + \log\mid \Gamma^T(\Mhat + \Uhat)^{-1}\Gamma \mid,
\end{equation}
where $\Mhat$ and $\Uhat$ are $\sqrt{n}$-consistent estimators of $M$ and $U$. The motivations for \eqref{Jn} comes its population counterpart, $J(\Gamma) = \log\mid \Gamma^TM\Gamma \mid + \log\mid \Gamma^T(M + U)^{-1}\Gamma \mid$, where \citet{algo} showed that any $\Gamma$ which minimizes $J(\Gamma)$ must satisfy the envelope condition that span$(U) \subseteq \text{span}(\Gamma)$. Assuming that the true envelope dimension $u$ is supplied, the $\rootn$-consistency of the estimated envelope estimator that is constructed from a minimizer of \eqref{Jn} is established in \citet{algo}. The true envelope dimension is not known in practice.

\section{Weighted envelope methodology}

We introduce model-free weighted envelope estimation that offers a balance between variance reduction and model misspecification in finite samples. The idea is to weigh each envelope estimator evaluated at different candidate dimensions with respect to a measure of fit of the candidate envelope dimension. The weighted envelope estimators that we propose are of the form
\begin{equation} \label{genwtenv}
  \hat{\theta}_w = \sum_{k=0}^p w_k\hat{\theta}_k, 
    \qquad w_k = f_k(\In(1),\ldots, \In(p)),
\end{equation}
where $\hat{\theta}_k$, $\In(k)$, $f_k$ are, respectively, the envelope estimator, an information criteria that assesses the fit of the envelope dimension, and a function of all information criteria at proposed dimension $k$, and $n$ is the sample size. As is standard in model averaging, we require that $f_k$ and $\In(k)$ be chosen so that $w_k \geq 0$ and $\sum_k w_k = 1$. However, unlike typical model averaging contexts, $\hat{\theta}_k$ is consistent for all weight choices that satisfy $\sum_{k=u}^p w_k \to 1$ where $w_k \geq 0$ for $k \geq u$. Such weight choices induce a consistent estimator since the envelope estimator $\hat{\theta}_k$ is a consistent estimator for $\theta$ for all $k \geq u$. It is of course more desirable to select $\In(k)$ and $f_k$ so that $w_u \to 1$.

In this paper, we study two choices of $\In(\cdot)$ and one choice of $f_k$ so that $w_u \to 1$ 
at a fast enough rate to facilitate reliable estimation of the variability of $\hat{\theta}_u$, 
$u$ unknown, via a nonparametric bootstrap. The weights will be of the form 
\begin{equation} \label{genwtenvw}
  w_k = \frac{\exp\left\{-n\In(k)\right\}}{\sum_{j=0}^p\exp\left\{-n\In(j)\right\}}.
\end{equation}
The choice of $f_k$ that yields the weights \eqref{genwtenvw} is motivated by \cite{eck2017weighted}, where the choice of $\In(k)$ in \cite{eck2017weighted} reflected the BIC value of a multivariate linear regression model with the envelope estimator at dimension $k$ plugged in. The two choices of $\In(k)$ that we study here facilitate weighted envelope estimation within the general envelope estimation context \eqref{context}. The first choice of $\In(k)$, denoted $\IFG(k)$, allows for consistent estimation of the variability of $\hat{\theta}_u$ using the nonparametric bootstrap. This is achieved by setting $\IFG(k) = J_n(\hat{\Gamma}) + \text{pen}(k)$ and showing that 1) $J_n(\hat{\Gamma})$ can be cast a quasi-likelihood objective function that is optimized via a full Grassmannian envelope optimization routine \citep{zhangmai}; 2) this partially optimized quasi-likelihood objective function can be cast as an M-estimation problem. The second choice of $\In(k)$, denoted $\IoneD(k)$, corresponds to sequential 1 dimension (1D) optimization \citep{algo, found, zhangmai}. The choice $\IoneD(k)$ does not facilitate the same consistent estimation of the variability of $\hat{\theta}_u$. However, we demonstrate that the choice $\IoneD(k)$ allows for reliable estimation of the variability of the envelope estimator that is estimated using the 1D algorithm at the true $u$. Both our simulations, and the simulations presented in \citet{zhangmai} find that the choice of $\IoneD(k)$ exhibits greater empirical variance reduction than the choice of $\IFG(k)$. Moreover, the 1D optimization routine is faster, more stable, and less sensitive to initial values than the Full Grassmannian approach \citep{zeng2019TRES}.

\subsection{Weighted envelope estimation via quasi-likelihood optimization}
\label{section:FG}

In this section we construct $\hat{\theta}_w$ in \eqref{genwtenv} where the information criteria $\IFG(k)$ is derived from full Grassmannian optimization of the objective function \eqref{Jn}. The minimizer $\widehat{\Gamma}$ of the objective function \eqref{Jn} is the estimated basis of the envelope subspace at dimension $u$. After obtaining $\widehat{\Gamma}$, the envelope estimator is $\TFG = \widehat{\Gamma}\widehat{\Gamma}^T\Ttil$ where the superscript FG denotes envelope estimation with respect to full Grassmannian optimization. This envelope estimator is the original estimator projected into the estimated envelope subspace. When $u$ is known, $\TFG_u$ is $\rootn$-consistent and has been shown to have lower variability than $\Ttil$ in finite samples \citep{cook2010, found, cook2018introduction}. The weighted envelope estimator corresponding to $\IFG(k)$ is  
\begin{equation} \label{envFG}
  \EnvwFG = \sum_{k=0}^p \wFG_k\TFG_k, 
    \qquad \wFG_k = \frac
  {
    \exp\left\{-n\IFG(k)\right\}
  }
  {
    \sum_{j=0}^p\exp\left\{-n\IFG(j)\right\}
  },
\end{equation}
where $\TFG_k$ is the envelope estimator of $\theta$ constructed from full Grassmannian optimization of $J_n(\Gamhat_k)$ at dimension $k$. In the remaining part of this Section we make $\IFG(k)$ and demonstrate how our choices supplement Section~\ref{sec:bootFG} to yield consistent estimation of the variability of $\TFG_u$.

\citet{zhangmai} showed that optimization of $J_n$ in \eqref{Jn} is the same as optimization of a partially minimized quasi-likelihood function. Define this quasi-likelihood function as, 
\begin{equation} \label{ln}
  l_n(M, \theta) = \log\mid M\mid + \tr\left[M^{-1}\left\{\Mhat 
    + (\Ttil - \theta)(\Ttil - \theta)^T\right\}\right], 
\end{equation}
and, for some candidate dimension $k = 1$, $\ldots$, $p$, define the constraint set for the minimization of \eqref{ln} to be,
\begin{equation} \label{Ak}
  \A_k = \left\{(M,\theta) : 
    M = \Gamma\Omega\Gamma^T + \Gamma_o\Omega_o\Gamma_o^T > 0, \; 
    \theta = \Gamma\eta, \; \eta \in \R^k, \; 
    (\Gamma, \Gamma_o)^T(\Gamma, \Gamma_o) = I_p
  \right\}.
\end{equation}
Minimization of \eqref{ln} over the constraint set \eqref{Ak} is the same as minimizing $J_n$ in \eqref{Jn}. 

\begin{lem} \label{lem1:lemma1}
\citep[Lemma 3.1]{zhangmai}. The minimizer of $l_n(M,\theta)$ in \eqref{ln} under the envelope parameterization \eqref{Ak} is 
$
  \widehat{M}_{\text{Env}} 
    = \Gamhat\Gamhat^T\Omhat\Gamhat\Gamhat^T 
      + \Gamhat_o\Gamhat_o^T\Omhat_o\Gamhat_o\Gamhat_o^T
$
and $\That = \Gamhat\Gamhat^T\Ttil$ where $\Gamhat$ is the minimizer of the partially optimized objective function 
$
  l_n(\Gamma) = \min_{\Omega,\Omega_o,\eta} l_n(\Gamma,\Omega,\Omega_o,\eta) 
    = J_n(\Gamma) + \log\mid \widehat{M} + \widehat{U} \mid + p
$
where $\widehat{U} = \Ttil\Ttil^T$. 
\end{lem}

We show that optimization of the quasi-likelihood $l_n(M, \theta)$ can be cast as an M-estimation problem when $\Mhat = n^{-1}\sum_{i=1}^n h(\X_i, \Ttil)$ provided that $\Ttil$ is an optimal solution to some objective function, as in the same vein as \citet[pg. 29]{stefanski2002M}. The justification that a nonparametric bootstrap procedure will consistently estimate the variability of $\TFG_u$ follows from bootstrap theory for M-estimators in Section 2 of \cite{andrews}. Therefore recasting \eqref{Jn} and \eqref{ln} as an M-estimation problem is an important theoretical consideration for our proposed methodology. The requirement that $\Mhat = n^{-1}\sum_{i=1}^n h(\X_i, \Ttil)$ is mild and it holds in linear regression, maximum likelihood estimation, and M-estimation. Our parameterization of $\Mhat$ gives, 
\begin{align*}
  l_n(M, \theta) &= \log\mid M \mid + \tr\left[M^{-1}\left\{\Mhat 
    + (\Ttil - \theta)(\Ttil - \theta)^T\right\}\right] \\
  &= \log\mid M \mid + \tr\left[n^{-1}\sum_{j=i}^n M^{-1}\left\{
      h(\X_i, \Ttil) + (\Ttil - \theta)(\Ttil - \theta)^T\right\} 
    \right] \\
  &= \log\mid M \mid + n^{-1}\sum_{i=1}^n\tr\left[M^{-1}\left\{
      h(\X_i, \Ttil) + (\Ttil - \theta)(\Ttil - \theta)^T\right\} 
    \right] \\  
  &= n^{-1}\sum_{i=1}^n\left(\log\mid M \mid + \tr\left[M^{-1}\left\{ 
      h(\X_i, \Ttil) + (\Ttil - \theta)(\Ttil - \theta)^T\right\} 
    \right]\right) \\ 
  &= n^{-1}\sum_{i=1}^n f(\X_i, \theta, M).  
\end{align*}
Lemma~\ref{lem1:lemma1} then gives, 
\begin{equation} \label{Mest}
  l_n(\Gamma) = \min_{\Omega,\Omega_o,\eta} n^{-1}
    \sum_{i=1}^n f(\X_i, \Gamma, \Omega, \Omega_o, \eta) 
  = J_n(\Gamma) + \log\mid \Mhat + \Uhat \mid + p,
\end{equation}
where the minimization takes place over $\A_k$. The minimization of $n^{-1}\sum_{i=1}^n f(\X_i, \theta, M)$ over $\A_k$ provides estimates for both $\Gamma$ and $\eta$ and yields $\theta = \Gamma\eta$ and $M = \Gamma\Omega\Gamma^T + \Gamma_o\Omega_o\Gamma_o^T$. The proof of Lemma 3.1 in \citet{zhangmai} reveals that $\TFG_u = \widehat{\Gamma}\widehat{\Gamma}^T\Ttil$. In Section~\ref{sec:bootFG} we develop a nonparametric bootstrap to consistently estimate the variability of $\TFG_u$. 
Now define
\begin{equation} \label{FGcrit}
  \IFG(k) = J_n(\Gamhat_k) + \frac{Ck\log(n)}{n}, \qquad (k = 0,1,\ldots p), 
\end{equation}
as in \cite{zhangmai} where $C > 0$ is a constant and $\IFG(0) = 0$. The envelope dimension is selected as $\uFG = \text{arg}\min_{0\leq k\leq p} \In(k)$. Theorem 3.1 in \citet{zhangmai} showed that $\Prob(\uFG = u) \to 1$ as $n \to \infty$, provided that $C > 0$ and $\Mhat$ and $\Uhat$ are $\rootn$-consistent estimators of $M$ and $U$ respectively. 

\cite{zhangmai} provided evidence that selecting $\uFG$ in this manner leads to envelope estimators with lower variability than traditional estimators and that the correct dimension selection improves in $n$. However, correct dimension selection was far from perfect in their simulations, especially for small and moderate sample sizes. Moreover, dimension selection is being ignored in these simulations. We use $\IFG(k)$ in \eqref{FGcrit} to construct $\EnvwFG$ in \eqref{envFG}. This construction yields consistent estimation of $\theta$ as seen in Section~\ref{consisprop} and consistent estimation of the variability of $\TFG_u$ through the combination of Theorem~\ref{thm:TFG} and our formulation of $l_n(M, \theta)$ as an M-estimation problem.

\subsection{Weighted envelope estimation via the 1D algorithm}
\label{section:1D}

In this section we construct $\hat{\theta}_w$ in \eqref{genwtenv} where the information criteria $\IoneD(k)$ is derived from the 1D algorithm \citep{found, algo}. The 1D algorithm performs a sequence of optimizations that each return a basis vector of the envelope space (in the population) or a $\rootn$-consistent estimator of a basis vector for the envelope space (in finite-samples). The number of optimizations corresponds to the dimension of the envelope space and is provided by the user. The returned envelope estimator is $\TD_u = \widehat{\Gamma}\widehat{\Gamma}\Ttil$ where the estimated basis matrix $\widehat{\Gamma}$ is obtained from the 1D algorithm. The weighted envelope estimator corresponding to $\_oneD(k)$ is 
\begin{equation} \label{env1D}
  \EnvwoneD = \sum_{k=0}^p \woneD_k\TD_k, 
    \qquad \woneD_k = \frac
  {
    \exp\left\{-n\IoneD(k)\right\}
  }
  {
    \sum_{j=0}^p\exp\left\{-n\IoneD(j)\right\}
  }.
\end{equation}
When $C = 1$ in \eqref{1Dcrit}, the terms $n\IoneD(k)$ are BIC values corresponding to the asymptotic log likelihood the envelope model of dimension $k$. 

The algorithm is as follows: Set $u_o \leq p-1$ to be the user inputted number of optimizations. For step $k = 0$, $\ldots$, $u_o$, let $g_k \in \R^p$ denote the $k$-th direction to be obtained by the 1D algorithm. Define $G_k = (g_1$, $\ldots$, $g_k)$, and $(G_k$, $G_{0k})$ to be an orthogonal basis for $\R^p$ and set initial value $g_0 = G_{0} = 0$. Define $M_k = G_{0k}^TMG_{0k}$, $U_k = G_{0k}^TU G_{0k}$, and the objective function after $k$ steps 
$$
  \phi_k(v) = \log(v^T M_k v) + \log\{v^T(M_k + U_k)^{-1} v\}.
$$
The $(k+1)$-th envelope direction is $g_{k+1} = G_{0k}v_{k+1}$ where $v_{k+1} = \text{argmax}_{v}\phi_k(v)$ subject to $v^Tv = 1$. 

In the population, the 1D algorithm produces a nested solution path that contains the true envelope:
$$
  \text{span}\left(G_1\right) \subset \text{span}\left(G_{u-1}\right) 
  \subset \cdots \subset 
  \text{span}\left(G_u\right) = \mathcal{E}_M(U) 
  \subset \text{span}\left(G_{u+1}\right) \subset \cdots 
    \subset \text{span}\left(G_p\right) = \R^p.
$$
Replacing $M$ and $U$ with $\rootn$-consistent estimators $\Mhat$ and $\Uhat$ yields $\rootn$-consistent estimates $\Ghat_k = \left(\ghat_1, \ldots, \ghat_k\right) \in \R^{p \times k}$, $k = 1$, $\ldots$, $p$ by optimizing
$$
  \phi_{k,n}(v) = \log(v^T \Mhat_k v) + \log\{v^T(\Mhat_k + \Uhat_k)^{-1} v\}. 
$$
The resulting envelope estimator $\TD_u$ is therefore $\rootn$-consistent \citep{found}.

\citet{zhangmai} proposed a model selection criterion to estimate $u$ in practical settings. This criterion is, 
\begin{equation}
  \IoneD(k) = \sum_{j=1}^k \phi_{j,n}(\hat{v}_j) + \frac{Cj\log{n}}{n}, 
  \qquad 
  (k = 0, \ldots, p),
\label{1Dcrit}
\end{equation}
where $C > 0$ is a constant and $\IoneD(0) = 0$. The envelope dimension selected is given by $\uoneD = \text{arg}\min_{0\leq k\leq p} \IoneD(k)$. Theorem 3.2 in \citet{zhangmai} showed that $\Prob(\uoneD = u) \to 1$ as $n \to \infty$, provided that $C > 0$ and $\Mhat$ and $\Uhat$ are $\rootn$-consistent estimators of $M$ and $U$ respectively. We use $\IoneD(k)$ in \eqref{1Dcrit} to construct $\EnvwoneD$ in \eqref{env1D}. This construction yields consistent estimation of $\theta$ as seen in Section~\ref{consisprop} and allows for reliable estimation of the variability of $\TD_u$.

\subsection{Consistency properties of weighted envelope estimators}
\label{consisprop}

Weighted envelope estimators exhibit desirable consistency properties. First of all, the weights in \eqref{envFG} and \eqref{env1D} can be constructed so that they both satisfy $\wFG_u \to 1$ and $\woneD_u \to 1$ as $n \to \infty$. 

\begin{lem}
For any constant $C > 0$ and $\sqrt{n}$-consistent $\Mhat$ and $\Uhat$ in \eqref{FGcrit}, $\wFG_u \to 1$ as $n \to \infty$.
\label{lem:weightsFG}
\end{lem}

\begin{lem}
For any constant $C > 0$ and $\sqrt{n}$-consistent $\Mhat$ and $\Uhat$ in \eqref{1Dcrit}, $\woneD_u \to 1$ as $n \to \infty$.
\label{lem:weights1D}
\end{lem}

The proofs of both Lemmas are included in the Supplementary Materials. Lemmas~\ref{lem:weightsFG} and \ref{lem:weights1D} facilitate consistent estimation of $\theta$ using $\EnvwFG$ and $\EnvwoneD$.

\begin{lem}
For any constant $C > 0$ and $\sqrt{n}$-consistent $\Mhat$ and $\Uhat$ in \eqref{FGcrit} and \eqref{1Dcrit}, both $\EnvwFG \to \theta$ and $\EnvwoneD \to \theta$ as $n \to \infty$.
\label{lem:envcon}
\end{lem}

The proof of Lemma~\ref{lem:envcon} immediately follows from Lemmas~\ref{lem:weightsFG} and \ref{lem:weights1D} and \citet[Proposition 2.1]{zhangmai}. While consistency is desirable, Lemma~\ref{lem:envcon} does not provide knowledge about the asymptotic variability of $\EnvwFG$ or $\EnvwoneD$. Therefore, these estimators do not offer any assurance of variance reduction via model-free weighted envelope estimation. We expect that $\EnvwFG$ and $\EnvwoneD$ will have lower asymptotic variance than $\Ttil$ when $u < p$, but explicit computations of the asymptotic variance for both estimators are cumbersome. We will instead estimate the asymptotic variability of $\EnvwFG$ and $\EnvwoneD$ with a nonparametric bootstrap. Theoretical justification for these bootstrap procedures are provided in the next section.

\section{Bootstrapping for model-free weighted envelope estimators}
\label{sec:boot}

\subsection{Nonparametric bootstrap}

Let $\X_1$, $\ldots$, $\X_n$ be the original data. We will estimate the variability of $\TFG_u$ and $\TD_u$ by bootstrapping with respect to the weighted estimators $\TFG_w$ and $\TD_w$. We will also show that bootstrapping with respect to $\TFG_w$ can consistently estimate the variability of $\TFG_u$.  

For each iteration of this nonparametric bootstrap procedure we denote the resampled data by $\Xstar_1$, $\ldots$, $\Xstar_n$ where each $\Xstar_i$, $i = 1$, $\ldots$, $n$ is sampled, with replacement, from the original data with equal probability $1/n$. Define the bootstrapped envelope estimators $\TstarFG = \TFG\left(\Xstar_1, \ldots, \Xstar_n\right)$, $\TstarD = \TD\left(\Xstar_1, \ldots, \Xstar_n\right)$, and the bootstrapped version of the original estimator $\Ttil$ as $\Ttilstar = \Ttil(\Xstar_1, \ldots \Xstar_n)$. Furthermore, define $\Mstar$ and $\Ustar$ in the same manner as $\Mhat$ and $\Uhat$ with respect to the starred data.

\subsection{For quasi-likelihood optimization}
\label{sec:bootFG}

In this section we provide justification for the nonparametric bootstrap as a method to estimate the variability of $\TFG_w$. Define 
\begin{equation} \label{Jnstar}
  \Jstar_n(\Gamma) = \log\mid\Gamma^T\Mstar\Gamma\mid 
    + \mid\Gamma^T\left(\Mstar + \Ustar\right)^{-1}\Gamma\mid, 
\end{equation}
as the starred analog to $J_n$ in \eqref{Jn} and define, 
\begin{equation} \label{lnstar}
  \lstar_n(M, \theta) = \log\mid M\mid + \tr\left[M^{-1}\left\{\Mstar  
    + (\Ttilstar - \theta)(\Ttilstar - \theta)^T\right\}\right], 
\end{equation}
as the starred analog to $l_n$ in \eqref{ln}. Define $\Gamstar$ as the minimizer to \eqref{Jnstar}. When $\Mstar = n^{-1}\sum_{i=1}^nh(\Xstar_i, \Ttilstar)$ both Lemma~\ref{lem1:lemma1} and our likelihood derivation in Section~\ref{section:FG} give, 
$$
  \lstar_n(\Gamma) = \min_{\Omega,\Omega_o,\eta} n^{-1}
    \sum_{i=1}^n f(\Xstar_i, \Gamma, \Omega, \Omega_o, \eta) 
  = \Jstar_n(\Gamma) + \log\mid\Mstar + \Ustar\mid + p, 
$$
which is the starred analog to \eqref{Mest}. Thus $\Gamstar$ is an M-estimator, being the minimizer of the partially minimized objective function $\lstar_n(\Gamma)$. We then let $\EnvuFGstar = \Gamstar\GamstarT\Ttilstar$ where $\Ttilstar$ is obtained in the same minimization of $\sum_{i=1}^n f(\Xstar_i, \Gamma, \Omega, \Omega_o, \eta)$. The envelope estimator $\EnvuFGstar$ is a product of M-estimators obtained from the same objective function. Therefore we can use the nonparametric bootstrap to consistently estimate the variability of $\TFG_u$ \citep[Section 2]{andrews}. The problem with this setup is that $u$ is unknown and requires estimation. We show that bootstrapping with respect to our weighted envelope estimator consistently estimates the variability of the envelope estimator $\TFG_u$ at the true unknown dimension when $\Mstar = n^{-1}\sum_{i=1}^nh(\Xstar_i, \Ttilstar)$.

\begin{thm}
Let $\Ttil$ be a $\rootn$-consistent and asymptotically normal estimator. Let $\TFG_k$ be the envelope estimator obtained from full Grassmannian optimization at dimension $k = 0$, $\ldots$, $p$ and let $\TFG_w$ be the weighted envelope estimator with weights $\wFG$. Let $\TstarFG_k$ and $\TstarFG_w$ denote the corresponding quantities obtained by resampled data. Then as $n$ tends to $\infty$,
\begin{equation}
\begin{split}
  \rootn\left(\TstarFG_w - \TFG_w\right) 
    = \rootn\left(\TstarFG_u - \TFG_u\right) 
    + O_P\left\{n^{\left(1/2 - C\right)}\right\}
    + O_P\left[n^{\left\{Cu + 1/2\right\}}\right]
      \exp\left\{-n|O_P(1)|\right\}.
\end{split}
\label{TFGterms}
\end{equation} 
\label{thm:TFG}
\end{thm}

\noindent {\bf Remarks}: 

\begin{itemize}
\item[1.] Theorem~\ref{thm:TFG} shows that our bootstrap procedure consistently estimates the asymptotic variability of $\TFG_u$ when $u$ is unknown. We see that the second $O_P$ term in \eqref{TFGterms} vanishes quickly in $n$. These terms are associated with under selecting the true envelope dimension. Therefore it is more likely that our bootstrap procedures will conservatively estimate the variability of $\TFG_u$ in finite samples.
\item[2.] We advocate for the case with $C = 1$ because of the close connection that $\IFG(k)$ has with the Bayesian Information Criterion, similar reasoning was given in \citet{zhangmai}. The $Ck\log(n)/n$ penalty term in $\IFG(k)$ facilitates the decaying bias in $n$ represented by the $O_P$ terms in \eqref{TFGterms}. Redefining $\IFG(k)$ to have a penalty term that is fixed in $n$, similar to that of Akaike Information Criterion, fundamentally changes the $O_P$ terms in \eqref{TFGterms}. Specifically, the $O_P(n^{-1/2})$ term (when $C = 1$) disappears and the weights $\wFG_k$ fail to vanish for $k > u$. Therefore unknown non-zero asymptotic weight is given to candidate models with dimension $k > u$. Weighting in this manner is therefore suboptimal and is not advised.
\item[3.] The weights $\wFG_k$ have a similar form to the weights which appear in the model averaging literature \citep{buckland, burnham, hjort, claeskens, tsague}. These weights are of the form
\begin{equation} \label{weights-post}
  w_k = \frac
    {
      \exp\left\{-n\IFG(k)/2\right\}
    }
    {
      \sum_{j=0}^p\exp\left\{-n\IFG(p)/2\right\}
    }
\end{equation}
and they correspond to a posterior probability approximation for model $k$ under the prior that assigns equal weight to all candidate models, given the observed data. The weights \eqref{weights-post} do not have the same asymptotic properties as our weights. The difference between the two is a rescaling of $C$. Weights \eqref{weights-post} replace the constant $C$ in \eqref{TFGterms} with $C/2$. When $C = 1$, nonzero asymptotic weight would be placed on the envelope model with dimension $k = u + 1$. Therefore, weighting according to \eqref{weights-post} leads to higher estimated variability than is necessary in practice. 
\end{itemize}

\subsection{For the 1D algorithm}

In this section we provide justification for the nonparametric bootstrap as a method to estimate the variability of $\TD$. It is important to note that the M-estimation argument which justified using $\TstarFG_u$ as a consistent estimator for the variability of $\TFG_u$ does not hold here, we cannot cast the 1D objective function $\phi_k(v)$ as an M-estimation problem. The best we can do is verify that bootstrapping with respect to $\TD_w$ is asymptotically the same as bootstrapping with respect to $\TD_u$. We first define the quantities of the 1D algorithm applied to the starred data.

For step $k = 0$, $\ldots$, $p-1$ of the 1D algorithm applied to starred data, let $\gstar_k \in \R^p$ denote the $k$-th direction to be obtained. Define $\Gstar_k = (\gstar_1$, $\ldots$, $\gstar_k)$, and $(\Gstar_k$, $\Gstar_{0k})$ to be an orthogonal basis for $\R^p$ and set initial value $\gstar_0 = \Gstar_{0} = 0$. Define $\Mstar_k = \GstarT_{0k}\Mstar\Gstar_{0k}$, $\Ustar_k = \GstarT_{0k}\Ustar\Gstar_{0k}$, and the objective function after $k$ steps 
$$
  \phistar_{k,n}(v) = \log(v^T \Mstar_k v) 
    + \log\{v^T(\Mstar_k + \Ustar_k)^{-1} v\}.
$$
The $(k+1)$-th envelope direction is $\gstar_{k+1} = \Gstar_{0k}\vstar_{k+1}$ where $\vstar_{k+1} = \text{argmax}_{v^Tv}\phistar_k(v)$. The estimated projection into envelope space is then $\Pstar = \Gstar\GstarT$. We then arrive at the bootstrapped envelope estimator $\TstarD_u = \Pstar_u\Ttilstar$. We show that bootstrapping $\TD_w$ estimates the variability of the envelope estimator $\TD_u$ at the true unknown dimension.

\begin{thm}
Let $\Ttil$ be a $\rootn$-consistent and asymptotically normal estimator. Let $\TD_k$ be the envelope estimator obtained from the 1D algorithm at dimension $k = 1$, $\ldots$, $p$ and let $\TD_w$ be the weighted envelope estimator with weights $\woneD$. Let $\TstarD_k$ and $\TstarD_w$ denote the corresponding quantities obtained by resampled data. Then as $n$ tends to $\infty$,
\begin{equation}
\begin{split}
  \rootn\left(\TstarD_w - \TD_w\right) 
    = \rootn\left(\TstarD_u - \TD_u\right) 
    + O_P\left\{n^{\left(1/2 - C\right)}\right\}
    + O_P\left[n^{\left\{Cu + 1/2\right\}}\right]
      \exp\left\{-n|O_P(1)|\right\}.
\end{split}
\label{TDterms}
\end{equation} 
\label{thm:TD}
\end{thm}

The remarks to Theorem~\ref{thm:TD} are similar to those for Theorem~\ref{thm:TFG}. We see that the second $O_P$ term in \eqref{TDterms} vanishes quickly in $n$. These terms are associated with under selecting the true envelope dimension. Therefore it is more likely that our bootstrap procedures will conservatively estimate the variability of $\TD_u$ in finite samples. As previously mentioned, we advocate for the case with $C = 1$ because of the close connection that $\IoneD(k)$ has with the Bayesian Information Criterion, and we note that the weights $\woneD$ have a similar form to the weights which appear in the model averaging literature with similar weights as those in \eqref{weights-post}. Manuals for available software recommend use of one-directional optimizations, such as the 1D algorithm or the ECD algorithm \citep{cook2018fast}, because they are faster, stable, and less sensitive to initial values \citep{zeng2019TRES}.

\section{Examples}

\subsection{Exponential family generalized linear models (GLMs) simulations} 

\citet{found} showed that envelope estimation can provide variance reduction for parameter estimation in exponential family GLMs when predictors are normally distributed. We demonstrate that model free weighted envelope estimation can also achieve variance reduction in this context while accounting for model selection variability. Model free envelope estimation techniques and maximum likelihood estimation are then used to estimate the canonical parameter vector corresponding to GLMs (the regression coefficient vector with canonical link function). Estimation is performed using functionality in the \texttt{TRES} R package \citep{zeng2019TRES}. Following the recommendations in the \texttt{TRES} R package manuals, we use the 1D algorithm. We will therefore compare the performance of $\TD_w$ to $\TD_{\hat{u}_{\text{1D}}}$, the envelope estimator of $\theta$ evaluated at estimated dimension $\hat{u}_{\text{1D}}$. A nonparametric bootstrap with sample size $5000$ is then used to estimate the variability of these estimators. Our bootstrap simulation will consider two model selection regimes for obtaining $\hat{u}_{\text{1D}}$. In one regime, we estimate the envelope dimension at every iteration of the bootstrap (variable $u$ regime, estimated dimension denoted as $\hat{u}_{\text{1D}}^{*}$). In the other regime, we estimate the dimension of the envelope space in the original data set and then condition on this estimated dimension as if it were the true dimension (fixed $u$ regime, estimated dimension denoted as $\hat{u}_{\text{1D}}$). The fixed $u$ regime ignores the variability associated with model selection. Theorem~\ref{thm:TD} provides some guidance for the performance of the nonparametric bootstrap for estimating the variability of $\TD_w$. A formal analog does not exist for the other envelope estimators, although empirical evidence in \citet{zhangmai} and \citet{eck2020aster} suggest that the variable $u$ regime will provide some robustness to variability in dimension selection.

We simulate four different exponential family GLM settings where $p = 8$ and the true envelope dimension is $u = 2$. These four settings are divided into two GLM regression models and two settings within these GLM regression models. The GLM models considered are the logistic and Poisson regression models. Within these models, one simulation setting is designed to be favorable to envelope modeling and one is not. Nonignorable model selection variability is present in all of these simulation settings. Predictors are generated $X \sim N(0, \Sigma_X)$, where $\Sigma_X = \Gamma\Omega\Gamma^T + \Gamma_0\Omega_0\Gamma_0^T$. We construct the canonical parameter vector (regression coefficient vector as $\theta = \Gamma\Gamma^T v$ where $\Gamma$ and $v$ are provided in the supplement materials. In the logistic regression simulations we generate $Y_i \sim \text{Bernoulli}(\text{logit}(\theta^T X_i))$, and in the Poisson simulations we generate $Y_i \sim \text{Poisson}(\exp(\theta^T X_i))$. Our logistic regression simulation settings are Setting A: $\Omega$ has diagonal elements 2 and 3, $\Omega_0$ has diagonal elements $\exp(-4), \exp(-3), \ldots, \exp(1)$; Setting B: $\Omega$ has diagonal elements -4 and -5, $\Omega_0$ has diagonal elements $\exp(-3), \exp(-2), \ldots, \exp(2)$. Our Poisson regression simulation settings are Setting A: $\Omega$ has diagonal elements 1 and 10, $\Omega_0$ has diagonal elements $\exp(-6), \exp(-5), \ldots, \exp(-1)$; Setting B: $\Omega$ has diagonal elements $-3$ and $-2$, $\Omega_0$ has diagonal elements $\exp(-4), \exp(-3), \ldots, \exp(1)$.

In both logistic and Poisson models, the configurations of $\Omega$ and $\Omega_0$ is designed to be favorable (unfavorable) to envelope estimation in setting A (setting B). Ratios of bootstrap standard deviations for estimators of the first component of the canonical parameter vector across all simulation settings are depicted in Table~\ref{Tab:GLMratios}. These ratios are of the form $r(\widetilde{\theta},\TD_w) = \widehat{\text{sd}}^{*}(\widetilde{\theta})/\widehat{\text{sd}}^{*}(\TD_w)$
where the standard deviations $\widehat{\text{sd}}^{*}(\widetilde{\theta})$ and $\widehat{\text{sd}}^{*}(\TD_w)$ are the element in the first row and the first column of 
$$
	\left(\frac{1}{B}\sum_{b=1}^B(\Ttilstar_b - \widetilde{\theta})
    	(\Ttilstar_b - \widetilde{\theta})^T\right)^{1/2}, 
    \qquad
    \left(\frac{1}{B}\sum_{b=1}^B(\TstarD_{w_b} - \TD_w)
    	(\TstarD_{w_b} - \TD_w)^T\right)^{1/2}, 
$$ 
respectively, and $B$ is the bootstrap sample size. Envelope estimation provides variance reduction in all settings. However, notice that the fixed $u$ regime exhibits erratic performance across $n$. This is due to the large variability in the estimated dimension across $n$, details of which are in the Supplementary Materials. Weighted envelope estimation and envelope estimation under the variable $u$ regime perform similarly with the weighted estimator performing slightly better. These simulations demonstrate the utility of weighted envelope estimation in the presence of model selection variability. Similar results are observed for other components of the canonical parameter vector as seen in the Supplementary Materials.

\begin{table}
\centering
\begin{tabular}{llccc|ccc}  
   & & \multicolumn{3}{c}{Setting A} & \multicolumn{3}{c}{Setting B} \\
 model & $n$ & $r(\widetilde{\theta},\TD_w)$ & 
 $r(\widetilde{\theta},\TD_{\hat{u}^{*}_{\text{1D}}})$ & 
   $r(\widetilde{\theta},\TD_{\hat{u}_{\text{1D}}})$ & 
   $r(\widetilde{\theta},\TD_w)$  &
   $r(\widetilde{\theta},\TD_{\hat{u}^{*}_{\text{1D}}})$ & 
   $r(\widetilde{\theta},\TD_{\hat{u}_{\text{1D}}})$ \\
 & 300  & 1.22 & 1.16 & 3.26 & 0.99 & 0.97 & 1.13 \\
 & 500  & 1.88 & 1.76 & 2.88 & 1.04 & 1.02 & 1.14 \\
Logistic & 750 & 1.03 & 0.95 & 3.64 & 1.29 & 1.27 & 1.48 \\
 & 1000 & 1.00 & 0.88 & 3.50 & 1.04 & 1.02 & 1.14 \\
\hline
 & 300  & 1.15 & 1.11 & 2.09 & 1.15 & 1.13 & 1.41 \\
 & 500  & 2.58 & 2.29 & 17.29 & 1.35 & 1.21 & 2.38 \\
Poisson & 750  & 3.81 & 3.48 & 61.88 & 1.12 & 1.10 & 1.17 \\
 & 1000 & 4.09 & 3.62 & 93.11 & 1.58 & 1.49 & 2.85 
\end{tabular}
\caption{Performance of envelope estimators of the first component of the regression coefficients for the logistic and Poisson regression models in comparison to the MLE. The first and fourth ratio columns display the ratio of the bootstrap standard deviation of the MLE to that of the weighted envelope estimator. The second and fifth ratio columns display the ratio of the bootstrap standard deviation of the MLE to that of the envelope estimator under the variable $u$ dimension selection regime. The third and sixth ratio columns display the ratio of the bootstrap standard deviation of the MLE to that of the envelope estimator under the fixed $u$ dimension selection regime. 
}
\label{Tab:GLMratios}
\end{table}

\subsection{Real data illustration}

Diabetes is a group of metabolic diseases associated with long-term damage, dysfunction, and failure of different organs, especially the eyes, kidneys, nerves, heart, and blood vessels \citep{american2010diagnosis}. In 2017 approximately 5 million adult deaths worldwide were attributable to diabetes; global healthcare expenditures on people with diabetes are estimated USD 850 billion \citep{cho2018idf}. Diabetes remains undiagnosed for an estimated 30\% of the people who have the disease \citep{heikes2008diabetes}. One way to address the problem of undiagnosed diabetes is to develop simple, inexpensive diagnostic tools that can identify people who are at high risk of pre-diabetes or diabetes using only readily-available clinical or demographic information \citep{heikes2008diabetes}. 

We examine the influence of several variables on a positive diagnosis of diabetes. We will let a positive diagnosis of diabetes be when an individual's hemoglobin percentage (also known as HbA1c) exceeds a value of 6.5\% \citep{world2011use}. We will consider an individual's height, weight, age, hip size, waist size, and gender, all of which are easy to measure, inexpensive, and do not require any laboratory testing, and a measure of their stabilized glucose as predictors for a positive diagnosis of diabetes. The data in this analysis come from a population-based sample of 403 rural African-Americans in Virginia \citep{willems1997prevalence}, and is taken from the \texttt{faraway} R package \citep{faraway2016R}. We considered a logistic regression model with response variable denoting a diagnosis of diabetes (1 when HbA1c $> 6.5\%$ and $0$ otherwise) that includes log transformed values for each continuous covariate and a main effect for gender. The log transformation was used to transform these variables to univariate normality while maintaining a scale that is interpretable. 

Model free envelope estimation techniques and maximum likelihood estimation are then used to estimate the canonical parameter vector corresponding to this logistic regression model. Estimation is performed using functionality in the \texttt{TRES} R package \citep{zeng2019TRES}. We will compare the performance of $\TD_w$ to $\TD_{\hat{u}_{\text{1D}}}$. A nonparametric bootstrap with sample size $5000$ is then used to estimate the variability of these estimators. Our bootstrap simulation will consider both the variable $u$ and fixed $u$ model selection regimes. Performance results are displayed in Table~\ref{Tab:diabetesperform}. We see that $\TD_w$ and $\TD_{\hat{u}_{\text{1D}}}$ are very similar to each other and both are very different than the MLE $\widetilde{\theta}$. Similarity of $\TD_w$ and $\TD_{\hat{u}_{\text{1D}}}$ follows from empirical weights $w_1 = 0.982$, $w_2 = 0.0176$, and $w_k \approx 0$ for all $3 \leq k \leq 7$. Also observe that the bootstrap standard deviation estimates vary across the model selection procedures. Most notably, the fixed $u$ regime provides massive variance reduction while the weighted estimator and variable $u$ regime provide similar modest but appreciable variance reduction. The variance reduction discrepancy between the fixed $u$ regime and the the weighted estimator and variable $u$ regime is due to large model selection variability. Specifically, the selected dimension probabilities across our nonparametric bootstrap procedure are $p(\hat{u}_{\text{1D}} = 1) = 0.568$, $p(\hat{u}_{\text{1D}} = 2) = 0.358$, $p(\hat{u}_{\text{1D}} = 3) = 0.067$, and $p(\hat{u}_{\text{1D}} = 4) = 0.007$. It is clear that unaccounted model selection variability may lead users astray when they use the fixed $u$ regime in estimating standard deviations via bootstrapping. This example shows how difficult it can be to report reliable variance reduction in practice, and how tempting it can be to ignore model selection variability.

\begin{table} 
\footnotesize
\centering       
\begin{tabular}{lcccccccccc}   
  & $\TD_w$ & $\widehat{\text{sd}}^{*}(\TD_w)$ & $\TD_{\hat{u}_{\text{1D}}}$ & 
   $\widehat{\text{sd}}^{*}(\TD_{\hat{u}^{*}_{\text{1D}}}$) &
   $\widehat{\text{sd}}^{*}(\TD_{\hat{u}_{\text{1D}}}$) &
   $\widetilde{\theta}$ & $\widehat{\text{sd}}^{*}(\widetilde{\theta}$) & 
   $r(\widetilde{\theta},\TD_w)$  &
   $r(\widetilde{\theta},\TD_{\hat{u}^{*}_{\text{1D}}})$ & 
   $r(\widetilde{\theta},\TD_{\hat{u}_{\text{1D}}})$ \\   
$\log(\text{Age})$ & 1.78 & 0.66 & 1.78 & 0.68 & 0.69 & 2.03 & 0.82 & 1.24 & 1.21 & 1.19 \\ 
$\log(\text{Weight})$ & 0.70 & 1.69 & 0.70 & 1.81 & 0.38 & 1.26 & 2.67 & 1.58 & 1.48 & 7.07 \\
$\log(\text{Height})$ & 0.03 & 4.39 & 0.03 & 4.73 & 0.09 & -4.39 & 5.07 & 1.16 & 1.07 & 54.76 \\   
$\log(\text{Waist})$& 0.68 & 1.93 & 0.68 & 2.06 & 0.26 & 2.65 & 2.98 & 1.54 & 1.45 & 11.60 \\
$\log(\text{Hip})$ & 0.49 & 3.26 & 0.49 & 3.47 & 0.24 & -2.64 & 4.23 & 1.30 & 1.22 & 17.80 \\   
$\text{Female}$ & 0.40 & 0.62 & 0.40 & 0.64 & 0.62 & 0.17 & 0.74 & 1.19 & 1.17 & 1.20 \\
$\log(\text{Stab. Gluc.})$ & 5.11 & 0.93 & 5.11 & 0.93 & 0.83 & 5.03 & 1.07 & 1.15 & 1.15 & 1.29
\end{tabular}
\caption{Performance of envelope estimates of the regression coefficients for the logistic regression of diabetes diagnosis on seven predictors. The first, third, and sixth column display the weighted envelope estimator, the envelope estimator with $\hat{u}_{\text{1D}} = 1$, and the MLE respectively. The second column displays the bootstrap standard deviation of the weighted envelope estimator. The fourth and fifth columns display the bootstrap standard deviation for the envelope estimator under the variable $u$ and fixed $u$ regimes respectively. The seventh column displays the bootstrap standard deviation of the MLE. The last three columns displays the ratio of bootstrap standard deviations of all envelope estimators to the those of the MLE.}
\label{Tab:diabetesperform}
\end{table}

\subsection{Reproducing and extending Monte Carlo simulations of \citet{zhangmai}}
\label{sec:repro}

Here, we compare the performance of $\TFG_w$ and $\TD_w$ to the consistent envelope estimators $\TFG_{\hat{u}_{\text{FG}}}$ and $\TD_{\hat{u}_{\text{1D}}}$ using the simulation settings in \citet{zhangmai}. For our first comparison we reproduce the Monte Carlo simulations in Section 4.2 of \citet{zhangmai} and add both $\TFG_w$ and $\TD_w$ to the list of estimators under comparison. Performance of all estimators at a sample size of $n = 75$ is also assessed. The data generating models that are considered are a single predictor linear regression model with 10 responses, a logistic regression model with a 10 predictors, and a Cox proportional hazards model with 10 predictors. In all three modeling setups, the true dimension of the envelope space is set at $u = 2$. In-depth details about this simulation setup are presented in \citet{zhangmai}. It is important to note $\TFG_{\hat{u}_{\text{FG}}}$ and $\TD_{\hat{u}_{\text{1D}}}$ are estimated according to the variable $u$ regime. The Monte Carlo sample size is 200, as in \citet{zhangmai}.

Table~\ref{tab:tab3inzhangmai} displays the results. From Table~\ref{tab:tab3inzhangmai} we see that the weighted envelope estimators perform very similarly to the consistent envelope estimators. This suggests that the variability in model selection is captured by all envelope estimators. This finding is expected in larger samples when the correct dimension selected percentage approaches 1, and it is a direct consequence of Lemmas~\ref{lem:weightsFG} and \ref{lem:weights1D} and Theorem 3.2 in \citet{zhangmai}. On the other hand, this finding is illuminating for sample sizes where the correct dimension selected percentages are nowhere near 1. Some variability in selection of $u$ which was used to construct both $\TFG_{\hat{u}_{\text{FG}}}$ and $\TD_{\hat{u}_{\text{1D}}}$ is incorporated into these simulations since $u$ is estimated at every iteration. 

\begin{table}
\begin{center}
\begin{tabular}{lccc|cccccc}
    & \multicolumn{3}{c}{Correct Selection \%} 
    & \multicolumn{5}{c}{Estimation Error $\|\hat{\theta} - \theta\|_F$} & \\
    & & & & Standard & \multicolumn{5}{c}{Envelope} \\
  Model  & $n$ & 1D & FG & & true $u$ & 1D & FG & W1D & WFG \\ 
  \hline
  		 &  75 & 74 & 63.5 & 0.69 & 0.50 & 0.55 & 0.55 & 0.54 & 0.55 \\
  Linear & 150 & 93 & 81 & 0.49 & 0.31 & 0.33 & 0.33 & 0.34 & 0.33 \\
         & 300 & 99 & 92 & 0.33 & 0.19 & 0.19 & 0.20 & 0.19 & 0.19 \\
  		 & 600 & 99 & 92.5 & 0.23 & 0.13 & 0.14 & 0.14 & 0.14 & 0.14 \\
  		 \hline
  		  &  75 & 22.5 & 42   & 4.04 & 1.04 & 1.06 & 1.00 & 1.09 & 1.08 \\
 Logistic & 150 & 72   & 77.5 & 2.16 & 0.56 & 0.67 & 0.60 & 0.67 & 0.64 \\
          & 300 & 92   & 89.5 & 1.40 & 0.34 & 0.35 & 0.34 & 0.37 & 0.36 \\
  		  & 600 & 98   & 94   & 0.98 & 0.22 & 0.22 & 0.24 & 0.24 & 0.24 \\
  		  \hline
 		  &  75 & 35   & 38   & 2.07 & 1.99 & 1.95 & 1.96 & 2.04 & 2.05 \\
      Cox & 150 & 57.5 & 53.5 & 1.33 & 1.24 & 1.21 & 1.22 & 1.27 & 1.28 \\
          & 300 & 83   & 75.5 & 0.98 & 0.90 & 0.89 & 0.90 & 0.93 & 0.93 \\
  		  & 600 & 100  & 93   & 0.79 & 0.72 & 0.72 & 0.72 & 0.75 & 0.75 
\end{tabular}
\end{center}
\caption{Table of Monte Carlo simulation results for different envelope estimators with respect to three different envelope models in the spirit of Table 3 from \citet{zhangmai}. Left panel includes percentages of correct selection for these envelope estimators. Right panel includes means and standard errors of $\|\hat{\theta} - \theta\|_F$ for the standard estimator and the envelope estimators with either true or estimated dimensions.}
\label{tab:tab3inzhangmai}
\end{table}

We now demonstrate the small sample performance of weighted envelope estimation in the linear and logistic cases of \citet{zhangmai}. As before, this simulation uses the exact specifications in \citet{zhangmai} which were not designed to showcase weighted envelope estimation techniques. We ignore the Cox proportional hazards model case because appreciable envelope estimation was not observed in the original Monte Carlo simulation. For this bootstrap procedure, we generated one data set corresponding to the linear and logistic regression models in the previous simulation at sample sizes $n = 75, 150, 300$. We then perform a nonparametric bootstrap to estimate the variability of each envelope estimator using a bootstrap sample size of 200 iterations. We repeat this process 25 times, and report the average ratios of standard deviations relative to the standard estimator across these 25 Monte Carlo samples. Note that estimates of $u$ are allowed to (and do) vary across the iterations of the 25 Monte Carlos samples. 

Table~\ref{tab:bootstrap} displays the results with respect to the first component of the parameter vector (other components behave similarly) in both regression settings. In Table~\ref{tab:bootstrap} we see that weighted envelope estimation provides larger variance reduction then given by $\TD_{\hat{u}_{\text{1D}}^*}$ and $\TFG_{\hat{u}_{\text{FG}}^*}$ and is comparable to oracle estimation in most settings. The estimators $\TFG_{\hat{u}_{\text{FG}}}$ and $\TD_{\hat{u}_{\text{1D}}}$ outperform weighted envelope estimation. However, this variance reduction is due to underestimation of $u$ in many of the original samples. Thus, weighted envelope estimation provides a desirable balance between model variance reduction and robustness to model misspecification.

\begin{table}
\small
\begin{center}
\begin{tabular}{lc|ccccccc}    
  Model  & $n$ & $r(\tilde{\theta},\hat{\theta}_u)$ &
  $r(\tilde{\theta},\hat{\theta}^{\text{1D}}_{\hat{u}_{\text{1D}}})$ &
  $r(\tilde{\theta},\hat{\theta}^{\text{FG}}_{\hat{u}_{\text{FG}}})$ &  
  $r(\tilde{\theta},\hat{\theta}^{\text{1D}}_{\hat{u}^{*}_{\text{1D}}})$ & 
  $r(\tilde{\theta},\hat{\theta}^{\text{FG}}_{\hat{u}^{*}_{\text{FG}}})$ &   
  $r(\tilde{\theta},\hat{\theta}^{\text{1D}}_w)$ & 
  $r(\tilde{\theta},\hat{\theta}^{\text{FG}}_w)$ \\  
  \hline
  		 & 75  & 0.992 & 2.024 & 1.768 & 0.991 & 0.947 & 1.094 & 1.024 \\
  Linear & 150 & 1.076 & 1.592 & 1.524 & 1.033 & 1.008 & 1.105 & 1.046 \\
  		 & 300 & 1.236 & 2.219 & 2.108 & 1.173 & 1.102 & 1.264 & 1.171 \\
  		 \hline
  		   &  75 & 1.013 & 1.054 & 1.022 & 0.978 & 0.966 & 1.079 & 1.033 \\
  Logistic & 150 & 1.548 & 2.741 & 2.459 & 1.231 & 1.008 & 1.374 & 1.079 \\
           & 300 & 4.525 & 7.338 & 5.738 & 1.331 & 1.003 & 1.450 & 1.042 \\
  		  \hline
\end{tabular}
\end{center}
\caption{Ratios of standard deviations for envelope estimators relative to the MLE.}
\label{tab:bootstrap}
\end{table}

\section{Discussion}

We proposed two weighted envelope estimators that properly account for model selection uncertainty in general envelope estimation settings. These estimators are a unification of the weighted envelope estimators proposed in \citet{eck2017weighted} which only account for model selection variability in the context of multivariate linear regression, and the generic algorithms (FG and 1D algorithms) in \citet{zhangmai} which provide consistent envelope dimension selection in general problems but have no finite-sample guarantees. Our weighted envelope estimators are theoretically justified, intuitive, and easy to implement. Our numerical examples show that our estimators possess desirable properties, especially when the sample size is prohibitively small for consistent envelope estimation techniques that do not properly account for variability in model selection.

\citet{efron} provided a double bootstrap procedure that aims to incorporate variability in model selection. Their formulation is applicable for exponential families and it has been applied to envelope methodology \citep{eck2020aster, eck2018supporting}. Useful variance reduction was found empirically in this context. Neither \citet{efron} or \citet{eck2020aster} provided formal asymptotic justification for the bootstrap procedures that are implemented within. In this paper we provide formal justification for the bootstrap procedures that are developed within. Moreover, our weighted envelope estimators are appropriate for a more general class of envelope models than either \citet{efron}, \citet{eck2020aster}, or \citet{eck2017weighted} can claim. The idea of a model free weighting of envelope estimators across all candidate dimensions extends to partial envelopes \citep{su}, inner envelopes \citep{su-inner}, scaled envelopes \citep{cook-scale}, predictor envelopes \citep{cook-scale-pls}, sparse response envelopes \citep{su2016sparse}, tensor response regression \citep{li2017parsimonious}, matrix-variate response regression \citep{ding2018matrix} which is explicitly mentioned in \cite{ding2018matrixsupplement}, and envelopes regression models with nonlinearity and heteroscedasticity \citep{zhang2020envelopes}.

One noted limitation of bootstrapping a weighted envelope estimation is that it can be computationally expensive, especially when $p$ is large \citep{yau2019hypothesis}. In such settings, we recommend investigating if the range of candidate dimensions can reasonably be reduced to a less computationally burdensome set of values or using the variable $u$ approach when estimating the envelope dimension at every iteration of the nonparametric bootstrap. Existing envelope software implements the former approach in the context of multivariate linear regression \citep{lee2019Renvlp}. Our simulations provide some empirical justification for the performance of the latter approach.

\citet{yau2019hypothesis} developed a novel hypothesis testing procedure with respect to the multivariate linear envelope model. They showed that model averaging as in \citet{eck2017weighted} is very successful and is comparable in performance to their proposed methodology. They dismissed the model averaging technique by saying, ``there is an intuitive justification for why the model average estimator is not that viable. We may recall that the original motivation for applying the envelope model is to achieve dimension reduction. When one obtains $\hat{\theta}_w$, it is true that this estimator accounts for the variability for selecting $u$, however, because all possible envelope models are involved in \eqref{envFG} and \eqref{env1D}, it becomes unclear which subspace is being projected to as a result.'' The motivation for envelope methodology is not to ``achieve dimension reduction,'' rather the motivation for envelope methodology is to increase efficiency in multivariate analyses without altering traditional objectives \citep[first sentence of page 1]{cook2018introduction}. Dimension reduction is at the core of envelope methodology, but it is just a means to an end for achieving useful variance reduction. The reporting of a specific subspace is not of foundational importance to practitioners seeking variance reduction, especially when there is both uncertainty in the subspace selected and its dimension. 

When there is uncertainty about the correct envelope dimension, model averaging with our weighted envelope estimator provides a desirable balance between massive variance reduction and correct model specification.

\section*{Acknowldegement}
The author would like to thank R. Dennis Cook, Forrest W. Crawford, Karl Oskar Ekvall, Dootika Vats, and Xin Zhang for valuable feedback that improved the presentation of this paper.

\section*{Supplementary Materials}
Supplementary materials are available with this paper. This supplement includes proofs of all technical results and it doubles as a fully reproducible technical report that makes all R based analyses transparent. The simulations in Section~\ref{sec:repro} are not included in the supplementary materials. These simulations are adopted from MatLab code that accompanied \citet{zhangmai}. This code is readily available upon request.

\newpage
\begin{center}
\huge{Supplementary Materials for "General model-free weighted envelope estimation"}
\end{center}

This supplement contains the proofs of all technical results that appear in the main text. It is also a fully reproducible technical report which makes the R based simulations and diabetes data analysis fully transparent. The Matlab based simulations adopted from \citet{zhangmai} are not fully reproduced here, however the code to produce those simulations is available upon request. This supplement begins with the proof of technical results followed by numerical examples and the R functions used to create these examples.

\section*{Proofs of technical results in main text}

\begin{proof}[Proof of Lemma 2]
Note that 
$$
  \wFG_k = \frac
  {
    \exp\left\{-n\IFG(k)\right\}
  }
  {
    \sum_{j=0}^p\exp\left\{-n\IFG(j)\right\}
  }
  = \frac
  {
    \exp\left[n\left\{\IFG(u) - \IFG(k)\right\}\right]
  }
  {
    \sum_{j=0}^p\exp\left[n\left\{\IFG(u) - \IFG(j)\right\}\right]
  }.
$$

By definition of $\IFG(k)$, we have that 
\begin{equation}
  n\left\{\IFG(k) - \IFG(u)\right\} 
    = n\left\{J_n(\Gamhat_k) - J_n(\Gamhat_u)\right\} + C(k-u)\log(n).  
\label{derivFG}
\end{equation}
We show that $\wFG_k \to 0$ as $n \to \infty$ for all $k \neq u$ by following 
a similar argument as the proof of Theorems 3.1 and 3.2 in \cite{zhangmai}.  
Lemma 2 in \citet{zhangmai} states that $J(\Gamma_u) < J(\Gamma_k) < 0$ for 
all $k = 0$, $\ldots$, $u-1$, and $J(\Gamma_k) = J(\Gamma_u)$ for all 
$k = u+1$, $\ldots$, $p$.  First suppose that $k = 0$, $\ldots$, $u-1$.  
In this setting, we have that \eqref{derivFG} tends to $\infty$ 
as $n \to \infty$.  Now suppose that $k = u$, $\ldots$, $p$.  In this setting, 
we have that 
$
  n\left\{J_n(\Gamhat_k) - J_n(\Gamhat_u)\right\} = O_P(1)
$ 
in \eqref{derivFG}.  Therefore \eqref{derivFG} tends to $\infty$ as 
$n \to \infty$ when $k = u+1$, $\ldots$, $p$. 
Putting this together implies that $\wFG_k \to 0$ for all $k \neq u$ 
and $\wFG_u \to 1$ as $n \to \infty$.   
\end{proof}

\begin{proof}[Proof of Lemma 3]
Note that 
\begin{equation}
  \woneD_k = \frac
  {
    \exp\left\{-n\IoneD(k)\right\}
  }
  {
    \sum_{j=0}^p\exp\left\{-n\IoneD(j)\right\}
  }
  = \frac
  {
    \exp\left[n\left\{\IoneD(u) - \IoneD(k)\right\}\right]
  }
  {
    \sum_{j=0}^p\exp\left[n\left\{\IoneD(u) - \IoneD(j)\right\}\right]
  }.
\label{int}
\end{equation}
We show that $\woneD_k \to 0$ as $n \to \infty$ for all $k \neq u$ by 
following a similar argument as the proof of Theorems 3.1 and 3.2 
in \cite{zhangmai}.  First suppose that $k > u$ and observe that 
$$
  n\left\{\IoneD(u) - \IoneD(k)\right\} 
    = n\left\{\sum_{j=u+1}^k \phi_{j,n}(\hat{v}_j) 
      + \frac{C(u-k)\log(n)}{n}\right\}.
$$
We have that $\phi_{j,n}(\hat{v}_j) = O_P\left(n^{-1}\right)$. Therefore 
$$
  n\left\{\IoneD(u) - \IoneD(k)\right\} \to -\infty
$$ 
as $n \to \infty$.  
From \eqref{int} we can conclude that $\woneD_k \to 0$ as $n \to \infty$ 
for all $k > u$. 

Now suppose that $k < u$. Let 
$$
  n\left\{\IoneD(u) - \IoneD(k)\right\} 
    = n\left\{\sum_{j=k+1}^u \phi_{j,n}(\hat{v}_j) 
      + \frac{C(u-k)\log(n)}{n}\right\}.
$$
The function $\phi_{j,n}(\hat{v}_j) \to \phi_{j}(v_j) < 0$ in probability as 
shown in the proof of Propositions 5 and 6 in \cite{algo}. Therefore 
$n\left\{\IoneD(u) - \IoneD(k)\right\} \to -\infty$ as $n \to \infty$.  
From \eqref{int} we can conclude that $\woneD_k \to 0$ as $n \to \infty$ 
for all $k < u$. Therefore $\woneD_k \to 0$ as $n \to \infty$ for all 
$k \neq u$ which implies that $\woneD_u \to 1$ as $n \to \infty$.
\end{proof}

<!-- 
\begin{proof}[Proof of Theorem 1]
Notice that 
\begin{align*}
  &\rootn\left(\TstarFG_w - \TFG_w\right)  
    = \rootn\left(\wstarFG_u\TstarFG_u - \wFG_u\TFG_u\right) 
      + \rootn\left(\sum_{k\neq u}^p\wstarFG_k\TstarFG_k 
        - \sum_{k \neq u}^p \wFG_k\TFG_k\right) \\
  &\qquad = \rootn\left(\TstarFG_u - \TFG_u\right) 
      + \rootn\left\{\sum_{k\neq u}^p\wstarFG_k
        \left(\TstarFG_k - \TFG_u\right) 
        - \sum_{k \neq u}^p \wFG_k\left(\TFG_k - \TFG_u\right)\right\}
\end{align*}
We show that $\wFG_k$, $\wstarFG_k \to 0$ for all $k \neq u$ such that 
\begin{align*}
  &\rootn\|\sum_{k\neq u}^p\wstarFG_k\left(\TstarFG_k - \TstarFG_u\right) 
    - \sum_{k \neq u}^p \wFG_k\left(\TFG_k - \TFG_u\right)\| \\
  &\qquad\leq \sum_{k\neq u}^p\left(\rootn\wstarFG_k
    \|\TstarFG_k - \TstarFG_u\| 
    + \rootn \wFG_k\|\TFG_k - \TFG_u\|\right) \to 0
\end{align*}
as $n\to\infty$ where the rates of the bound are given by 
\eqref{TFGterms}. We have that 
\begin{equation}
\begin{split}
  &\rootn \wFG_k\|\TFG_k - \TFG_u\| = \frac
    {
      \rootn\exp\left\{-n\IFG(k)\right\}
    }
    {
      \sum_{j=0}^p\exp\left\{-n\IFG(j)\right\}
    }\|\TFG_j - \TFG_u\| \\
  &\qquad \leq \rootn\exp\left\{n\IFG(u) - n\IFG(k)\right\}
    \|\TFG_k - \TFG_u\| \\
  &\qquad= O_P\left(\sqrt{n}\right)\exp
    \left[n\left\{\IFG(u) - n\IFG(k)\right\}\right] \\
  &\qquad= O_P\left(\sqrt{n}\right)
    \exp\left[n\left\{J_n(\Gamhat_u) - J_n(\Gamhat_k)\right\} 
      + C(u-k)\log(n)\right] \\
  &\qquad= O_P\left(n^{C(u-k) + 1/2}\right)
    \exp\left[n\left\{J_n(\Gamhat_u) - J_n(\Gamhat_k)\right\}\right].    
\end{split}
\label{intFG}
\end{equation}
The same steps as \eqref{intFG} yield
\begin{equation}
  \rootn \wstarFG_k\|\TstarFG_k - \TstarFG_u\| \leq 
    O_P\left(n^{C(u-k) + 1/2}\right)
      \exp\left[n\left\{\Jstar_n(\Gamstar_u) 
        - \Jstar_n(\Gamstar_k)\right\}\right].
\label{intFG2}
\end{equation}
For $0 \leq k < u$, we have that 
$
  J_n(\Gamhat_u) - J_n(\Gamhat_k) 
    = J(\Gamma_u) - J(\Gamma_k) + o_p(1)
$
where $J(\Gamma_u) < J(\Gamma_k)$ as in the proof of 
\citet[Theorem 3.1]{zhangmai}.  Similarly we have that 
$$
  \Jstar_n(\Gamstar_u) - \Jstar_n(\Gamstar_k) 
    = J(\Gamma_u) - J(\Gamma_k) + o_p(1).
$$
Therefore the rates for the exponent in the last line of \eqref{intFG} and 
the right hand side of \eqref{intFG2} are $-n|O_P(1)|$.  Notice that the rates 
in the last line of \eqref{intFG} and the right hand side of \eqref{intFG2} 
are upper bounded when $k = 0$.  Putting this together yields 
\begin{align*}
  \rootn \wFG_k\|\TFG_k - \TFG_u\| 
    &= O_P\left(n^{Cu + 1/2}\right)\exp\left\{-n|O_P(1)|\right\}, \\
  \rootn \wstarFG_k\|\TstarFG_k - \TstarFG_u\| 
    &= O_P\left(n^{Cu + 1/2}\right)\exp\left\{-n|O_P(1)|\right\};    
\end{align*}
for all $0 \leq k < u$. 

Now consider $u < k \leq p$. From the proof of \citet[Theorem 3.1]{zhangmai} 
we have that $J_n(\Gamhat_u) - J_n(\Gamhat_k) = O_p(n^{-1})$. Combining this 
result with the steps in \eqref{intFG} yields 
\begin{equation}
  \rootn \wFG_k\|\TFG_k - \TFG_u\| \leq O_P\left(n^{C(u-k) + 1/2}\right).
\label{intFG3}
\end{equation}
A similar argument applied to the starred data gives
\begin{equation}
  \rootn \wstarFG_k\|\TstarFG_k - \TstarFG_u\| 
    \leq O_P\left(n^{C(u-k) + 1/2}\right). 
\label{intFG4}
\end{equation}
The rates in both \eqref{intFG3} and \eqref{intFG4} are upper bounded when 
$k = u - 1$.  Putting this together yields 
\begin{align*}
  \rootn \wFG_k\|\TFG_k - \TFG_u\| 
    &= O_P\left(n^{1/2 - C}\right), \\
  \rootn \wstarFG_k\|\TstarFG_k - \TstarFG_u\| 
    &= O_P\left(n^{1/2 - C}\right);    
\end{align*}
for all $u < k \leq p$. Therefore 
\begin{align*}
  &\rootn\left\{\sum_{k\neq u}^p\wstarFG_k\left(\TstarFG_k - \TFG_u\right) 
        - \sum_{k \neq u}^p \wFG_k\left(\TFG_k - \TFG_u\right)\right\} \\
  &\qquad= O_P\left(n^{1/2 - C}\right) 
    + O_P\left(n^{Cu + 1/2}\right)\exp\left\{-n|O_P(1)|\right\},   
\end{align*}
as desired and the conclusion follows.
\end{proof}

<!-- 
\begin{proof}[Proof of Theorem 2]
Notice that 
\begin{align*}
  &\rootn\left(\TstarD_w - \TD_w\right)  
    = \rootn\left(\wstar_u\TstarD_u - w_u\TD_u\right) 
      + \rootn\left(\sum_{k\neq u}^p\wstar_k\Tstar_k 
        - \sum_{k \neq u}^p w_k\That_k\right) \\
  &\qquad = \rootn\left(\Tstar_u - \That_u\right) 
      + \rootn\left\{\sum_{k\neq u}^p\wstar_k\left(\Tstar_k - \Tstar_u\right)
        - \sum_{k \neq u}^p w_k\left(\That_k - \That_u\right)\right\}.
\end{align*}
We show that $w_k$, $\wstar_k \to 0$ such that 
$$
  \rootn\|\sum_{k\neq u}^p\wstar_k\left(\Tstar_k - \Tstar_u\right) 
    - \sum_{k \neq u}^p w_k\left(\That_k - \That_u\right)\| 
  \leq \sum_{k=1}^p\left(\rootn\wstar_k\|\Tstar_k - \Tstar_u\| 
    + \rootn w_k\|\That_k - \That_u\|\right) \to 0
$$
as $n\to\infty$ for all $k \neq u$ and find the rates at which they vanish. 
We have that 
\begin{equation}
\begin{split}
  &\rootn w_k\|\That_k - \That_u\| = \frac
    {
      \rootn\exp\left\{-n\IoneD(k)\right\}
    }
    {
      \sum_{j=0}^p\exp\left\{-n\IoneD(j)\right\}
    }\|\That_k - \That_u\| \\
  &\qquad \leq \rootn\exp\left\{n\IoneD(u) - n\IoneD(k)\right\}
    \|\That_k - \That_u\| \\
  &\qquad= \rootn\exp\left\{
      n\sum_{j=1}^u \phi_{j,n}(\hat{v}_j) - n\sum_{j=1}^k\phi_{j,n}(\hat{v}_j) 
        + (u - k)\frac{C\log{n}}{n}
    \right\}\|\That_k - \That_u\| \\
  &\qquad= n^{\left\{C(u - k) + 1/2\right\}}\exp\left\{
      n\sum_{j=1}^u \phi_{j,n}(\hat{v}_j) - n\sum_{j=1}^k\phi_{j,n}(\hat{v}_j) 
    \right\}\|\That_k - \That_u\| \\
  &\qquad= O_P\left[n^{\left\{C(u - k) + 1/2\right\}}\right]\exp\left\{
      n\sum_{j=1}^u \phi_{j,n}(\hat{v}_j) - n\sum_{j=1}^k\phi_{j,n}(\hat{v}_j) 
    \right\}
\end{split}
\label{foo}
\end{equation}
where the last equality follows from the fact that 
$\|\That_k - \That_u\| = |O_P(1)|$ for all $k = 1$, $\ldots$, $p$. This is 
because $\That_k \to \theta$ for all $k = u$, $\ldots$, $p$ and 
$\|\That_k\| \to a \leq \|\theta\|$ for all $k = 1$, $\ldots$, $u-1$ since the 
envelope estimator exhibits shrinkage when $k = 1$, $\ldots$, $u-1$. First 
suppose that $k = 1$, $\ldots$, $u-1$. In this setting   
$\phi_{k,n}(\hat{v}_k) \to \phi_k(v_k) < 0$ as $n \to \infty$ 
\citep[proof of Theorems 5 and 6]{algo}. From \eqref{foo} we have 
\begin{equation}
\begin{split}
  &\rootn w_k\|\That_k - \That_u\| 
    \leq O_P\left[n^{\left\{C(u - k) + 1/2\right\}}\right] 
      \exp\left\{n\sum_{j=k-1}^u \phi_{j,n}(\hat{v}_j)\right\} \\
  &\qquad= O_P\left[n^{\left\{C(u - k) + 1/2\right\}}\right]
    \exp\left\{-n|O_P(1)|\right\}.
\end{split}
\label{bar-1}
\end{equation}
Now suppose that $k = u+1$, $\ldots$, $p$. In this setting,  
$\phi_{k,n}(\hat{v}_k) = O_p\left(n^{-1}\right)$ 
\citep[proof of Theorem 3.1]{zhangmai}. From \eqref{foo} we have 
\begin{equation}
\begin{split}
  &\rootn w_k\|\That_k - \That_u\| 
    \leq O_p\left[n^{\left\{C(u - k) + 1/2\right\}}\right] 
      \exp\left\{-n\sum_{j=u+1}^k \phi_{j,n}(\hat{v}_j)\right\} \\
  &\qquad= O_p\left[n^{\left\{C(u - k) + 1/2\right\}}\right].
\end{split}
\label{bar-2}
\end{equation}
The same steps in \eqref{bar-1} and \eqref{bar-2} apply to the starred data 
so that  
\begin{equation}
  \rootn \wstar_k\|\Tstar_k - \Tstar_u\| \leq 
    O_P\left[n^{\left\{C(u - k) + 1/2\right\}}\right]
      \exp\left\{-n|O_P(1)|\right\},
    \qquad (k = 1,...,u-1),
\label{baz-1}
\end{equation}
and
\begin{equation}
  \rootn \wstar_k\|\Tstar_k - \Tstar_u\| = 
    O_p\left[n^{\left\{C(u - k) + 1/2\right\}}\right],
    \qquad (k = u+1,...,p).
\label{baz-2}
\end{equation}
Our conclusion follows by noting that \eqref{bar-1}, \eqref{bar-2}, 
\eqref{baz-1}, and \eqref{baz-2} implies that 
\begin{align*}
  \rootn w_k\|\That_k - \That_u\| &\leq 
    O_P\left[n^{\left\{Cu + 1/2\right\}}\right]
      \exp\left\{-n|O_P(1)|\right\},
  \qquad (k = 1,...,u-1); \\
  \rootn w_k\|\That_k - \That_u\| 
    &\leq O_p\left\{n^{\left(1/2 - C\right)}\right\}, 
  \qquad (k = u+1,...,p); \\
  \rootn \wstar_k\|\Tstar_k - \Tstar_u\| &\leq 
    O_P\left[n^{\left\{Cu + 1/2\right\}}\right]
      \exp\left\{-n|O_P(1)|\right\},
  \qquad (k = 1,...,u-1); \\
  \rootn \wstar_k\|\Tstar_k - \Tstar_u\| 
    &\leq O_p\left\{n^{\left(1/2 - C\right)}\right\}, 
  \qquad (k = u+1,...,p); 
\end{align*}
respectively.
\end{proof}

\section*{Numerical examples}

The following software packages are needed to reproduce the analyses in the main text.

\begin{knitrout}
\definecolor{shadecolor}{rgb}{0.969, 0.969, 0.969}\color{fgcolor}\begin{kframe}
\begin{alltt}
\hlkwd{rm}\hlstd{(}\hlkwc{list} \hlstd{=} \hlkwd{ls}\hlstd{())}
\hlkwd{library}\hlstd{(tidyverse)}
\hlkwd{library}\hlstd{(TRES)}
\hlkwd{library}\hlstd{(MASS)}
\hlkwd{library}\hlstd{(foreach)}
\hlkwd{library}\hlstd{(doParallel)}
\hlkwd{library}\hlstd{(xtable)}
\hlkwd{library}\hlstd{(faraway)}
\end{alltt}
\end{kframe}
\end{knitrout}

To register doParallel to be used with foreach, we call the registerDoParallel function and specify the number of cores to be used for parallel computing.

\begin{knitrout}
\definecolor{shadecolor}{rgb}{0.969, 0.969, 0.969}\color{fgcolor}\begin{kframe}
\begin{alltt}
\hlstd{numCores} \hlkwb{<-} \hlkwd{detectCores}\hlstd{()} \hlopt{-} \hlnum{1}\hlstd{; numCores}
\end{alltt}
\begin{verbatim}
## [1] 15
\end{verbatim}
\begin{alltt}
\hlkwd{registerDoParallel}\hlstd{(}\hlkwc{cores} \hlstd{= numCores)}
\hlkwd{RNGkind}\hlstd{(}\hlstr{"L'Ecuyer-CMRG"}\hlstd{)}
\end{alltt}
\end{kframe}
\end{knitrout}

\section*{Logistic regression simulations}

\subsection*{Setting A:}

We reproduce the logistic regression simulation in the main text. We first create the basis matrix $\Gamma$ for the true envelope space and the basis matrix for its orthogonal complement $\Gamma_0$.

\begin{knitrout}
\definecolor{shadecolor}{rgb}{0.969, 0.969, 0.969}\color{fgcolor}\begin{kframe}
\begin{alltt}
\hlstd{p} \hlkwb{<-} \hlnum{8}\hlstd{; u} \hlkwb{<-} \hlnum{2}
\hlstd{v1} \hlkwb{<-} \hlkwd{matrix}\hlstd{(}\hlkwd{rep}\hlstd{(}\hlnum{1}\hlopt{/}\hlkwd{sqrt}\hlstd{(p),p),} \hlkwc{nrow} \hlstd{= p)}
\hlstd{O} \hlkwb{<-}\hlkwd{qr.Q}\hlstd{(}\hlkwd{qr}\hlstd{(v1),} \hlkwc{complete} \hlstd{=} \hlnum{TRUE}\hlstd{)}
\hlstd{Gamma} \hlkwb{<-} \hlstd{O[,} \hlnum{1}\hlopt{:}\hlstd{u]}
\hlstd{Gamma0} \hlkwb{<-} \hlstd{O[, (u}\hlopt{+}\hlnum{1}\hlstd{)}\hlopt{:}\hlstd{p]}
\end{alltt}
\end{kframe}
\end{knitrout}

We next create the core of the material and immaterial variation, denoted as $\Omega$ and $\Omega_0$ respectively.

\begin{knitrout}
\definecolor{shadecolor}{rgb}{0.969, 0.969, 0.969}\color{fgcolor}\begin{kframe}
\begin{alltt}
\hlstd{Omega} \hlkwb{<-} \hlkwd{diag}\hlstd{(}\hlnum{2}\hlopt{:}\hlnum{3}\hlstd{)}
\hlstd{Omega0} \hlkwb{<-} \hlkwd{diag}\hlstd{(}\hlkwd{exp}\hlstd{(}\hlkwd{c}\hlstd{(}\hlopt{-}\hlnum{4}\hlopt{:}\hlnum{1}\hlstd{)))}
\end{alltt}
\end{kframe}
\end{knitrout}

We now build the variance matrix of the predictor variables and construct the true canonical parameter vector (regression coefficient vector) as an element contained in $\text{span}(\Gamma)$.

\begin{knitrout}
\definecolor{shadecolor}{rgb}{0.969, 0.969, 0.969}\color{fgcolor}\begin{kframe}
\begin{alltt}
\hlstd{SigmaX} \hlkwb{<-} \hlstd{Gamma}\hlopt{%*%}\hlstd{Omega}\hlopt{%*%}\hlkwd{t}\hlstd{(Gamma)} \hlopt{+} \hlstd{Gamma0}\hlopt{%*%}\hlstd{Omega0}\hlopt{%*%}\hlkwd{t}\hlstd{(Gamma0)}
\hlstd{beta} \hlkwb{<-} \hlstd{Gamma} \hlopt{%*%} \hlkwd{t}\hlstd{(Gamma)} \hlopt{%*%} \hlstd{(}\hlnum{1} \hlopt{+} \hlstd{(}\hlnum{1}\hlopt{:}\hlstd{p} \hlopt{-} \hlstd{p}\hlopt{/}\hlnum{2}\hlstd{)}\hlopt{/}\hlstd{(p}\hlopt{^}\hlnum{2}\hlstd{))}
\hlstd{eig} \hlkwb{<-} \hlkwd{eigen}\hlstd{(SigmaX)}
\hlstd{SigmaX.half} \hlkwb{<-} \hlstd{eig}\hlopt{$}\hlstd{vec} \hlopt{%*%} \hlkwd{diag}\hlstd{(}\hlkwd{sqrt}\hlstd{(eig}\hlopt{$}\hlstd{val))} \hlopt{%*%} \hlkwd{t}\hlstd{(eig}\hlopt{$}\hlstd{vec)}
\end{alltt}
\end{kframe}
\end{knitrout}

We now perform the nonparametric bootstrap procedure for all envelope estimators of the canonical parameter vector mentioned in the main text and the MLE. This nonparametric bootstrap has a bootstrap sample size of $5000$.  Our bootstrap simulation will consider two model selection regimes for determining the envelope dimension $\hat{u}_{1D}$ at every iteration.  In one scheme, we estimate the envelope dimension at every iteration of the bootstrap (variable $u$). In the other scheme, we estimate the dimension of the envelope space in the original data set and then treat this estimated dimension as the true dimension when we resample our data and calculate these envelope estimators (fixed $u$), thus ignoring the variability associated with model selection.  Theorem 3 in the main text provides some guidance for the performance of the nonparametric bootstrap for estimating the variability of $\TD_w$, an analog does not exist for the other envelope estimators. 

The logistic\_sim function below generates a logistic regression model that incorporates the above envelope structure that is stored in your global environment. The function then calls on model\_boot (code is in the Appendix) to perform the nonparametric bootstrap with respect to all considered estimators. This function also returns the estimated envelope dimension at every iteration of the nonparametric bootstrap.

\begin{knitrout}
\definecolor{shadecolor}{rgb}{0.969, 0.969, 0.969}\color{fgcolor}\begin{kframe}
\begin{alltt}
\hlstd{logistic_sim} \hlkwb{<-} \hlkwa{function}\hlstd{(}\hlkwc{n}\hlstd{,} \hlkwc{p} \hlstd{= p)\{}
  \hlstd{X} \hlkwb{<-} \hlkwd{matrix}\hlstd{(}\hlkwd{rnorm}\hlstd{(n}\hlopt{*}\hlstd{p),} \hlkwc{nrow} \hlstd{= n)} \hlopt{%*%} \hlstd{SigmaX.half}
  \hlstd{gx} \hlkwb{<-} \hlkwd{as.numeric}\hlstd{(X} \hlopt{%*%} \hlstd{beta)}
  \hlstd{Y} \hlkwb{<-} \hlkwd{rbinom}\hlstd{(n,} \hlkwc{size} \hlstd{=} \hlnum{1}\hlstd{,} \hlkwc{prob} \hlstd{=} \hlnum{1} \hlopt{/} \hlstd{(}\hlnum{1} \hlopt{+} \hlkwd{exp}\hlstd{(}\hlopt{-}\hlstd{gx)))}
  \hlstd{data_sim} \hlkwb{<-} \hlkwd{as.data.frame}\hlstd{(}\hlkwd{cbind}\hlstd{(Y, X))}
  \hlstd{m1} \hlkwb{<-} \hlkwd{glm}\hlstd{(Y} \hlopt{~ -}\hlnum{1} \hlopt{+} \hlstd{.,} \hlkwc{family} \hlstd{=} \hlstr{"binomial"}\hlstd{,} \hlkwc{data} \hlstd{= data_sim)}
  \hlkwd{model_boot}\hlstd{(}\hlkwc{model} \hlstd{= m1,} \hlkwc{nboot} \hlstd{= nboot,} \hlkwc{cores} \hlstd{= numCores)}
\hlstd{\}}
\end{alltt}
\end{kframe}
\end{knitrout}

We perform the nonparametric bootstrap at five different sample sizes.

\begin{knitrout}
\definecolor{shadecolor}{rgb}{0.969, 0.969, 0.969}\color{fgcolor}\begin{kframe}
\begin{alltt}
\hlkwd{set.seed}\hlstd{(}\hlnum{13}\hlstd{)}
\hlstd{n} \hlkwb{<-} \hlkwd{c}\hlstd{(}\hlnum{300}\hlstd{,} \hlnum{500}\hlstd{,} \hlnum{750}\hlstd{,} \hlnum{1000}\hlstd{)}
\hlstd{nboot} \hlkwb{<-} \hlnum{5000}
\hlkwd{system.time}\hlstd{(\{}
  \hlstd{lsims} \hlkwb{<-} \hlkwd{lapply}\hlstd{(n,} \hlkwa{function}\hlstd{(}\hlkwc{j}\hlstd{)} \hlkwd{logistic_sim}\hlstd{(}\hlkwc{n} \hlstd{= j,} \hlkwc{p} \hlstd{= p))}
\hlstd{\})}
\end{alltt}
\begin{verbatim}
##      user    system   elapsed 
## 11838.292   100.072   830.252
\end{verbatim}
\end{kframe}
\end{knitrout}

The distribution of the estimated dimension across bootstrap iterations and sample sizes is depicted below:

\begin{knitrout}
\definecolor{shadecolor}{rgb}{0.969, 0.969, 0.969}\color{fgcolor}\begin{kframe}
\begin{alltt}
\hlstd{u_boot_l} \hlkwb{<-} \hlkwd{lapply}\hlstd{(}\hlnum{1}\hlopt{:}\hlkwd{length}\hlstd{(lsims),} \hlkwa{function}\hlstd{(}\hlkwc{j}\hlstd{)\{}
  \hlkwd{round}\hlstd{(}\hlkwd{table}\hlstd{(lsims[[j]][,} \hlnum{4}\hlopt{*}\hlstd{p}\hlopt{+}\hlnum{1}\hlstd{])} \hlopt{/} \hlstd{nboot,} \hlnum{3}\hlstd{)}
\hlstd{\})}

\hlcom{## n = 300}
\hlstd{u_boot_l[[}\hlnum{1}\hlstd{]]}
\end{alltt}
\begin{verbatim}
## 
##     1     2     3     4     5     6 
## 0.407 0.410 0.158 0.022 0.002 0.000
\end{verbatim}
\begin{alltt}
\hlcom{## n = 500}
\hlstd{u_boot_l[[}\hlnum{2}\hlstd{]]}
\end{alltt}
\begin{verbatim}
## 
##     1     2     3     4 
## 0.659 0.298 0.040 0.003
\end{verbatim}
\begin{alltt}
\hlcom{## n = 750}
\hlstd{u_boot_l[[}\hlnum{3}\hlstd{]]}
\end{alltt}
\begin{verbatim}
## 
##     1     2     3     4     5 
## 0.312 0.436 0.212 0.037 0.003
\end{verbatim}
\begin{alltt}
\hlcom{## n = 1000}
\hlstd{u_boot_l[[}\hlnum{4}\hlstd{]]}
\end{alltt}
\begin{verbatim}
## 
##     1     2     3     4     5 
## 0.456 0.436 0.101 0.007 0.000
\end{verbatim}
\end{kframe}
\end{knitrout}

The Frobenius norm of all boostrapped covariance matrices for all estimators across sample sizes is depicted below:

\begin{kframe}
\begin{alltt}
\hlstd{volume_boot_l} \hlkwb{<-} \hlkwd{do.call}\hlstd{(rbind,} \hlkwd{lapply}\hlstd{(}\hlnum{1}\hlopt{:}\hlkwd{length}\hlstd{(lsims),} \hlkwa{function}\hlstd{(}\hlkwc{j}\hlstd{)\{}
  \hlkwd{unlist}\hlstd{(}\hlkwd{lapply}\hlstd{(}\hlkwd{normvar}\hlstd{(lsims[[j]]),} \hlkwa{function}\hlstd{(}\hlkwc{x}\hlstd{)} \hlkwd{norm}\hlstd{(x,} \hlkwc{type}\hlstd{=}\hlstr{"F"}\hlstd{)))}
\hlstd{\}))}
\hlkwd{rownames}\hlstd{(volume_boot_l)} \hlkwb{<-} \hlstd{n}
\hlkwd{xtable}\hlstd{(volume_boot_l,} \hlkwc{digits} \hlstd{=} \hlnum{3}\hlstd{)}
\end{alltt}
\end{kframe}
\begin{table}[ht]
\centering
\begin{tabular}{rrrrr}
  \hline
 & se\_wt & se\_env & se\_env\_fixedu & se\_MLE \\ 
  \hline
300 & 3.705 & 4.102 & 0.366 & 4.297 \\ 
  500 & 0.477 & 0.569 & 0.132 & 2.153 \\ 
  750 & 1.064 & 1.198 & 0.068 & 1.026 \\ 
  1000 & 0.784 & 1.033 & 0.053 & 0.792 \\ 
   \hline
\end{tabular}
\end{table}

The estimated efficiency gains for envelope estimators ($\text{se}^*(\hat{\theta}) / \text{se}^*(\hat{\theta}_{\text{env}})$) across sample sizes is depicted below:

\begin{knitrout}
\definecolor{shadecolor}{rgb}{0.969, 0.969, 0.969}\color{fgcolor}\begin{kframe}
\begin{alltt}
\hlstd{SEs_boot_l} \hlkwb{<-} \hlkwd{lapply}\hlstd{(}\hlnum{1}\hlopt{:}\hlkwd{length}\hlstd{(lsims),} \hlkwa{function}\hlstd{(}\hlkwc{j}\hlstd{)\{}
  \hlkwd{do.call}\hlstd{(cbind,} \hlkwd{lapply}\hlstd{(}\hlkwd{normvar}\hlstd{(lsims[[j]]),} \hlkwa{function}\hlstd{(}\hlkwc{x}\hlstd{)} \hlkwd{sqrt}\hlstd{(}\hlkwd{diag}\hlstd{(x))))}
\hlstd{\})}
\hlstd{ratios_boot_l} \hlkwb{<-} \hlkwd{lapply}\hlstd{(}\hlnum{1}\hlopt{:}\hlkwd{length}\hlstd{(lsims),} \hlkwa{function}\hlstd{(}\hlkwc{j}\hlstd{)\{}
  \hlcom{# ratio of SE(MLE) to SE(wtEnv)}
  \hlstd{out} \hlkwb{<-} \hlkwd{cbind}\hlstd{(SEs_boot_l[[j]][,} \hlnum{4}\hlstd{]} \hlopt{/} \hlstd{SEs_boot_l[[j]][,} \hlnum{1}\hlstd{],}
    \hlcom{# ratio of SE(MLE) to SE(Env)}
    \hlstd{SEs_boot_l[[j]][,} \hlnum{4}\hlstd{]} \hlopt{/} \hlstd{SEs_boot_l[[j]][,} \hlnum{2}\hlstd{],}
    \hlcom{# ratio of SE(MLE) to SE(Env_hat(u))}
    \hlstd{SEs_boot_l[[j]][,} \hlnum{4}\hlstd{]} \hlopt{/} \hlstd{SEs_boot_l[[j]][,} \hlnum{3}\hlstd{])}
  \hlkwd{colnames}\hlstd{(out)} \hlkwb{<-} \hlkwd{c}\hlstd{(}\hlstr{"se(MLE)/se(env_wt)"}\hlstd{,} \hlstr{"se(MLE)/se(env_varu)"}\hlstd{,}
                     \hlstr{"se(MLE)/se(env_fixu)"}\hlstd{)}
  \hlkwd{round}\hlstd{(out,} \hlnum{3}\hlstd{)}
\hlstd{\})}

\hlcom{## n = 300}
\hlstd{ratios_boot_l[[}\hlnum{1}\hlstd{]]}
\end{alltt}
\begin{verbatim}
##  se(MLE)/se(env_wt) se(MLE)/se(env_varu) se(MLE)/se(env_fixu)
##               1.224                1.158                3.263
##               1.142                1.112                1.646
##               1.013                0.965                8.304
##               1.160                1.091                3.225
##               1.439                1.331                1.816
##               1.154                1.137                1.319
##               1.057                1.034                1.142
##               1.183                1.152                1.596
\end{verbatim}
\begin{alltt}
\hlcom{## n = 500}
\hlstd{ratios_boot_l[[}\hlnum{2}\hlstd{]]}
\end{alltt}
\begin{verbatim}
##  se(MLE)/se(env_wt) se(MLE)/se(env_varu) se(MLE)/se(env_fixu)
##               1.879                1.763                2.883
##               1.291                1.270                1.563
##               2.141                1.951                9.651
##               2.177                1.969                5.369
##               2.205                2.064                2.811
##               1.460                1.434                1.547
##               1.107                1.091                1.138
##               1.209                1.186                1.319
\end{verbatim}
\begin{alltt}
\hlcom{## n = 750}
\hlstd{ratios_boot_l[[}\hlnum{3}\hlstd{]]}
\end{alltt}
\begin{verbatim}
##  se(MLE)/se(env_wt) se(MLE)/se(env_varu) se(MLE)/se(env_fixu)
##               1.028                0.951                3.642
##               0.977                0.921                1.475
##               1.067                0.999               11.346
##               0.774                0.732                5.263
##               0.632                0.592                2.721
##               1.138                1.117                1.301
##               0.941                0.919                0.963
##               1.018                0.972                1.415
\end{verbatim}
\begin{alltt}
\hlcom{## n = 1000}
\hlstd{ratios_boot_l[[}\hlnum{4}\hlstd{]]}
\end{alltt}
\begin{verbatim}
##  se(MLE)/se(env_wt) se(MLE)/se(env_varu) se(MLE)/se(env_fixu)
##               0.997                0.881                3.498
##               1.001                0.919                1.134
##               1.008                0.924               10.073
##               1.490                1.350                5.495
##               0.502                0.378                2.191
##               1.481                1.420                1.740
##               0.936                0.859                1.211
##               1.012                0.914                1.305
\end{verbatim}
\end{kframe}
\end{knitrout}

\subsection*{Setting B:}

We next create the core of the material and immaterial variation, denoted as $\Omega$ and $\Omega_0$ respectively, build the variance matrix of the predictor variables, and construct the true canonical parameter vector (regression coefficient vector) as an element contained in $\text{span}(\Gamma)$.

\begin{knitrout}
\definecolor{shadecolor}{rgb}{0.969, 0.969, 0.969}\color{fgcolor}\begin{kframe}
\begin{alltt}
\hlstd{Omega} \hlkwb{<-} \hlkwd{diag}\hlstd{(}\hlkwd{exp}\hlstd{(}\hlopt{-}\hlkwd{c}\hlstd{(}\hlnum{4}\hlopt{:}\hlnum{5}\hlstd{)))}
\hlstd{Omega0} \hlkwb{<-} \hlkwd{diag}\hlstd{(}\hlkwd{exp}\hlstd{(}\hlkwd{c}\hlstd{(}\hlopt{-}\hlnum{3}\hlopt{:}\hlnum{2}\hlstd{)))}
\hlstd{SigmaX} \hlkwb{<-} \hlstd{Gamma}\hlopt{%*%}\hlstd{Omega}\hlopt{%*%}\hlkwd{t}\hlstd{(Gamma)} \hlopt{+} \hlstd{Gamma0}\hlopt{%*%}\hlstd{Omega0}\hlopt{%*%}\hlkwd{t}\hlstd{(Gamma0)}
\hlstd{beta} \hlkwb{<-} \hlstd{Gamma} \hlopt{%*%} \hlkwd{t}\hlstd{(Gamma)} \hlopt{%*%} \hlstd{(}\hlnum{1} \hlopt{+} \hlstd{(}\hlnum{1}\hlopt{:}\hlstd{p} \hlopt{-} \hlstd{p}\hlopt{/}\hlnum{2}\hlstd{)}\hlopt{/}\hlstd{(p}\hlopt{^}\hlnum{2}\hlstd{))}
\hlstd{eig} \hlkwb{<-} \hlkwd{eigen}\hlstd{(SigmaX)}
\hlstd{SigmaX.half} \hlkwb{<-} \hlstd{eig}\hlopt{$}\hlstd{vec} \hlopt{%*%} \hlkwd{diag}\hlstd{(}\hlkwd{sqrt}\hlstd{(eig}\hlopt{$}\hlstd{val))} \hlopt{%*%} \hlkwd{t}\hlstd{(eig}\hlopt{$}\hlstd{vec)}
\end{alltt}
\end{kframe}
\end{knitrout}

We perform the nonparametric bootstrap at five different sample sizes.

\begin{knitrout}
\definecolor{shadecolor}{rgb}{0.969, 0.969, 0.969}\color{fgcolor}\begin{kframe}
\begin{alltt}
\hlkwd{set.seed}\hlstd{(}\hlnum{13}\hlstd{)}
\hlkwd{RNGkind}\hlstd{(}\hlstr{"L'Ecuyer-CMRG"}\hlstd{)}
\hlstd{n} \hlkwb{<-} \hlkwd{c}\hlstd{(}\hlnum{300}\hlstd{,} \hlnum{500}\hlstd{,} \hlnum{750}\hlstd{,} \hlnum{1000}\hlstd{)}
\hlstd{nboot} \hlkwb{<-} \hlnum{5000}
\hlkwd{system.time}\hlstd{(\{}
  \hlstd{lsimsB} \hlkwb{<-} \hlkwd{lapply}\hlstd{(n,} \hlkwa{function}\hlstd{(}\hlkwc{j}\hlstd{)} \hlkwd{logistic_sim}\hlstd{(}\hlkwc{n} \hlstd{= j,} \hlkwc{p} \hlstd{= p))}
\hlstd{\})}
\end{alltt}
\begin{verbatim}
##      user    system   elapsed 
## 11804.324    86.218   807.427
\end{verbatim}
\end{kframe}
\end{knitrout}

The distribution of the estimated dimension across bootstrap iterations and sample sizes is depicted below:

\begin{knitrout}
\definecolor{shadecolor}{rgb}{0.969, 0.969, 0.969}\color{fgcolor}\begin{kframe}
\begin{alltt}
\hlstd{u_boot_lB} \hlkwb{<-} \hlkwd{lapply}\hlstd{(}\hlnum{1}\hlopt{:}\hlkwd{length}\hlstd{(lsimsB),} \hlkwa{function}\hlstd{(}\hlkwc{j}\hlstd{)\{}
  \hlkwd{round}\hlstd{(}\hlkwd{table}\hlstd{(lsimsB[[j]][,} \hlnum{4}\hlopt{*}\hlstd{p}\hlopt{+}\hlnum{1}\hlstd{])} \hlopt{/} \hlstd{nboot,} \hlnum{3}\hlstd{)}
\hlstd{\})}
\hlstd{u_boot_lB}
\end{alltt}
\begin{verbatim}
## [[1]]
## 
##     1     2     3     4     5     6     7 
## 0.019 0.126 0.328 0.332 0.155 0.036 0.003 
## 
## [[2]]
## 
##     1     2     3     4     5     6     7 
## 0.003 0.083 0.275 0.363 0.212 0.056 0.007 
## 
## [[3]]
## 
##     1     2     3     4     5     6     7 
## 0.006 0.073 0.243 0.358 0.243 0.070 0.008 
## 
## [[4]]
## 
##     1     2     3     4     5     6     7     8 
## 0.002 0.041 0.166 0.324 0.300 0.131 0.032 0.003
\end{verbatim}
\end{kframe}
\end{knitrout}

The Frobenius norm of all boostrapped covariance matrices for all estimators across sample sizes is depicted below:

\begin{knitrout}
\definecolor{shadecolor}{rgb}{0.969, 0.969, 0.969}\color{fgcolor}\begin{kframe}
\begin{alltt}
\hlstd{volume_boot_lB} \hlkwb{<-} \hlkwd{do.call}\hlstd{(rbind,} \hlkwd{lapply}\hlstd{(}\hlnum{1}\hlopt{:}\hlkwd{length}\hlstd{(lsimsB),} \hlkwa{function}\hlstd{(}\hlkwc{j}\hlstd{)\{}
  \hlkwd{unlist}\hlstd{(}\hlkwd{lapply}\hlstd{(}\hlkwd{normvar}\hlstd{(lsimsB[[j]]),} \hlkwa{function}\hlstd{(}\hlkwc{x}\hlstd{)} \hlkwd{norm}\hlstd{(x,} \hlkwc{type}\hlstd{=}\hlstr{"F"}\hlstd{)))}
\hlstd{\}))}
\hlkwd{rownames}\hlstd{(volume_boot_lB)} \hlkwb{<-} \hlstd{n}
\hlstd{volume_boot_lB}
\end{alltt}
\begin{verbatim}
##          se_wt    se_env se_env_fixedu    se_MLE
## 300  2.7301055 2.8562090     1.9606413 2.6950735
## 500  1.1832098 1.2586726     0.8433347 1.4049298
## 750  0.4918665 0.5098347     0.4319702 0.9439191
## 1000 0.6538492 0.6800432     0.5299688 0.7114760
\end{verbatim}
\end{kframe}
\end{knitrout}

The estimated efficiency gains for envelope estimators ($\text{se}^*(\hat{\theta}) / \text{se}^*(\hat{\theta}_{\text{env}})$) across sample sizes is depicted below:

\begin{knitrout}
\definecolor{shadecolor}{rgb}{0.969, 0.969, 0.969}\color{fgcolor}\begin{kframe}
\begin{alltt}
\hlstd{SEs_boot_lB} \hlkwb{<-} \hlkwd{lapply}\hlstd{(}\hlnum{1}\hlopt{:}\hlkwd{length}\hlstd{(lsimsB),} \hlkwa{function}\hlstd{(}\hlkwc{j}\hlstd{)\{}
  \hlkwd{do.call}\hlstd{(cbind,} \hlkwd{lapply}\hlstd{(}\hlkwd{normvar}\hlstd{(lsimsB[[j]]),} \hlkwa{function}\hlstd{(}\hlkwc{x}\hlstd{)} \hlkwd{sqrt}\hlstd{(}\hlkwd{diag}\hlstd{(x))))}
\hlstd{\})}
\hlstd{ratios_boot_lB} \hlkwb{<-} \hlkwd{lapply}\hlstd{(}\hlnum{1}\hlopt{:}\hlkwd{length}\hlstd{(lsimsB),} \hlkwa{function}\hlstd{(}\hlkwc{j}\hlstd{)\{}
  \hlcom{# ratio of SE(MLE) to SE(wtEnv)}
  \hlstd{out} \hlkwb{<-} \hlkwd{cbind}\hlstd{(SEs_boot_lB[[j]][,} \hlnum{4}\hlstd{]} \hlopt{/} \hlstd{SEs_boot_lB[[j]][,} \hlnum{1}\hlstd{],}
    \hlcom{# ratio of SE(MLE) to SE(Env)}
    \hlstd{SEs_boot_lB[[j]][,} \hlnum{4}\hlstd{]} \hlopt{/} \hlstd{SEs_boot_lB[[j]][,} \hlnum{2}\hlstd{],}
    \hlcom{# ratio of SE(MLE) to SE(Env_hat(u))}
    \hlstd{SEs_boot_lB[[j]][,} \hlnum{4}\hlstd{]} \hlopt{/} \hlstd{SEs_boot_lB[[j]][,} \hlnum{3}\hlstd{])}
  \hlkwd{colnames}\hlstd{(out)} \hlkwb{<-} \hlkwd{c}\hlstd{(}\hlstr{"se(MLE)/se(env_wt)"}\hlstd{,} \hlstr{"se(MLE)/se(env_varu)"}\hlstd{,}
                     \hlstr{"se(MLE)/se(env_fixu)"}\hlstd{)}
  \hlkwd{round}\hlstd{(out,} \hlnum{3}\hlstd{)}
\hlstd{\})}

\hlcom{## n = 300}
\hlstd{ratios_boot_lB[[}\hlnum{1}\hlstd{]]}
\end{alltt}
\begin{verbatim}
##  se(MLE)/se(env_wt) se(MLE)/se(env_varu) se(MLE)/se(env_fixu)
##               0.991                0.970                1.129
##               1.014                0.992                1.356
##               1.051                1.041                1.120
##               0.995                0.986                0.992
##               0.898                0.890                0.864
##               0.958                0.942                0.856
##               0.924                0.915                0.851
##               0.947                0.937                0.854
\end{verbatim}
\begin{alltt}
\hlcom{## n = 500}
\hlstd{ratios_boot_lB[[}\hlnum{2}\hlstd{]]}
\end{alltt}
\begin{verbatim}
##  se(MLE)/se(env_wt) se(MLE)/se(env_varu) se(MLE)/se(env_fixu)
##               1.045                1.025                1.135
##               1.111                1.065                1.559
##               0.967                0.960                0.927
##               0.971                0.967                0.916
##               1.023                1.020                1.034
##               0.995                0.993                0.994
##               0.994                0.992                0.991
##               1.005                1.004                1.000
\end{verbatim}
\begin{alltt}
\hlcom{## n = 750}
\hlstd{ratios_boot_lB[[}\hlnum{3}\hlstd{]]}
\end{alltt}
\begin{verbatim}
##  se(MLE)/se(env_wt) se(MLE)/se(env_varu) se(MLE)/se(env_fixu)
##               1.293                1.274                1.475
##               1.557                1.507                2.798
##               0.981                0.980                0.816
##               0.993                0.936                1.060
##               0.990                0.987                1.069
##               1.054                1.044                1.117
##               1.063                1.060                1.086
##               1.063                1.062                1.076
\end{verbatim}
\begin{alltt}
\hlcom{## n = 1000}
\hlstd{ratios_boot_lB[[}\hlnum{4}\hlstd{]]}
\end{alltt}
\begin{verbatim}
##  se(MLE)/se(env_wt) se(MLE)/se(env_varu) se(MLE)/se(env_fixu)
##               1.044                1.025                1.140
##               1.042                1.019                1.183
##               1.028                1.022                1.064
##               1.022                1.013                1.064
##               0.990                0.987                0.977
##               1.002                0.999                0.998
##               1.009                1.019                1.032
##               1.009                1.005                1.019
\end{verbatim}
\end{kframe}
\end{knitrout}

\section*{Poisson regression simulations}

\subsection*{Setting A:}

We reproduce the Poisson regression simulation in the main text. The setup is essesntially the same as the logistic regression simulations. We first create the basis matrix $\Gamma$ for the true envelope space and the basis matrix for its orthogonal complement $\Gamma_0$.

\begin{knitrout}
\definecolor{shadecolor}{rgb}{0.969, 0.969, 0.969}\color{fgcolor}\begin{kframe}
\begin{alltt}
\hlstd{p} \hlkwb{<-} \hlnum{8}\hlstd{; u} \hlkwb{<-} \hlnum{2}
\hlstd{v1} \hlkwb{<-} \hlkwd{matrix}\hlstd{(}\hlkwd{rep}\hlstd{(}\hlnum{1}\hlopt{/}\hlkwd{sqrt}\hlstd{(p),p),} \hlkwc{nrow} \hlstd{= p)}
\hlstd{O} \hlkwb{<-}\hlkwd{qr.Q}\hlstd{(}\hlkwd{qr}\hlstd{(v1),} \hlkwc{complete} \hlstd{=} \hlnum{TRUE}\hlstd{)}
\hlstd{Gamma} \hlkwb{<-} \hlstd{O[, (p}\hlopt{-}\hlstd{u}\hlopt{+}\hlnum{1}\hlstd{)}\hlopt{:}\hlstd{p]}
\hlstd{Gamma0} \hlkwb{<-} \hlstd{O[,} \hlnum{1}\hlopt{:}\hlstd{(p}\hlopt{-}\hlstd{u)]}
\end{alltt}
\end{kframe}
\end{knitrout}

We next create the core of the material and immaterial variation, denoted as $\Omega$ and $\Omega_0$ respectively.

\begin{knitrout}
\definecolor{shadecolor}{rgb}{0.969, 0.969, 0.969}\color{fgcolor}\begin{kframe}
\begin{alltt}
\hlstd{Omega} \hlkwb{<-} \hlkwd{diag}\hlstd{(}\hlkwd{c}\hlstd{(}\hlnum{1}\hlstd{,}\hlnum{10}\hlstd{))}
\hlstd{Omega0} \hlkwb{<-}  \hlkwd{diag}\hlstd{(}\hlkwd{exp}\hlstd{(}\hlopt{-}\hlnum{6}\hlopt{:-}\hlnum{1}\hlstd{))}
\end{alltt}
\end{kframe}
\end{knitrout}

We now build the variance matrix of the predictor variables and construct the true canonical parameter vector (regression coefficient vector) as an element contained in $\text{span}(\Gamma)$.

\begin{knitrout}
\definecolor{shadecolor}{rgb}{0.969, 0.969, 0.969}\color{fgcolor}\begin{kframe}
\begin{alltt}
\hlstd{SigmaX} \hlkwb{<-} \hlstd{Gamma}\hlopt{%*%}\hlstd{Omega}\hlopt{%*%}\hlkwd{t}\hlstd{(Gamma)} \hlopt{+} \hlstd{Gamma0}\hlopt{%*%}\hlstd{Omega0}\hlopt{%*%}\hlkwd{t}\hlstd{(Gamma0)}
\hlstd{beta} \hlkwb{<-} \hlstd{Gamma} \hlopt{%*%} \hlkwd{t}\hlstd{(Gamma)} \hlopt{%*%} \hlstd{(}\hlnum{1} \hlopt{+} \hlstd{(}\hlnum{1}\hlopt{:}\hlstd{p} \hlopt{-} \hlstd{p}\hlopt{/}\hlnum{2}\hlstd{)}\hlopt{/}\hlstd{(p}\hlopt{^}\hlnum{2}\hlstd{))}
\hlstd{eig} \hlkwb{<-} \hlkwd{eigen}\hlstd{(SigmaX)}
\hlstd{SigmaX.half} \hlkwb{<-} \hlstd{eig}\hlopt{$}\hlstd{vec} \hlopt{%*%} \hlkwd{diag}\hlstd{(}\hlkwd{sqrt}\hlstd{(eig}\hlopt{$}\hlstd{val))} \hlopt{%*%} \hlkwd{t}\hlstd{(eig}\hlopt{$}\hlstd{vec)}
\end{alltt}
\end{kframe}
\end{knitrout}

The poisson\_sim function below generates a Poisson regression model that incorporates the above envelope structure that is stored in your global environment. The function then calls on model\_boot (code is in the Appendix) to perform the nonparametric bootstrap with respect to all considered estimators. This function also returns the estimated envelope dimension at every iteration of the nonparametric bootstrap.

\begin{knitrout}
\definecolor{shadecolor}{rgb}{0.969, 0.969, 0.969}\color{fgcolor}\begin{kframe}
\begin{alltt}
\hlstd{poisson_sim} \hlkwb{<-} \hlkwa{function}\hlstd{(}\hlkwc{n}\hlstd{,} \hlkwc{p} \hlstd{= p)\{}
  \hlstd{X} \hlkwb{<-} \hlkwd{matrix}\hlstd{(}\hlkwd{rnorm}\hlstd{(n}\hlopt{*}\hlstd{p),} \hlkwc{nrow} \hlstd{= n)} \hlopt{%*%} \hlstd{SigmaX.half}
  \hlstd{gx} \hlkwb{<-} \hlkwd{as.numeric}\hlstd{(X} \hlopt{%*%} \hlstd{beta)}
  \hlstd{Y} \hlkwb{<-} \hlkwd{rpois}\hlstd{(n,} \hlkwc{lambda} \hlstd{=} \hlkwd{exp}\hlstd{(gx))}
  \hlstd{data_sim} \hlkwb{<-} \hlkwd{as.data.frame}\hlstd{(}\hlkwd{cbind}\hlstd{(Y, X))}
  \hlstd{m1} \hlkwb{<-} \hlkwd{glm}\hlstd{(Y} \hlopt{~ -}\hlnum{1} \hlopt{+} \hlstd{.,} \hlkwc{family} \hlstd{=} \hlstr{"poisson"}\hlstd{,} \hlkwc{data} \hlstd{= data_sim)}
  \hlkwd{model_boot}\hlstd{(}\hlkwc{model} \hlstd{= m1,} \hlkwc{nboot} \hlstd{= nboot,} \hlkwc{cores} \hlstd{= numCores)}
\hlstd{\}}
\end{alltt}
\end{kframe}
\end{knitrout}

We perform the nonparametric bootstrap at five different sample sizes.

\begin{knitrout}
\definecolor{shadecolor}{rgb}{0.969, 0.969, 0.969}\color{fgcolor}\begin{kframe}
\begin{alltt}
\hlkwd{set.seed}\hlstd{(}\hlnum{13}\hlstd{)}
\hlkwd{RNGkind}\hlstd{(}\hlstr{"L'Ecuyer-CMRG"}\hlstd{)}
\hlstd{n} \hlkwb{<-} \hlkwd{c}\hlstd{(}\hlnum{300}\hlstd{,} \hlnum{500}\hlstd{,} \hlnum{750}\hlstd{,} \hlnum{1000}\hlstd{)}
\hlstd{nboot} \hlkwb{<-} \hlnum{5000}
\hlkwd{system.time}\hlstd{(\{}
  \hlstd{psims} \hlkwb{<-} \hlkwd{lapply}\hlstd{(n,} \hlkwa{function}\hlstd{(}\hlkwc{j}\hlstd{)} \hlkwd{poisson_sim}\hlstd{(}\hlkwc{n} \hlstd{= j,} \hlkwc{p} \hlstd{= p))}
\hlstd{\})}
\end{alltt}
\begin{verbatim}
##      user    system   elapsed 
## 10135.183    81.872   691.811
\end{verbatim}
\end{kframe}
\end{knitrout}

The distribution of the estimated dimension across bootstrap iterations and sample sizes is depicted below:

\begin{knitrout}
\definecolor{shadecolor}{rgb}{0.969, 0.969, 0.969}\color{fgcolor}\begin{kframe}
\begin{alltt}
\hlstd{u_boot_p} \hlkwb{<-} \hlkwd{lapply}\hlstd{(}\hlnum{1}\hlopt{:}\hlkwd{length}\hlstd{(psims),} \hlkwa{function}\hlstd{(}\hlkwc{j}\hlstd{)\{}
  \hlkwd{round}\hlstd{(}\hlkwd{table}\hlstd{(psims[[j]][,} \hlnum{4}\hlopt{*}\hlstd{p}\hlopt{+}\hlnum{1}\hlstd{])} \hlopt{/} \hlstd{nboot,} \hlnum{3}\hlstd{)}
\hlstd{\})}

\hlcom{## n = 300}
\hlstd{u_boot_p[[}\hlnum{1}\hlstd{]]}
\end{alltt}
\begin{verbatim}
## 
##     1     2     3     4 
## 0.446 0.425 0.117 0.012
\end{verbatim}
\begin{alltt}
\hlcom{## n = 500}
\hlstd{u_boot_p[[}\hlnum{2}\hlstd{]]}
\end{alltt}
\begin{verbatim}
## 
##     1     2     3     4 
## 0.911 0.086 0.003 0.000
\end{verbatim}
\begin{alltt}
\hlcom{## n = 750}
\hlstd{u_boot_p[[}\hlnum{3}\hlstd{]]}
\end{alltt}
\begin{verbatim}
## 
##     1     2     3     4 
## 0.522 0.436 0.040 0.002
\end{verbatim}
\begin{alltt}
\hlcom{## n = 1000}
\hlstd{u_boot_p[[}\hlnum{4}\hlstd{]]}
\end{alltt}
\begin{verbatim}
## 
##     1     2     3     4 
## 0.901 0.097 0.002 0.000
\end{verbatim}
\end{kframe}
\end{knitrout}

The Frobenius norm of all boostrapped covariance matrices for all estimators across sample sizes is depicted below:

\begin{kframe}
\begin{alltt}
\hlstd{volume_boot_p} \hlkwb{<-} \hlkwd{do.call}\hlstd{(rbind,} \hlkwd{lapply}\hlstd{(}\hlnum{1}\hlopt{:}\hlkwd{length}\hlstd{(psims),} \hlkwa{function}\hlstd{(}\hlkwc{j}\hlstd{)\{}
  \hlkwd{unlist}\hlstd{(}\hlkwd{lapply}\hlstd{(}\hlkwd{normvar}\hlstd{(psims[[j]]),} \hlkwa{function}\hlstd{(}\hlkwc{x}\hlstd{)} \hlkwd{norm}\hlstd{(x,} \hlkwc{type}\hlstd{=}\hlstr{"F"}\hlstd{)))}
\hlstd{\}))}
\hlkwd{rownames}\hlstd{(volume_boot_p)} \hlkwb{<-} \hlstd{n}
\hlkwd{xtable}\hlstd{(volume_boot_p,} \hlkwc{digits} \hlstd{=} \hlnum{3}\hlstd{)}
\end{alltt}
\end{kframe}
\begin{table}[ht]
\centering
\begin{tabular}{rrrrr}
  \hline
 & se\_wt & se\_env & se\_env\_fixedu & se\_MLE \\ 
  \hline
300 & 0.814 & 0.904 & 0.286 & 1.192 \\ 
  500 & 0.123 & 0.160 & 0.003 & 0.748 \\ 
  750 & 0.025 & 0.030 & 0.000 & 0.525 \\ 
  1000 & 0.025 & 0.031 & 0.000 & 0.522 \\ 
   \hline
\end{tabular}
\end{table}

The estimated efficiency gains for envelope estimators ($\text{se}^*(\hat{\theta}) / \text{se}^*(\hat{\theta}_{\text{env}})$) across sample sizes is depicted below:

\begin{knitrout}
\definecolor{shadecolor}{rgb}{0.969, 0.969, 0.969}\color{fgcolor}\begin{kframe}
\begin{alltt}
\hlstd{SEs_boot_p} \hlkwb{<-} \hlkwd{lapply}\hlstd{(}\hlnum{1}\hlopt{:}\hlkwd{length}\hlstd{(psims),} \hlkwa{function}\hlstd{(}\hlkwc{j}\hlstd{)\{}
  \hlkwd{do.call}\hlstd{(cbind,} \hlkwd{lapply}\hlstd{(}\hlkwd{normvar}\hlstd{(psims[[j]]),} \hlkwa{function}\hlstd{(}\hlkwc{x}\hlstd{)} \hlkwd{sqrt}\hlstd{(}\hlkwd{diag}\hlstd{(x))))}
\hlstd{\})}
\hlstd{ratios_boot_p} \hlkwb{<-} \hlkwd{lapply}\hlstd{(}\hlnum{1}\hlopt{:}\hlkwd{length}\hlstd{(psims),} \hlkwa{function}\hlstd{(}\hlkwc{j}\hlstd{)\{}
  \hlcom{# ratio of SE(MLE) to SE(wtEnv)}
  \hlstd{out} \hlkwb{<-} \hlkwd{cbind}\hlstd{(SEs_boot_p[[j]][,} \hlnum{4}\hlstd{]} \hlopt{/} \hlstd{SEs_boot_p[[j]][,} \hlnum{1}\hlstd{],}
    \hlcom{# ratio of SE(MLE) to SE(Env)}
    \hlstd{SEs_boot_p[[j]][,} \hlnum{4}\hlstd{]} \hlopt{/} \hlstd{SEs_boot_p[[j]][,} \hlnum{2}\hlstd{],}
    \hlcom{# ratio of SE(MLE) to SE(Env_hat(u))}
    \hlstd{SEs_boot_p[[j]][,} \hlnum{4}\hlstd{]} \hlopt{/} \hlstd{SEs_boot_p[[j]][,} \hlnum{3}\hlstd{])}
  \hlkwd{colnames}\hlstd{(out)} \hlkwb{<-} \hlkwd{c}\hlstd{(}\hlstr{"se(MLE)/se(env_wt)"}\hlstd{,} \hlstr{"se(MLE)/se(env_varu)"}\hlstd{,}
                     \hlstr{"se(MLE)/se(env_fixu)"}\hlstd{)}
  \hlkwd{round}\hlstd{(out,} \hlnum{3}\hlstd{)}
\hlstd{\})}

\hlcom{## n = 300}
\hlstd{ratios_boot_p[[}\hlnum{1}\hlstd{]]}
\end{alltt}
\begin{verbatim}
##  se(MLE)/se(env_wt) se(MLE)/se(env_varu) se(MLE)/se(env_fixu)
##               1.152                1.115                2.086
##               0.893                0.837                1.470
##               0.999                0.937                1.563
##               2.480                2.381                4.364
##               2.279                2.127                3.707
##               2.344                2.157                3.452
##               2.785                2.712                4.875
##               2.783                2.698                5.668
\end{verbatim}
\begin{alltt}
\hlcom{## n = 500}
\hlstd{ratios_boot_p[[}\hlnum{2}\hlstd{]]}
\end{alltt}
\begin{verbatim}
##  se(MLE)/se(env_wt) se(MLE)/se(env_varu) se(MLE)/se(env_fixu)
##               2.579                2.287               17.290
##               3.670                3.179               39.262
##               2.820                2.486               21.889
##               2.679                2.347               18.994
##               2.376                2.089               16.124
##               2.403                2.104               15.983
##               2.390                2.094               15.526
##               2.353                2.073               12.627
\end{verbatim}
\begin{alltt}
\hlcom{## n = 750}
\hlstd{ratios_boot_p[[}\hlnum{3}\hlstd{]]}
\end{alltt}
\begin{verbatim}
##  se(MLE)/se(env_wt) se(MLE)/se(env_varu) se(MLE)/se(env_fixu)
##               3.810                3.479               61.878
##               5.573                5.277              341.419
##               2.975                2.711              235.859
##               4.420                3.991               47.244
##               3.407                3.059              247.080
##               4.870                4.487              180.729
##               3.545                2.984               24.367
##               4.794                4.439               23.493
\end{verbatim}
\begin{alltt}
\hlcom{## n = 1000}
\hlstd{ratios_boot_p[[}\hlnum{4}\hlstd{]]}
\end{alltt}
\begin{verbatim}
##  se(MLE)/se(env_wt) se(MLE)/se(env_varu) se(MLE)/se(env_fixu)
##               4.092                3.623               93.113
##               3.368                2.990              480.452
##               4.228                3.764              398.744
##               5.218                4.722              330.380
##               3.714                3.378              301.965
##               4.782                4.297              285.126
##               4.918                4.432              240.507
##               4.934                4.455               30.992
\end{verbatim}
\end{kframe}
\end{knitrout}

\subsection*{Setting B:}

We first create the basis matrix $\Gamma$ for the true envelope space and the basis matrix for its orthogonal complement $\Gamma_0$, build the variance matrix of the predictor variables, and construct the true canonical parameter vector (regression coefficient vector) as an element contained in $\text{span}(\Gamma)$.

\begin{knitrout}
\definecolor{shadecolor}{rgb}{0.969, 0.969, 0.969}\color{fgcolor}\begin{kframe}
\begin{alltt}
\hlstd{Omega} \hlkwb{<-} \hlkwd{diag}\hlstd{(}\hlkwd{exp}\hlstd{(}\hlkwd{c}\hlstd{(}\hlopt{-}\hlnum{3}\hlstd{,}\hlopt{-}\hlnum{2}\hlstd{)))}
\hlstd{Omega0} \hlkwb{<-}  \hlkwd{diag}\hlstd{(}\hlkwd{exp}\hlstd{(}\hlopt{-}\hlnum{4}\hlopt{:}\hlnum{1}\hlstd{))}
\hlstd{SigmaX} \hlkwb{<-} \hlstd{Gamma}\hlopt{%*%}\hlstd{Omega}\hlopt{%*%}\hlkwd{t}\hlstd{(Gamma)} \hlopt{+} \hlstd{Gamma0}\hlopt{%*%}\hlstd{Omega0}\hlopt{%*%}\hlkwd{t}\hlstd{(Gamma0)}
\hlstd{beta} \hlkwb{<-} \hlstd{Gamma} \hlopt{%*%} \hlkwd{t}\hlstd{(Gamma)} \hlopt{%*%} \hlkwd{rowMeans}\hlstd{(Gamma)}
\hlstd{eig} \hlkwb{<-} \hlkwd{eigen}\hlstd{(SigmaX)}
\hlstd{SigmaX.half} \hlkwb{<-} \hlstd{eig}\hlopt{$}\hlstd{vec} \hlopt{%*%} \hlkwd{diag}\hlstd{(}\hlkwd{sqrt}\hlstd{(eig}\hlopt{$}\hlstd{val))} \hlopt{%*%} \hlkwd{t}\hlstd{(eig}\hlopt{$}\hlstd{vec)}
\end{alltt}
\end{kframe}
\end{knitrout}

We perform the nonparametric bootstrap at five different sample sizes.

\begin{knitrout}
\definecolor{shadecolor}{rgb}{0.969, 0.969, 0.969}\color{fgcolor}\begin{kframe}
\begin{alltt}
\hlkwd{set.seed}\hlstd{(}\hlnum{13}\hlstd{)}
\hlkwd{RNGkind}\hlstd{(}\hlstr{"L'Ecuyer-CMRG"}\hlstd{)}
\hlstd{n} \hlkwb{<-} \hlkwd{c}\hlstd{(}\hlnum{300}\hlstd{,} \hlnum{500}\hlstd{,} \hlnum{750}\hlstd{,} \hlnum{1000}\hlstd{)}
\hlstd{nboot} \hlkwb{<-} \hlnum{5000}
\hlkwd{system.time}\hlstd{(\{}
  \hlstd{psimsB} \hlkwb{<-} \hlkwd{lapply}\hlstd{(n,} \hlkwa{function}\hlstd{(}\hlkwc{j}\hlstd{)} \hlkwd{poisson_sim}\hlstd{(}\hlkwc{n} \hlstd{= j,} \hlkwc{p} \hlstd{= p))}
\hlstd{\})}
\end{alltt}
\begin{verbatim}
##      user    system   elapsed 
## 12486.444    88.840   853.086
\end{verbatim}
\end{kframe}
\end{knitrout}

The distribution of the estimated dimension across bootstrap iterations and sample sizes is depicted below:

\begin{knitrout}
\definecolor{shadecolor}{rgb}{0.969, 0.969, 0.969}\color{fgcolor}\begin{kframe}
\begin{alltt}
\hlstd{u_boot_pB} \hlkwb{<-} \hlkwd{lapply}\hlstd{(}\hlnum{1}\hlopt{:}\hlkwd{length}\hlstd{(psimsB),} \hlkwa{function}\hlstd{(}\hlkwc{j}\hlstd{)\{}
  \hlkwd{round}\hlstd{(}\hlkwd{table}\hlstd{(psimsB[[j]][,} \hlnum{4}\hlopt{*}\hlstd{p}\hlopt{+}\hlnum{1}\hlstd{])} \hlopt{/} \hlstd{nboot,} \hlnum{3}\hlstd{)}
\hlstd{\})}

\hlcom{## n = 300}
\hlstd{u_boot_pB[[}\hlnum{1}\hlstd{]]}
\end{alltt}
\begin{verbatim}
## 
##     1     2     3     4 
## 0.521 0.387 0.086 0.005
\end{verbatim}
\begin{alltt}
\hlcom{## n = 500}
\hlstd{u_boot_pB[[}\hlnum{2}\hlstd{]]}
\end{alltt}
\begin{verbatim}
## 
##     1     2     3 
## 0.785 0.200 0.015
\end{verbatim}
\begin{alltt}
\hlcom{## n = 750}
\hlstd{u_boot_pB[[}\hlnum{3}\hlstd{]]}
\end{alltt}
\begin{verbatim}
## 
##     1     2     3     4 
## 0.081 0.785 0.129 0.004
\end{verbatim}
\begin{alltt}
\hlcom{## n = 1000}
\hlstd{u_boot_pB[[}\hlnum{4}\hlstd{]]}
\end{alltt}
\begin{verbatim}
## 
##     1     2     3     4 
## 0.730 0.245 0.025 0.000
\end{verbatim}
\end{kframe}
\end{knitrout}

The Frobenius norm of all boostrapped covariance matrices for all estimators across sample sizes is depicted below:

\begin{kframe}
\begin{alltt}
\hlstd{volume_boot_pB} \hlkwb{<-} \hlkwd{do.call}\hlstd{(rbind,} \hlkwd{lapply}\hlstd{(}\hlnum{1}\hlopt{:}\hlkwd{length}\hlstd{(psimsB),} \hlkwa{function}\hlstd{(}\hlkwc{j}\hlstd{)\{}
  \hlkwd{unlist}\hlstd{(}\hlkwd{lapply}\hlstd{(}\hlkwd{normvar}\hlstd{(psimsB[[j]]),} \hlkwa{function}\hlstd{(}\hlkwc{x}\hlstd{)} \hlkwd{norm}\hlstd{(x,} \hlkwc{type}\hlstd{=}\hlstr{"F"}\hlstd{)))}
\hlstd{\}))}
\hlkwd{rownames}\hlstd{(volume_boot_pB)} \hlkwb{<-} \hlstd{n}
\hlkwd{xtable}\hlstd{(volume_boot_pB,} \hlkwc{digits} \hlstd{=} \hlnum{3}\hlstd{)}
\end{alltt}
\end{kframe}
\begin{table}[ht]
\centering
\begin{tabular}{rrrrr}
  \hline
 & se\_wt & se\_env & se\_env\_fixedu & se\_MLE \\ 
  \hline
300 & 0.199 & 0.214 & 0.195 & 0.167 \\ 
  500 & 0.058 & 0.073 & 0.025 & 0.106 \\ 
  750 & 0.062 & 0.064 & 0.058 & 0.076 \\ 
  1000 & 0.037 & 0.045 & 0.009 & 0.073 \\ 
   \hline
\end{tabular}
\end{table}

The estimated efficiency gains for envelope estimators ($\text{se}^*(\hat{\theta}) / \text{se}^*(\hat{\theta}_{\text{env}})$) across sample sizes is depicted below:

\begin{knitrout}
\definecolor{shadecolor}{rgb}{0.969, 0.969, 0.969}\color{fgcolor}\begin{kframe}
\begin{alltt}
\hlstd{SEs_boot_pB} \hlkwb{<-} \hlkwd{lapply}\hlstd{(}\hlnum{1}\hlopt{:}\hlkwd{length}\hlstd{(psimsB),} \hlkwa{function}\hlstd{(}\hlkwc{j}\hlstd{)\{}
  \hlkwd{do.call}\hlstd{(cbind,} \hlkwd{lapply}\hlstd{(}\hlkwd{normvar}\hlstd{(psimsB[[j]]),} \hlkwa{function}\hlstd{(}\hlkwc{x}\hlstd{)} \hlkwd{sqrt}\hlstd{(}\hlkwd{diag}\hlstd{(x))))}
\hlstd{\})}
\hlstd{ratios_boot_pB} \hlkwb{<-} \hlkwd{lapply}\hlstd{(}\hlnum{1}\hlopt{:}\hlkwd{length}\hlstd{(psimsB),} \hlkwa{function}\hlstd{(}\hlkwc{j}\hlstd{)\{}
  \hlcom{# ratio of SE(MLE) to SE(wtEnv)}
  \hlstd{out} \hlkwb{<-} \hlkwd{cbind}\hlstd{(SEs_boot_pB[[j]][,} \hlnum{4}\hlstd{]} \hlopt{/} \hlstd{SEs_boot_pB[[j]][,} \hlnum{1}\hlstd{],}
    \hlcom{# ratio of SE(MLE) to SE(Env) }
    \hlstd{SEs_boot_pB[[j]][,} \hlnum{4}\hlstd{]} \hlopt{/} \hlstd{SEs_boot_pB[[j]][,} \hlnum{2}\hlstd{],}
    \hlcom{# ratio of SE(MLE) to SE(Env_hat(u))}
    \hlstd{SEs_boot_pB[[j]][,} \hlnum{4}\hlstd{]} \hlopt{/} \hlstd{SEs_boot_pB[[j]][,} \hlnum{3}\hlstd{])}
  \hlkwd{colnames}\hlstd{(out)} \hlkwb{<-} \hlkwd{c}\hlstd{(}\hlstr{"se(MLE)/se(env_wt)"}\hlstd{,} \hlstr{"se(MLE)/se(env_varu)"}\hlstd{,}
                     \hlstr{"se(MLE)/se(env_fixu)"}\hlstd{)}
  \hlkwd{round}\hlstd{(out,} \hlnum{3}\hlstd{)}
\hlstd{\})}

\hlcom{## n = 300}
\hlstd{ratios_boot_pB[[}\hlnum{1}\hlstd{]]}
\end{alltt}
\begin{verbatim}
##  se(MLE)/se(env_wt) se(MLE)/se(env_varu) se(MLE)/se(env_fixu)
##               1.153                1.133                1.407
##               0.704                0.669                0.817
##               0.950                0.911                0.840
##               1.492                1.470                1.804
##               1.556                1.479                1.939
##               1.603                1.512                1.899
##               1.120                1.088                1.361
##               0.810                0.769                0.650
\end{verbatim}
\begin{alltt}
\hlcom{## n = 500}
\hlstd{ratios_boot_pB[[}\hlnum{2}\hlstd{]]}
\end{alltt}
\begin{verbatim}
##  se(MLE)/se(env_wt) se(MLE)/se(env_varu) se(MLE)/se(env_fixu)
##               1.346                1.211                2.377
##               1.780                1.600                4.014
##               1.115                1.102                1.165
##               2.162                2.020                3.292
##               2.156                1.982                4.672
##               2.387                2.189                5.502
##               1.007                0.872                3.387
##               1.234                1.202                1.373
\end{verbatim}
\begin{alltt}
\hlcom{## n = 750}
\hlstd{ratios_boot_pB[[}\hlnum{3}\hlstd{]]}
\end{alltt}
\begin{verbatim}
##  se(MLE)/se(env_wt) se(MLE)/se(env_varu) se(MLE)/se(env_fixu)
##               1.124                1.104                1.168
##               1.089                1.068                1.064
##               1.204                1.188                1.196
##               1.941                1.851                1.969
##               1.973                1.885                1.977
##               2.274                2.176                2.323
##               0.815                0.798                0.873
##               1.226                1.224                1.250
\end{verbatim}
\begin{alltt}
\hlcom{## n = 1000}
\hlstd{ratios_boot_pB[[}\hlnum{4}\hlstd{]]}
\end{alltt}
\begin{verbatim}
##  se(MLE)/se(env_wt) se(MLE)/se(env_varu) se(MLE)/se(env_fixu)
##               1.584                1.486                2.852
##               1.611                1.463                7.292
##               1.651                1.611                1.793
##               2.615                2.446                4.653
##               2.605                2.377                7.169
##               2.723                2.495                7.324
##               0.949                0.849                8.528
##               1.454                1.427                1.615
\end{verbatim}
\end{kframe}
\end{knitrout}

\subsection*{Build Table in main text}

We now build Table 1 in the main text.

\begin{knitrout}
\definecolor{shadecolor}{rgb}{0.969, 0.969, 0.969}\color{fgcolor}\begin{kframe}
\begin{alltt}
\hlstd{tab_sim} \hlkwb{<-} \hlkwd{rbind}\hlstd{(}\hlkwd{cbind}\hlstd{(n,} \hlkwd{do.call}\hlstd{(rbind,} \hlkwd{lapply}\hlstd{(ratios_boot_l,}
  \hlkwa{function}\hlstd{(}\hlkwc{xx}\hlstd{) xx[}\hlnum{1}\hlstd{, ])),}
  \hlkwd{do.call}\hlstd{(rbind,} \hlkwd{lapply}\hlstd{(ratios_boot_lB,} \hlkwa{function}\hlstd{(}\hlkwc{xx}\hlstd{) xx[}\hlnum{1}\hlstd{, ]))),}
  \hlkwd{cbind}\hlstd{(n,} \hlkwd{do.call}\hlstd{(rbind,} \hlkwd{lapply}\hlstd{(ratios_boot_p,} \hlkwa{function}\hlstd{(}\hlkwc{xx}\hlstd{) xx[}\hlnum{1}\hlstd{, ])),}
    \hlkwd{do.call}\hlstd{(rbind,} \hlkwd{lapply}\hlstd{(ratios_boot_pB,} \hlkwa{function}\hlstd{(}\hlkwc{xx}\hlstd{) xx[}\hlnum{1}\hlstd{, ]))))}
\hlkwd{xtable}\hlstd{(tab_sim,} \hlkwc{digits} \hlstd{=} \hlnum{2}\hlstd{)}
\end{alltt}
\begin{verbatim}
## % latex table generated in R 3.6.1 by xtable 1.8-4 package
## % Mon Feb  3 13:49:00 2020
## \begin{table}[ht]
## \centering
## \begin{tabular}{rrrrrrrr}
##   \hline
##  & n & se(MLE)/se(env\_wt) & se(MLE)/se(env\_varu) & se(MLE)/se(env\_fixu) & se(MLE)/se(env\_wt) & se(MLE)/se(env\_varu) & se(MLE)/se(env\_fixu) \\ 
##   \hline
## 1 & 300.00 & 1.22 & 1.16 & 3.26 & 0.99 & 0.97 & 1.13 \\ 
##   2 & 500.00 & 1.88 & 1.76 & 2.88 & 1.04 & 1.02 & 1.14 \\ 
##   3 & 750.00 & 1.03 & 0.95 & 3.64 & 1.29 & 1.27 & 1.48 \\ 
##   4 & 1000.00 & 1.00 & 0.88 & 3.50 & 1.04 & 1.02 & 1.14 \\ 
##   5 & 300.00 & 1.15 & 1.11 & 2.09 & 1.15 & 1.13 & 1.41 \\ 
##   6 & 500.00 & 2.58 & 2.29 & 17.29 & 1.35 & 1.21 & 2.38 \\ 
##   7 & 750.00 & 3.81 & 3.48 & 61.88 & 1.12 & 1.10 & 1.17 \\ 
##   8 & 1000.00 & 4.09 & 3.62 & 93.11 & 1.58 & 1.49 & 2.85 \\ 
##    \hline
## \end{tabular}
## \end{table}
\end{verbatim}
\end{kframe}
\end{knitrout}

\section*{Diabetes example}

Model free envelope estimation techniques and maximum likelihood estimation are then used to estimate the canonical parameter vector corresponding to a logistic regression model (the regression coefficient vector with inverse logit link function) for diabetes diagnoses. 

We load in the diabetes data from the faraway package.  Log transformations are used for several predictor variables as a means to transform the variable to approximate normality while maintaining a scale that is interpretable. The respose variable is a diagnosis of diabetes based in an individual's hemoglobin percentage (HbA1c). A positive diagnosis is when HbA1c $> 6.5\%$. Only complete records are kept for this analysis.

\begin{knitrout}
\definecolor{shadecolor}{rgb}{0.969, 0.969, 0.969}\color{fgcolor}\begin{kframe}
\begin{alltt}
\hlkwd{data}\hlstd{(diabetes)}
\hlcom{## add diagnosis and roughly transform predictors to univarite normality}
\hlstd{dat} \hlkwb{<-} \hlstd{diabetes} \hlopt{%>%} \hlkwd{mutate}\hlstd{(}\hlkwc{diagnose} \hlstd{=} \hlkwd{ifelse}\hlstd{(glyhb} \hlopt{>} \hlnum{6.5}\hlstd{,}\hlnum{1}\hlstd{,}\hlnum{0}\hlstd{))} \hlopt{%>%}
  \hlkwd{mutate}\hlstd{(}\hlkwc{l.stab.glu} \hlstd{=} \hlkwd{log}\hlstd{(stab.glu),} \hlkwc{l.weight} \hlstd{=} \hlkwd{log}\hlstd{(weight),}
         \hlkwc{l.age} \hlstd{=} \hlkwd{log}\hlstd{(age),} \hlkwc{l.hip} \hlstd{=} \hlkwd{log}\hlstd{(hip),} \hlkwc{l.waist} \hlstd{=} \hlkwd{log}\hlstd{(waist),}
         \hlkwc{l.height} \hlstd{=} \hlkwd{log}\hlstd{(height))} \hlopt{%>%}
    \hlstd{dplyr}\hlopt{::}\hlkwd{select}\hlstd{(diagnose, l.age, l.weight, l.height, l.waist, l.hip,}
                  \hlstd{gender, l.stab.glu)}
\hlcom{## exclude missing observations}
\hlstd{dat} \hlkwb{<-} \hlkwd{na.omit}\hlstd{(dat)}
\hlcom{## turn gender to factor variable}
\hlstd{dat}\hlopt{$}\hlstd{gender} \hlkwb{<-} \hlkwd{as.factor}\hlstd{(dat}\hlopt{$}\hlstd{gender)}
\end{alltt}
\end{kframe}
\end{knitrout}

Here are density plots of the log transformed predictor variables.

\begin{knitrout}
\definecolor{shadecolor}{rgb}{0.969, 0.969, 0.969}\color{fgcolor}\begin{kframe}
\begin{alltt}
\hlstd{dat} \hlopt{%>%} \hlstd{dplyr}\hlopt{::}\hlkwd{select}\hlstd{(}\hlopt{-}\hlstd{diagnose)} \hlopt{%>%} \hlkwd{keep}\hlstd{(is.numeric)} \hlopt{%>%}
  \hlkwd{gather}\hlstd{()} \hlopt{%>%} \hlkwd{ggplot}\hlstd{(}\hlkwd{aes}\hlstd{(value))} \hlopt{+}
    \hlkwd{facet_wrap}\hlstd{(}\hlopt{~} \hlstd{key,} \hlkwc{scales} \hlstd{=} \hlstr{"free"}\hlstd{)} \hlopt{+}
    \hlkwd{theme_minimal}\hlstd{()} \hlopt{+}
    \hlkwd{geom_density}\hlstd{()}
\end{alltt}
\end{kframe}
\includegraphics[width=\maxwidth]{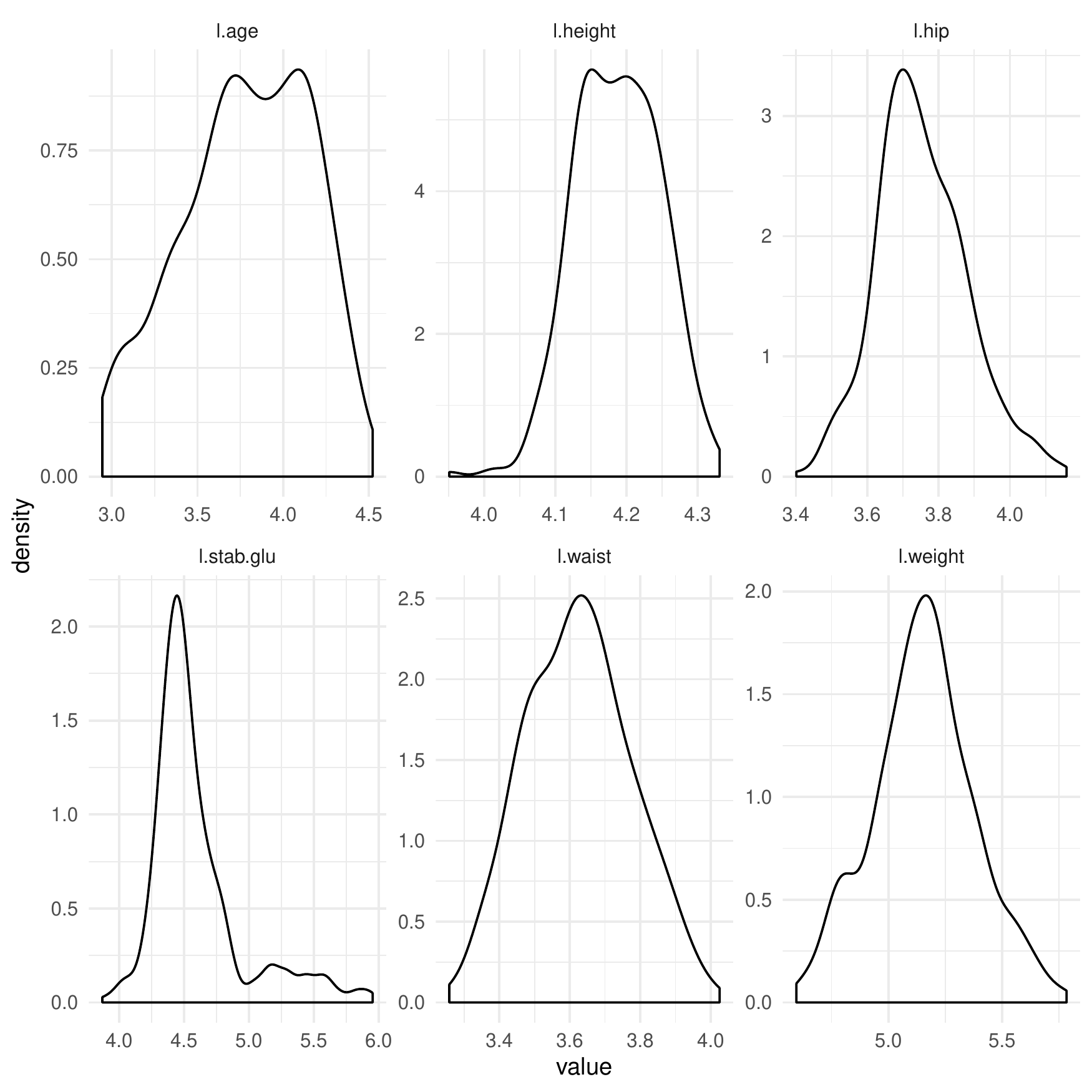} 

\end{knitrout}

We now fit the model with diagnosis as a response variable and log transformed versions of age, weight, height, waist, hip, and stabilized glucose as predictors.  

\begin{kframe}
\begin{alltt}
\hlstd{m1} \hlkwb{<-} \hlkwd{glm}\hlstd{(diagnose} \hlopt{~} \hlstd{.,} \hlkwc{family} \hlstd{=} \hlstr{"binomial"}\hlstd{,} \hlkwc{data} \hlstd{= dat,}
          \hlkwc{x} \hlstd{=} \hlnum{TRUE}\hlstd{,} \hlkwc{y} \hlstd{=} \hlnum{TRUE}\hlstd{)}
\hlstd{betahat} \hlkwb{<-} \hlstd{m1}\hlopt{$}\hlstd{coefficients}
\hlkwd{xtable}\hlstd{(}\hlkwd{summary}\hlstd{(m1))}
\end{alltt}
\end{kframe}
\begin{table}[ht]
\centering
\begin{tabular}{rrrrr}
  \hline
 & Estimate & Std. Error & z value & Pr($>$$|$z$|$) \\ 
  \hline
(Intercept) & -21.3575 & 23.0558 & -0.93 & 0.3543 \\ 
  l.age & 2.0285 & 0.7583 & 2.67 & 0.0075 \\ 
  l.weight & 1.2546 & 2.4216 & 0.52 & 0.6044 \\ 
  l.height & -4.3927 & 5.2798 & -0.83 & 0.4054 \\ 
  l.waist & 2.6525 & 3.0997 & 0.86 & 0.3922 \\ 
  l.hip & -2.6424 & 3.6213 & -0.73 & 0.4656 \\ 
  genderfemale & 0.1744 & 0.6025 & 0.29 & 0.7722 \\ 
  l.stab.glu & 5.0333 & 0.6602 & 7.62 & 0.0000 \\ 
   \hline
\end{tabular}
\end{table}

We now perform the nonparametric bootstrap procedure for all envelope estimators mentioned in the main text and the MLE. This nonparametric bootstrap has a bootstrap sample size of $5000$.  

\begin{knitrout}
\definecolor{shadecolor}{rgb}{0.969, 0.969, 0.969}\color{fgcolor}\begin{kframe}
\begin{alltt}
\hlkwd{set.seed}\hlstd{(}\hlnum{13}\hlstd{)}
\hlstd{numCores} \hlkwb{<-} \hlkwd{detectCores}\hlstd{()}
\hlstd{nboot} \hlkwb{<-} \hlnum{5000}
\hlkwd{RNGkind}\hlstd{(}\hlstr{"L'Ecuyer-CMRG"}\hlstd{)}
\hlstd{boot_sample_diabetes} \hlkwb{<-} \hlkwd{model_boot}\hlstd{(}\hlkwc{model} \hlstd{= m1,} \hlkwc{nboot} \hlstd{= nboot,}
  \hlkwc{cores} \hlstd{= numCores,} \hlkwc{intercept} \hlstd{=} \hlnum{TRUE}\hlstd{)}
\end{alltt}
\end{kframe}
\end{knitrout}

The distribution of the estimated dimension across bootstrap iterations is depicted below. A non-trivial amount of model selection volitility exists across iterations of our nonparametric bootstrap.

\begin{knitrout}
\definecolor{shadecolor}{rgb}{0.969, 0.969, 0.969}\color{fgcolor}\begin{kframe}
\begin{alltt}
\hlkwd{round}\hlstd{(}\hlkwd{table}\hlstd{(boot_sample_diabetes[,} \hlkwd{ncol}\hlstd{(boot_sample_diabetes)])} \hlopt{/} \hlstd{nboot,} \hlnum{3}\hlstd{)}
\end{alltt}
\begin{verbatim}
## 
##     1     2     3     4     5 
## 0.568 0.358 0.067 0.007 0.000
\end{verbatim}
\end{kframe}
\end{knitrout}

The Frobenius norm of all boostrapped covariance matrices for all estimators is depicted below:

\begin{knitrout}
\definecolor{shadecolor}{rgb}{0.969, 0.969, 0.969}\color{fgcolor}\begin{kframe}
\begin{alltt}
\hlkwd{unlist}\hlstd{(}\hlkwd{lapply}\hlstd{(}\hlkwd{normvar}\hlstd{(boot_sample_diabetes),} \hlkwa{function}\hlstd{(}\hlkwc{x}\hlstd{)} \hlkwd{norm}\hlstd{(x,} \hlkwc{type}\hlstd{=}\hlstr{"F"}\hlstd{)))}
\end{alltt}
\begin{verbatim}
##         se_wt        se_env se_env_fixedu        se_MLE 
##     26.762481     30.571294      1.040725     37.316460
\end{verbatim}
\end{kframe}
\end{knitrout}

Ratios of bootstrap standard errors are reported below. These ratios compare the MLE to the envelope estimators under study. The bootstrap standard error for the MLE is displayed in the numerator, so that a value greater than 1 indicate variance reduction via envelope methodology.

\begin{kframe}
\begin{alltt}
\hlstd{SEs_diabetes} \hlkwb{<-} \hlkwd{do.call}\hlstd{(cbind,} \hlkwd{lapply}\hlstd{(}\hlkwd{normvar}\hlstd{(boot_sample_diabetes),}
  \hlkwa{function}\hlstd{(}\hlkwc{x}\hlstd{)} \hlkwd{sqrt}\hlstd{(}\hlkwd{diag}\hlstd{(x))))}
\hlstd{ratios_diabetes} \hlkwb{<-} \hlkwd{cbind}\hlstd{(SEs_diabetes[,} \hlnum{4}\hlstd{]} \hlopt{/}
  \hlstd{SEs_diabetes[,} \hlnum{1}\hlstd{],} \hlcom{# ratio of SE(MLE) to SE(wtEnv)}
  \hlcom{# ratio of SE(MLE) to SE(Env)}
  \hlstd{SEs_diabetes[,} \hlnum{4}\hlstd{]} \hlopt{/} \hlstd{SEs_diabetes[,} \hlnum{2}\hlstd{],}
  \hlcom{# ratio of SE(MLE) to SE(Env_hat(u))}
  \hlstd{SEs_diabetes[,} \hlnum{4}\hlstd{]} \hlopt{/} \hlstd{SEs_diabetes[,} \hlnum{3}\hlstd{])}
\hlkwd{colnames}\hlstd{(ratios_diabetes)} \hlkwb{<-} \hlkwd{c}\hlstd{(}\hlstr{"se(MLE)/se(env_wt)"}\hlstd{,}
  \hlstr{"se(MLE)/se(env_varu)"}\hlstd{,} \hlstr{"se(MLE)/se(env_fixu)"}\hlstd{)}
\hlkwd{rownames}\hlstd{(ratios_diabetes)} \hlkwb{<-} \hlkwd{names}\hlstd{(m1}\hlopt{$}\hlstd{coefficients)[}\hlopt{-}\hlnum{1}\hlstd{]}
\hlkwd{xtable}\hlstd{(ratios_diabetes,} \hlkwc{digits} \hlstd{=} \hlnum{3}\hlstd{)}
\end{alltt}
\end{kframe}
\begin{table}[ht]
\centering
\begin{tabular}{rrrr}
  \hline
 & se(MLE)/se(env\_wt) & se(MLE)/se(env\_varu) & se(MLE)/se(env\_fixu) \\ 
  \hline
l.age & 1.241 & 1.208 & 1.191 \\ 
  l.weight & 1.581 & 1.475 & 7.070 \\ 
  l.height & 1.156 & 1.073 & 54.761 \\ 
  l.waist & 1.541 & 1.448 & 11.599 \\ 
  l.hip & 1.296 & 1.220 & 17.800 \\ 
  genderfemale & 1.192 & 1.170 & 1.197 \\ 
  l.stab.glu & 1.147 & 1.146 & 1.287 \\ 
   \hline
\end{tabular}
\end{table}

The estimated envelope dimension and weights vector are displayed in the code below. We can see that the estimated envelope dimension is 1 in the original sample, and the vector of weights is close to a point mass at 1. The weighted envelope estimator and the envelope estimator at the estimated dimension are very similar.  That being said, the bootstrap distribution for the estimated envelope dimension $u$ is far from a point mass at 1 and the bootstrap standard errors for the weighted envelope estimator are lower than those obtained by using the variable $u$ approach which selects $u$ at every dimension.

\begin{knitrout}
\definecolor{shadecolor}{rgb}{0.969, 0.969, 0.969}\color{fgcolor}\begin{kframe}
\begin{alltt}
\hlstd{Y} \hlkwb{<-} \hlstd{dat}\hlopt{$}\hlstd{diagnose}
\hlstd{X} \hlkwb{<-} \hlkwd{as.matrix}\hlstd{(m1}\hlopt{$}\hlstd{x)[,}\hlopt{-}\hlnum{1}\hlstd{]}
\hlstd{n} \hlkwb{<-} \hlkwd{nrow}\hlstd{(X)}
\hlstd{a} \hlkwb{<-} \hlstd{betahat[}\hlnum{1}\hlstd{]}
\hlstd{b} \hlkwb{<-} \hlstd{betahat[}\hlopt{-}\hlnum{1}\hlstd{]}
\hlstd{model_cov} \hlkwb{<-} \hlkwd{Logistic_cov}\hlstd{(}\hlkwc{Y} \hlstd{= Y,} \hlkwc{X} \hlstd{= X,} \hlkwc{a} \hlstd{= a,} \hlkwc{b} \hlstd{= b)}
\hlstd{M} \hlkwb{<-} \hlstd{model_cov}\hlopt{$}\hlstd{M; U} \hlkwb{<-} \hlstd{model_cov}\hlopt{$}\hlstd{U}
\hlstd{bic_val} \hlkwb{<-} \hlkwd{bic_compute}\hlstd{(}\hlkwc{M} \hlstd{= M,} \hlkwc{U} \hlstd{= U,} \hlkwc{n} \hlstd{= n)}

\hlcom{## estimated dimension in the original sample}
\hlstd{u} \hlkwb{<-} \hlkwd{which.min}\hlstd{(bic_val)}
\hlstd{u}
\end{alltt}
\begin{verbatim}
## [1] 1
\end{verbatim}
\begin{alltt}
\hlcom{## estimated weights for the weighted technique in }
\hlcom{## the original sample}
\hlstd{min_bic_val} \hlkwb{<-} \hlkwd{min}\hlstd{(bic_val)}
\hlstd{wt_bic} \hlkwb{<-} \hlkwd{exp}\hlstd{(min_bic_val} \hlopt{-} \hlstd{bic_val)} \hlopt{/} \hlkwd{sum}\hlstd{(}\hlkwd{exp}\hlstd{(min_bic_val} \hlopt{-} \hlstd{bic_val))}
\hlkwd{round}\hlstd{(wt_bic,} \hlnum{4}\hlstd{)}
\end{alltt}
\begin{verbatim}
## [1] 0.9921 0.0078 0.0001 0.0000 0.0000 0.0000 0.0000
\end{verbatim}
\end{kframe}
\end{knitrout}

We now replicate 
Table 2 in the main text. This table displays the performance of envelope estimates of the regression coefficients (canonical parameter vector) for the logistic regression of diabetes diagnosis on seven predictors. The first, third, and sixth column display the weighted envelope estimator, the envelope estimator with $\hat{u} = 1$, and the MLE respectively. The second column displays the bootstrap standard error of the weighted envelope estimator. The fourth and fifth columns display the bootstrap standard error for the envelope estimator under the variable $u$ and fixed $u$ regimes respectively. The seventh column displays the bootstrap standard error of the MLE. The last three columns displays the ratio of bootstrap standard errors of all envelope estimators to the those of the MLE.

\begin{kframe}
\begin{alltt}
\hlstd{Ghat} \hlkwb{<-} \hlkwd{manifold1D}\hlstd{(}\hlkwc{M} \hlstd{= M,} \hlkwc{U} \hlstd{= U,} \hlkwc{u} \hlstd{= u)}
\hlstd{Env_diabetes} \hlkwb{<-} \hlkwd{as.numeric}\hlstd{((Ghat} \hlopt{%*%} \hlkwd{t}\hlstd{(Ghat))} \hlopt{%*%} \hlstd{b)}
\hlstd{Env_wt_diabetes} \hlkwb{<-} \hlkwd{as.numeric}\hlstd{(}\hlkwd{wtenv}\hlstd{(M, U,} \hlkwc{wt} \hlstd{= wt_bic, b))}
\hlstd{inference_diabetes} \hlkwb{<-} \hlkwd{round}\hlstd{(}\hlkwd{cbind}\hlstd{(Env_wt_diabetes,}
  \hlstd{SEs_diabetes[,} \hlnum{1}\hlstd{],Env_diabetes, SEs_diabetes[,} \hlnum{2}\hlstd{],}
  \hlstd{SEs_diabetes[,} \hlnum{3}\hlstd{], b, SEs_diabetes[,} \hlnum{4}\hlstd{]),} \hlnum{4}\hlstd{)}
\hlkwd{rownames}\hlstd{(inference_diabetes)} \hlkwb{<-} \hlkwd{names}\hlstd{(betahat)[}\hlopt{-}\hlnum{1}\hlstd{]}
\hlstd{tab} \hlkwb{<-} \hlkwd{cbind}\hlstd{(inference_diabetes, ratios_diabetes)}
\hlkwd{colnames}\hlstd{(tab)} \hlkwb{<-} \hlkwa{NULL}
\hlkwd{xtable}\hlstd{(tab,} \hlkwc{digits} \hlstd{=} \hlnum{3}\hlstd{)}
\end{alltt}
\end{kframe}
\begin{table}[ht]
\centering
\begin{tabular}{rrrrrrrrrrr}
  \hline
 & 1 & 2 & 3 & 4 & 5 & 6 & 7 & 8 & 9 & 10 \\ 
  \hline
l.age & 1.775 & 0.660 & 1.775 & 0.679 & 0.688 & 2.029 & 0.820 & 1.241 & 1.208 & 1.191 \\ 
  l.weight & 0.702 & 1.685 & 0.702 & 1.806 & 0.377 & 1.255 & 2.665 & 1.581 & 1.475 & 7.070 \\ 
  l.height & 0.033 & 4.388 & 0.033 & 4.728 & 0.093 & -4.393 & 5.074 & 1.156 & 1.073 & 54.761 \\ 
  l.waist & 0.681 & 1.932 & 0.681 & 2.056 & 0.257 & 2.652 & 2.978 & 1.541 & 1.448 & 11.599 \\ 
  l.hip & 0.491 & 3.263 & 0.492 & 3.466 & 0.238 & -2.642 & 4.229 & 1.296 & 1.220 & 17.800 \\ 
  genderfemale & 0.401 & 0.624 & 0.402 & 0.636 & 0.621 & 0.174 & 0.744 & 1.192 & 1.170 & 1.197 \\ 
  l.stab.glu & 5.112 & 0.932 & 5.112 & 0.933 & 0.831 & 5.033 & 1.069 & 1.147 & 1.146 & 1.287 \\ 
   \hline
\end{tabular}
\end{table}

\bibliographystyle{plainnat}
\bibliography{envelopesources}

\newpage
\section*{Appendix: R code}

\begin{knitrout}
\definecolor{shadecolor}{rgb}{0.969, 0.969, 0.969}\color{fgcolor}\begin{kframe}
\begin{alltt}
\hlcom{##################################################}
\hlcom{#         1D optimization solve for gamma        #}
\hlcom{##################################################}

\hlcom{## stored in global environment}
\hlstd{ballGBB1D} \hlkwb{<-} \hlkwa{function}\hlstd{(}\hlkwc{M}\hlstd{,} \hlkwc{U}\hlstd{,} \hlkwc{opts}\hlstd{=}\hlkwa{NULL}\hlstd{) \{}
  \hlstd{W0} \hlkwb{<-} \hlkwd{get_ini1D}\hlstd{(M, U)}
  \hlkwa{if} \hlstd{(}\hlkwd{is.null}\hlstd{(opts}\hlopt{$}\hlstd{xtol))}
    \hlstd{opts}\hlopt{$}\hlstd{xtol} \hlkwb{=} \hlnum{1e-8} \hlkwa{else if} \hlstd{(opts}\hlopt{$}\hlstd{xtol} \hlopt{<} \hlnum{0} \hlopt{||} \hlstd{opts}\hlopt{$}\hlstd{xtol} \hlopt{>} \hlnum{1}\hlstd{)}
      \hlstd{opts}\hlopt{$}\hlstd{xtol} \hlkwb{=} \hlnum{1e-8}

    \hlkwa{if} \hlstd{(}\hlkwd{is.null}\hlstd{(opts}\hlopt{$}\hlstd{gtol))}
      \hlstd{opts}\hlopt{$}\hlstd{gtol} \hlkwb{=} \hlnum{1e-8} \hlkwa{else if} \hlstd{(opts}\hlopt{$}\hlstd{gtol} \hlopt{<} \hlnum{0} \hlopt{||} \hlstd{opts}\hlopt{$}\hlstd{gtol} \hlopt{>} \hlnum{1}\hlstd{)}
        \hlstd{opts}\hlopt{$}\hlstd{gtol} \hlkwb{=} \hlnum{1e-8}

      \hlkwa{if} \hlstd{(}\hlkwd{is.null}\hlstd{(opts}\hlopt{$}\hlstd{ftol))}
        \hlstd{opts}\hlopt{$}\hlstd{ftol} \hlkwb{=} \hlnum{1e-12} \hlkwa{else if} \hlstd{(opts}\hlopt{$}\hlstd{ftol} \hlopt{<} \hlnum{0} \hlopt{||} \hlstd{opts}\hlopt{$}\hlstd{ftol} \hlopt{>} \hlnum{1}\hlstd{)}
          \hlstd{opts}\hlopt{$}\hlstd{ftol} \hlkwb{=} \hlnum{1e-12}

        \hlkwa{if} \hlstd{(}\hlkwd{is.null}\hlstd{(opts}\hlopt{$}\hlstd{mxitr))}
          \hlstd{opts}\hlopt{$}\hlstd{mxitr} \hlkwb{=} \hlnum{800}

        \hlstd{X} \hlkwb{<-} \hlkwd{OptManiMulitBallGBB}\hlstd{(W0, opts, fun1D, M, U)}\hlopt{$}\hlstd{X}
        \hlkwd{return}\hlstd{(X)}
\hlstd{\}}

\hlcom{## compute M and U with normal predictors for logistic}
\hlcom{## regression model with normal predictors}
\hlstd{Logistic_cov} \hlkwb{<-} \hlkwa{function}\hlstd{(}\hlkwc{Y}\hlstd{,}\hlkwc{X}\hlstd{,}\hlkwc{a}\hlstd{,}\hlkwc{b}\hlstd{)\{}
  \hlstd{n} \hlkwb{<-} \hlkwd{nrow}\hlstd{(X); p} \hlkwb{<-} \hlkwd{ncol}\hlstd{(X)}
  \hlstd{theta} \hlkwb{<-} \hlstd{a} \hlopt{+} \hlstd{X}\hlopt{%*%}\hlstd{b}
  \hlstd{wts} \hlkwb{<-} \hlkwd{as.numeric}\hlstd{(}\hlkwd{exp}\hlstd{(theta)}\hlopt{/}\hlstd{((}\hlnum{1} \hlopt{+} \hlkwd{exp}\hlstd{(theta))}\hlopt{^}\hlnum{2}\hlstd{))}
  \hlstd{wts} \hlkwb{<-} \hlstd{wts}\hlopt{/}\hlkwd{mean}\hlstd{(wts)}
  \hlstd{Exw} \hlkwb{=} \hlkwd{t}\hlstd{(wts)} \hlopt{%*%} \hlstd{X} \hlopt{/} \hlstd{n}
  \hlstd{Sxw} \hlkwb{<-} \hlkwd{t}\hlstd{(X}\hlopt{-}\hlkwd{do.call}\hlstd{(rbind,} \hlkwd{replicate}\hlstd{(n, Exw,} \hlkwc{simplify}\hlstd{=}\hlnum{FALSE}\hlstd{)))} \hlopt{%*%}
    \hlkwd{diag}\hlstd{(wts)} \hlopt{%*%} \hlstd{(X}\hlopt{-}\hlkwd{do.call}\hlstd{(rbind,} \hlkwd{replicate}\hlstd{(n, Exw,} \hlkwc{simplify}\hlstd{=}\hlnum{FALSE}\hlstd{)))} \hlopt{/} \hlstd{n}
  \hlstd{Ys} \hlkwb{<-} \hlstd{theta} \hlopt{+} \hlstd{(Y}\hlopt{-}\hlkwd{exp}\hlstd{(theta)}\hlopt{/}\hlstd{(}\hlnum{1}\hlopt{+}\hlkwd{exp}\hlstd{(theta)))}\hlopt{/}\hlstd{wts}
  \hlstd{Eyw} \hlkwb{<-} \hlkwd{t}\hlstd{(wts)} \hlopt{%*%} \hlstd{Ys} \hlopt{/} \hlstd{n}
  \hlstd{Sxyw} \hlkwb{<-} \hlkwd{t}\hlstd{(X}\hlopt{-}\hlkwd{do.call}\hlstd{(rbind,} \hlkwd{replicate}\hlstd{(n, Exw,} \hlkwc{simplify}\hlstd{=}\hlnum{FALSE}\hlstd{)))} \hlopt{%*%}
    \hlkwd{diag}\hlstd{(wts)} \hlopt{%*%} \hlstd{(Ys}\hlopt{-}\hlkwd{do.call}\hlstd{(rbind,} \hlkwd{replicate}\hlstd{(n, Eyw,} \hlkwc{simplify}\hlstd{=}\hlnum{FALSE}\hlstd{)))} \hlopt{/} \hlstd{n}
  \hlstd{Syw} \hlkwb{=} \hlkwd{t}\hlstd{(Ys}\hlopt{-}\hlkwd{do.call}\hlstd{(rbind,} \hlkwd{replicate}\hlstd{(n, Eyw,} \hlkwc{simplify}\hlstd{=}\hlnum{FALSE}\hlstd{)))} \hlopt{%*%}
    \hlkwd{diag}\hlstd{(wts)} \hlopt{%*%} \hlstd{(Ys}\hlopt{-}\hlkwd{do.call}\hlstd{(rbind,} \hlkwd{replicate}\hlstd{(n, Eyw,} \hlkwc{simplify}\hlstd{=}\hlnum{FALSE}\hlstd{)))} \hlopt{/} \hlstd{n}
  \hlstd{M} \hlkwb{<-} \hlstd{Sxw}
  \hlstd{U} \hlkwb{<-}  \hlstd{Sxyw}\hlopt{%*%}\hlkwd{t}\hlstd{(Sxyw)} \hlopt{/} \hlkwd{as.numeric}\hlstd{(Syw)}
  \hlstd{out} \hlkwb{=} \hlkwd{list}\hlstd{(}\hlkwc{M} \hlstd{= M,} \hlkwc{U} \hlstd{= U)}
  \hlkwd{return}\hlstd{(out)}
\hlstd{\}}

\hlcom{## compute M and U with normal predictors for poisson }
\hlcom{## regression model with normal predictors}
\hlcom{## (Needs editing)}
\hlstd{Poisson_cov} \hlkwb{<-} \hlkwa{function}\hlstd{(}\hlkwc{Y}\hlstd{,}\hlkwc{X}\hlstd{,}\hlkwc{a}\hlstd{,}\hlkwc{b}\hlstd{)\{}
  \hlstd{n} \hlkwb{<-} \hlkwd{nrow}\hlstd{(X); p} \hlkwb{<-} \hlkwd{ncol}\hlstd{(X)}
  \hlstd{theta} \hlkwb{<-} \hlstd{a} \hlopt{+} \hlstd{X}\hlopt{%*%}\hlstd{b}
  \hlstd{wts} \hlkwb{<-} \hlkwd{as.numeric}\hlstd{(}\hlkwd{exp}\hlstd{(theta))}
  \hlstd{wts} \hlkwb{<-} \hlstd{wts}\hlopt{/}\hlkwd{mean}\hlstd{(wts)}
  \hlstd{Exw} \hlkwb{=} \hlkwd{t}\hlstd{(wts)} \hlopt{%*%} \hlstd{X} \hlopt{/} \hlstd{n}
  \hlstd{Sxw} \hlkwb{<-} \hlkwd{t}\hlstd{(X}\hlopt{-}\hlkwd{do.call}\hlstd{(rbind,} \hlkwd{replicate}\hlstd{(n, Exw,} \hlkwc{simplify}\hlstd{=}\hlnum{FALSE}\hlstd{)))} \hlopt{%*%}
    \hlkwd{diag}\hlstd{(wts)} \hlopt{%*%} \hlstd{(X}\hlopt{-}\hlkwd{do.call}\hlstd{(rbind,} \hlkwd{replicate}\hlstd{(n, Exw,} \hlkwc{simplify}\hlstd{=}\hlnum{FALSE}\hlstd{)))} \hlopt{/} \hlstd{n}
  \hlstd{Ys} \hlkwb{<-} \hlstd{theta} \hlopt{+} \hlstd{(Y}\hlopt{-}\hlkwd{exp}\hlstd{(theta))}\hlopt{/}\hlstd{wts}
  \hlstd{Eyw} \hlkwb{<-} \hlkwd{t}\hlstd{(wts)} \hlopt{%*%} \hlstd{Ys} \hlopt{/} \hlstd{n}
  \hlstd{Sxyw} \hlkwb{<-} \hlkwd{t}\hlstd{(X}\hlopt{-}\hlkwd{do.call}\hlstd{(rbind,} \hlkwd{replicate}\hlstd{(n, Exw,} \hlkwc{simplify}\hlstd{=}\hlnum{FALSE}\hlstd{)))} \hlopt{%*%}
    \hlkwd{diag}\hlstd{(wts)} \hlopt{%*%} \hlstd{(Ys}\hlopt{-}\hlkwd{do.call}\hlstd{(rbind,} \hlkwd{replicate}\hlstd{(n, Eyw,} \hlkwc{simplify}\hlstd{=}\hlnum{FALSE}\hlstd{)))} \hlopt{/} \hlstd{n}
  \hlstd{Syw} \hlkwb{=} \hlkwd{t}\hlstd{(Ys}\hlopt{-}\hlkwd{do.call}\hlstd{(rbind,} \hlkwd{replicate}\hlstd{(n, Eyw,} \hlkwc{simplify}\hlstd{=}\hlnum{FALSE}\hlstd{)))} \hlopt{%*%}
    \hlkwd{diag}\hlstd{(wts)} \hlopt{%*%} \hlstd{(Ys}\hlopt{-}\hlkwd{do.call}\hlstd{(rbind,} \hlkwd{replicate}\hlstd{(n, Eyw,} \hlkwc{simplify}\hlstd{=}\hlnum{FALSE}\hlstd{)))} \hlopt{/} \hlstd{n}
  \hlstd{M} \hlkwb{<-} \hlstd{Sxw}
  \hlstd{U} \hlkwb{<-}  \hlstd{Sxyw}\hlopt{%*%}\hlkwd{t}\hlstd{(Sxyw)} \hlopt{/} \hlkwd{as.numeric}\hlstd{(Syw)}
  \hlstd{out} \hlkwb{=} \hlkwd{list}\hlstd{(}\hlkwc{M} \hlstd{= M,} \hlkwc{U} \hlstd{= U)}
  \hlkwd{return}\hlstd{(out)}
\hlstd{\}}

\hlcom{## compute BIC scores }
\hlcom{# functionality is from the TRES package, but is editted to include }
\hlcom{# fitting when u = p}
\hlstd{bic_compute} \hlkwb{<-} \hlkwa{function}\hlstd{(}\hlkwc{M}\hlstd{,} \hlkwc{U}\hlstd{,} \hlkwc{n}\hlstd{,} \hlkwc{opts} \hlstd{=} \hlkwa{NULL}\hlstd{,} \hlkwc{multiD} \hlstd{=} \hlnum{1}\hlstd{)\{}
  \hlstd{p} \hlkwb{<-} \hlkwd{dim}\hlstd{(M)[}\hlnum{2}\hlstd{]}
  \hlstd{Mnew} \hlkwb{<-} \hlstd{M}
  \hlstd{Unew} \hlkwb{<-} \hlstd{U}
  \hlstd{G} \hlkwb{<-} \hlkwd{matrix}\hlstd{(}\hlnum{0}\hlstd{, p, p)}
  \hlstd{G0} \hlkwb{<-} \hlkwd{diag}\hlstd{(p)}
  \hlstd{phi} \hlkwb{<-} \hlkwd{rep}\hlstd{(}\hlnum{0}\hlstd{, p)}
  \hlkwa{for} \hlstd{(k} \hlkwa{in} \hlnum{1}\hlopt{:}\hlstd{p) \{}
    \hlkwa{if} \hlstd{(k} \hlopt{==} \hlstd{p)\{}
      \hlstd{gk} \hlkwb{<-} \hlkwd{ballGBB1D}\hlstd{(Mnew, Unew, opts)}
      \hlstd{phi[k]} \hlkwb{<-} \hlstd{n} \hlopt{*} \hlstd{(}\hlkwd{log}\hlstd{(}\hlkwd{t}\hlstd{(gk)} \hlopt{%*%} \hlstd{Mnew} \hlopt{%*%} \hlstd{gk)} \hlopt{+}
                       \hlkwd{log}\hlstd{(}\hlkwd{t}\hlstd{(gk)} \hlopt{%*%} \hlkwd{solve}\hlstd{(Mnew} \hlopt{+} \hlstd{Unew)} \hlopt{%*%} \hlstd{gk))} \hlopt{+}
        \hlkwd{log}\hlstd{(n)} \hlopt{*} \hlstd{multiD}
      \hlstd{G[, k]} \hlkwb{<-} \hlstd{G0} \hlopt{%*%} \hlstd{gk}
      \hlkwa{break}
    \hlstd{\}}
    \hlstd{gk} \hlkwb{<-} \hlkwd{ballGBB1D}\hlstd{(Mnew, Unew, opts)}
    \hlstd{phi[k]} \hlkwb{<-} \hlstd{n} \hlopt{*} \hlstd{(}\hlkwd{log}\hlstd{(}\hlkwd{t}\hlstd{(gk)} \hlopt{%*%} \hlstd{Mnew} \hlopt{%*%} \hlstd{gk)} \hlopt{+}
                     \hlkwd{log}\hlstd{(}\hlkwd{t}\hlstd{(gk)} \hlopt{%*%} \hlkwd{solve}\hlstd{(Mnew} \hlopt{+} \hlstd{Unew)} \hlopt{%*%} \hlstd{gk))} \hlopt{+}
      \hlkwd{log}\hlstd{(n)} \hlopt{*} \hlstd{multiD}
    \hlstd{G[, k]} \hlkwb{<-} \hlstd{G0} \hlopt{%*%} \hlstd{gk}
    \hlstd{G0} \hlkwb{<-} \hlkwd{qr.Q}\hlstd{(}\hlkwd{qr}\hlstd{(G[,} \hlnum{1}\hlopt{:}\hlstd{k]),} \hlkwc{complete} \hlstd{=} \hlnum{TRUE}\hlstd{)[, (k} \hlopt{+} \hlnum{1}\hlstd{)}\hlopt{:}\hlstd{p]}
    \hlstd{Mnew} \hlkwb{<-} \hlkwd{t}\hlstd{(G0)} \hlopt{%*%} \hlstd{M} \hlopt{%*%} \hlstd{G0}
    \hlstd{Unew} \hlkwb{<-} \hlkwd{t}\hlstd{(G0)} \hlopt{%*%} \hlstd{U} \hlopt{%*%} \hlstd{G0}
  \hlstd{\}}
  \hlstd{bic_val} \hlkwb{<-} \hlkwd{rep}\hlstd{(}\hlnum{0}\hlstd{, p)}
  \hlkwa{for} \hlstd{(k} \hlkwa{in} \hlnum{1}\hlopt{:}\hlstd{p) \{}
    \hlstd{bic_val[k]} \hlkwb{<-} \hlkwd{sum}\hlstd{(phi[}\hlnum{1}\hlopt{:}\hlstd{k])}
  \hlstd{\}}
  \hlstd{u} \hlkwb{=} \hlkwd{which.min}\hlstd{(bic_val)}
  \hlstd{bic_val}
\hlstd{\}}


\hlcom{## compute weighted envelope estimator from a weight vector }
\hlcom{## and the original estimator}
\hlstd{wtenv} \hlkwb{<-} \hlkwa{function}\hlstd{(}\hlkwc{M}\hlstd{,} \hlkwc{U}\hlstd{,} \hlkwc{wt}\hlstd{,} \hlkwc{b}\hlstd{)\{}
  \hlstd{p} \hlkwb{<-} \hlkwd{ncol}\hlstd{(M)}
  \hlstd{Env_wt} \hlkwb{<-} \hlkwd{rep}\hlstd{(}\hlnum{0}\hlstd{,p)}
  \hlkwa{for}\hlstd{(i} \hlkwa{in} \hlnum{1}\hlopt{:}\hlstd{p)\{}
    \hlkwa{if}\hlstd{(i} \hlopt{<} \hlstd{p)\{}
      \hlstd{G} \hlkwb{<-} \hlkwd{manifold1D}\hlstd{(}\hlkwc{M} \hlstd{= M,} \hlkwc{U} \hlstd{= U,} \hlkwc{u} \hlstd{= i)}
      \hlstd{Env_wt} \hlkwb{<-} \hlstd{Env_wt} \hlopt{+} \hlstd{wt[i]} \hlopt{*}
        \hlkwd{as.numeric}\hlstd{((G} \hlopt{%*%} \hlkwd{t}\hlstd{(G))} \hlopt{%*%} \hlstd{b)}
    \hlstd{\}}
    \hlkwa{if}\hlstd{(i} \hlopt{==} \hlstd{p) Env_wt} \hlkwb{<-} \hlstd{Env_wt} \hlopt{+} \hlstd{wt[i]} \hlopt{*} \hlstd{b}
  \hlstd{\}}
  \hlkwd{return}\hlstd{(Env_wt)}
\hlstd{\}}


\hlcom{## compute covariance matrices from bootstrap output}
\hlstd{covar} \hlkwb{<-} \hlkwa{function}\hlstd{(}\hlkwc{reg}\hlstd{)\{}
  \hlstd{p} \hlkwb{<-} \hlstd{(}\hlkwd{ncol}\hlstd{(reg)} \hlopt{-} \hlnum{1}\hlstd{)} \hlopt{/} \hlnum{4}
  \hlstd{covwt} \hlkwb{<-} \hlkwd{cov}\hlstd{(reg[,} \hlnum{1}\hlopt{:}\hlstd{p])}
  \hlstd{covenv} \hlkwb{<-} \hlkwd{cov}\hlstd{(reg[, (p}\hlopt{+}\hlnum{1}\hlstd{)}\hlopt{:}\hlstd{(}\hlnum{2}\hlopt{*}\hlstd{p)])}
  \hlstd{covenvfixedu} \hlkwb{<-} \hlkwd{cov}\hlstd{(reg[, (}\hlnum{2}\hlopt{*}\hlstd{p}\hlopt{+}\hlnum{1}\hlstd{)}\hlopt{:}\hlstd{(}\hlnum{3}\hlopt{*}\hlstd{p)])}
  \hlstd{covMLE} \hlkwb{<-} \hlkwd{cov}\hlstd{(reg[, (}\hlnum{3}\hlopt{*}\hlstd{p}\hlopt{+}\hlnum{1}\hlstd{)}\hlopt{:}\hlstd{(}\hlnum{4}\hlopt{*}\hlstd{p)])}
  \hlstd{out} \hlkwb{=} \hlkwd{list}\hlstd{(}\hlkwc{covwt}\hlstd{=covwt,} \hlkwc{covenv}\hlstd{=covenv,}
             \hlkwc{covenvfixedu} \hlstd{= covenvfixedu,} \hlkwc{covMLE}\hlstd{=covMLE)}
  \hlstd{out}
\hlstd{\}}


\hlcom{## compute normed matrices from bootstrap output}
\hlstd{normvar} \hlkwb{<-} \hlkwa{function}\hlstd{(}\hlkwc{reg}\hlstd{)\{}
  \hlstd{p} \hlkwb{<-} \hlstd{(}\hlkwd{ncol}\hlstd{(reg)} \hlopt{-} \hlnum{1}\hlstd{)} \hlopt{/} \hlnum{4}
  \hlstd{n} \hlkwb{<-} \hlkwd{nrow}\hlstd{(reg)}
  \hlstd{sewt} \hlkwb{<-} \hlkwd{crossprod}\hlstd{(reg[,} \hlnum{1}\hlopt{:}\hlstd{p])} \hlopt{/} \hlstd{n}
  \hlstd{seenv} \hlkwb{<-} \hlkwd{crossprod}\hlstd{(reg[, (p}\hlopt{+}\hlnum{1}\hlstd{)}\hlopt{:}\hlstd{(}\hlnum{2}\hlopt{*}\hlstd{p)])} \hlopt{/} \hlstd{n}
  \hlstd{seenvfixedu} \hlkwb{<-} \hlkwd{crossprod}\hlstd{(reg[, (}\hlnum{2}\hlopt{*}\hlstd{p}\hlopt{+}\hlnum{1}\hlstd{)}\hlopt{:}\hlstd{(}\hlnum{3}\hlopt{*}\hlstd{p)])} \hlopt{/} \hlstd{n}
  \hlstd{seMLE} \hlkwb{<-} \hlkwd{crossprod}\hlstd{(reg[, (}\hlnum{3}\hlopt{*}\hlstd{p}\hlopt{+}\hlnum{1}\hlstd{)}\hlopt{:}\hlstd{(}\hlnum{4}\hlopt{*}\hlstd{p)])} \hlopt{/} \hlstd{n}
  \hlstd{out} \hlkwb{=} \hlkwd{list}\hlstd{(}\hlkwc{se_wt}\hlstd{=sewt,} \hlkwc{se_env}\hlstd{=seenv,}
             \hlkwc{se_env_fixedu} \hlstd{= seenvfixedu,} \hlkwc{se_MLE}\hlstd{=seMLE)}
  \hlstd{out}
\hlstd{\}}


\hlcom{## Compute the bootstrap deviations for the weighted envelope }
\hlcom{## estimator, the envelope estimator with fixed dimension, the }
\hlcom{## envelope estimator with variable dimensions, and the MLE}
\hlcom{## Also report the selected envelope dimension at every }
\hlcom{## iteration of the nonparametric bootstrap}
\hlstd{model_boot} \hlkwb{<-} \hlkwa{function}\hlstd{(}\hlkwc{model}\hlstd{,} \hlkwc{nboot} \hlstd{=} \hlnum{1000}\hlstd{,} \hlkwc{cores} \hlstd{=} \hlnum{15}\hlstd{,}
  \hlkwc{intercept} \hlstd{=} \hlnum{FALSE}\hlstd{)\{}

  \hlcom{## important quantities an MLE of beta}
  \hlstd{dat} \hlkwb{<-} \hlstd{model}\hlopt{$}\hlstd{data}
  \hlstd{model} \hlkwb{<-} \hlkwd{update}\hlstd{(model,} \hlkwc{x} \hlstd{=} \hlnum{TRUE}\hlstd{,} \hlkwc{y} \hlstd{=} \hlnum{TRUE}\hlstd{,} \hlkwc{data} \hlstd{= dat)}
  \hlstd{Y} \hlkwb{<-} \hlstd{model}\hlopt{$}\hlstd{y}
  \hlstd{X} \hlkwb{<-} \hlstd{model}\hlopt{$}\hlstd{x}
  \hlstd{n} \hlkwb{<-} \hlkwd{nrow}\hlstd{(X)}
  \hlstd{b} \hlkwb{<-} \hlstd{betahat} \hlkwb{<-} \hlstd{model}\hlopt{$}\hlstd{coefficients}
  \hlstd{a} \hlkwb{<-} \hlnum{0}
  \hlkwa{if}\hlstd{(intercept)\{}
    \hlstd{a} \hlkwb{<-} \hlstd{betahat[}\hlnum{1}\hlstd{]}
    \hlstd{b} \hlkwb{<-} \hlstd{betahat[}\hlopt{-}\hlnum{1}\hlstd{]}
    \hlstd{X} \hlkwb{<-} \hlkwd{matrix}\hlstd{(X[,}\hlopt{-}\hlnum{1}\hlstd{],} \hlkwc{nrow} \hlstd{= n)}
  \hlstd{\}}
  \hlstd{p} \hlkwb{<-} \hlkwd{ncol}\hlstd{(X)}

  \hlcom{## intermediate envelope quanitites }
  \hlstd{fn} \hlkwb{<-} \hlkwa{NULL}
  \hlstd{fam} \hlkwb{<-} \hlstd{model}\hlopt{$}\hlstd{family}\hlopt{$}\hlstd{family}
  \hlkwa{if}\hlstd{(fam} \hlopt{==} \hlstr{"binomial"}\hlstd{)\{}
    \hlstd{fn} \hlkwb{<-} \hlkwa{function}\hlstd{(}\hlkwc{Y}\hlstd{,}\hlkwc{X}\hlstd{,}\hlkwc{a}\hlstd{,}\hlkwc{b}\hlstd{)} \hlkwd{Logistic_cov}\hlstd{(}\hlkwc{Y}\hlstd{=Y,}\hlkwc{X}\hlstd{=X,}\hlkwc{a}\hlstd{=a,}\hlkwc{b}\hlstd{=b)}
  \hlstd{\}}
  \hlkwa{if}\hlstd{(fam} \hlopt{==} \hlstr{"poisson"}\hlstd{)\{}
    \hlstd{fn} \hlkwb{<-} \hlkwa{function}\hlstd{(}\hlkwc{Y}\hlstd{,}\hlkwc{X}\hlstd{,}\hlkwc{a}\hlstd{,}\hlkwc{b}\hlstd{)} \hlkwd{Poisson_cov}\hlstd{(}\hlkwc{Y}\hlstd{=Y,}\hlkwc{X}\hlstd{=X,}\hlkwc{a}\hlstd{=a,}\hlkwc{b}\hlstd{=b)}
  \hlstd{\}}
  \hlstd{model_cov} \hlkwb{<-} \hlkwd{fn}\hlstd{(}\hlkwc{Y} \hlstd{= Y,} \hlkwc{X} \hlstd{= X,} \hlkwc{a} \hlstd{= a,} \hlkwc{b} \hlstd{= b)}
  \hlstd{M} \hlkwb{<-} \hlstd{model_cov}\hlopt{$}\hlstd{M; U} \hlkwb{<-} \hlstd{model_cov}\hlopt{$}\hlstd{U}
  \hlstd{bic_val} \hlkwb{<-} \hlkwd{bic_compute}\hlstd{(}\hlkwc{M} \hlstd{= M,} \hlkwc{U} \hlstd{= U,} \hlkwc{n} \hlstd{= n)}
  \hlstd{u} \hlkwb{<-} \hlkwd{which.min}\hlstd{(bic_val)}
  \hlstd{min_bic_val} \hlkwb{<-} \hlkwd{min}\hlstd{(bic_val)}
  \hlstd{wt_bic} \hlkwb{<-} \hlkwd{exp}\hlstd{(min_bic_val} \hlopt{-} \hlstd{bic_val)} \hlopt{/} \hlkwd{sum}\hlstd{(}\hlkwd{exp}\hlstd{(min_bic_val} \hlopt{-} \hlstd{bic_val))}

  \hlcom{## estimators}
  \hlstd{Ghat} \hlkwb{<-} \hlkwd{manifold1D}\hlstd{(}\hlkwc{M} \hlstd{= M,} \hlkwc{U} \hlstd{= U,} \hlkwc{u} \hlstd{= u)}
  \hlstd{Env} \hlkwb{<-} \hlkwd{as.numeric}\hlstd{((Ghat} \hlopt{%*%} \hlkwd{t}\hlstd{(Ghat))} \hlopt{%*%} \hlstd{b)}
  \hlstd{Env_wt} \hlkwb{<-} \hlkwd{as.numeric}\hlstd{(}\hlkwd{wtenv}\hlstd{(M, U,} \hlkwc{wt} \hlstd{= wt_bic, b))}

  \hlcom{## bootstrap routine}
  \hlkwd{registerDoParallel}\hlstd{(}\hlkwc{cores} \hlstd{= cores)}
  \hlstd{output} \hlkwb{<-} \hlkwd{foreach}\hlstd{(}\hlkwc{i}\hlstd{=}\hlnum{1}\hlopt{:}\hlstd{nboot,}\hlkwc{.combine}\hlstd{=rbind,}
    \hlkwc{.export}\hlstd{=}\hlkwd{c}\hlstd{(}\hlstr{"ballGBB1D"}\hlstd{,}\hlstr{"manifold1D"}\hlstd{,}\hlstr{"wtenv"}\hlstd{,}\hlstr{"fn"}\hlstd{))} \hlopt{%dopar%} \hlstd{\{}

    \hlcom{## refit model using bootstrap pairs       }
    \hlstd{m_boot} \hlkwb{<-} \hlkwd{update}\hlstd{(model,} \hlkwc{data} \hlstd{= dat[}\hlkwd{sample}\hlstd{(}\hlnum{1}\hlopt{:}\hlstd{n,} \hlkwc{replace} \hlstd{=} \hlnum{TRUE}\hlstd{), ],}
      \hlkwc{x} \hlstd{=} \hlnum{TRUE}\hlstd{,} \hlkwc{y} \hlstd{=} \hlnum{TRUE}\hlstd{)}

    \hlcom{## important quantities including the MLE of beta }
    \hlcom{## from the bootstrap sample}
    \hlstd{Y_boot} \hlkwb{<-} \hlstd{m_boot}\hlopt{$}\hlstd{y}
    \hlstd{X_boot} \hlkwb{<-} \hlstd{m_boot}\hlopt{$}\hlstd{x}
    \hlstd{b_boot} \hlkwb{<-} \hlstd{beta_boot} \hlkwb{<-} \hlstd{m_boot}\hlopt{$}\hlstd{coefficients}
    \hlstd{a_boot} \hlkwb{<-} \hlnum{0}
    \hlkwa{if}\hlstd{(intercept)\{}
      \hlstd{a_boot} \hlkwb{<-} \hlstd{beta_boot[}\hlnum{1}\hlstd{]}
      \hlstd{b_boot} \hlkwb{<-} \hlstd{beta_boot[}\hlopt{-}\hlnum{1}\hlstd{]}
      \hlstd{X_boot} \hlkwb{<-} \hlkwd{matrix}\hlstd{(X_boot[,}\hlopt{-}\hlnum{1}\hlstd{],} \hlkwc{nrow} \hlstd{= n)}
    \hlstd{\}}

    \hlcom{## intermediate envelope estimation quantities}
    \hlstd{model_cov_boot} \hlkwb{<-} \hlkwd{fn}\hlstd{(}\hlkwc{Y}\hlstd{=Y_boot,}\hlkwc{X}\hlstd{=X_boot,}\hlkwc{a}\hlstd{=a_boot,}\hlkwc{b}\hlstd{=b_boot)}
    \hlstd{M_boot} \hlkwb{<-} \hlstd{model_cov_boot}\hlopt{$}\hlstd{M; U_boot} \hlkwb{<-} \hlstd{model_cov_boot}\hlopt{$}\hlstd{U}
    \hlcom{#M <- vcov(m_boot); U <- beta_boot %o% beta_boot}
    \hlstd{bic_val_boot} \hlkwb{=} \hlkwd{bic_compute}\hlstd{(}\hlkwc{M}\hlstd{=M_boot,}\hlkwc{U}\hlstd{=U_boot,} \hlkwc{n}\hlstd{=n)}
    \hlstd{u_boot} \hlkwb{<-} \hlkwd{which.min}\hlstd{(bic_val_boot)}
    \hlstd{min_bic_val_boot} \hlkwb{<-} \hlkwd{min}\hlstd{(bic_val_boot)}
    \hlstd{wt_bic_boot} \hlkwb{<-} \hlkwd{exp}\hlstd{(min_bic_val_boot} \hlopt{-} \hlstd{bic_val_boot)} \hlopt{/}
      \hlkwd{sum}\hlstd{(}\hlkwd{exp}\hlstd{(min_bic_val_boot} \hlopt{-} \hlstd{bic_val_boot))}

    \hlcom{## envelope estimators from bootstrap sample}
    \hlstd{G_boot} \hlkwb{<-} \hlkwd{manifold1D}\hlstd{(}\hlkwc{M} \hlstd{= M_boot,} \hlkwc{U} \hlstd{= U_boot,} \hlkwc{u} \hlstd{= u_boot)}
    \hlstd{G_boot_fixedu} \hlkwb{<-} \hlkwd{manifold1D}\hlstd{(}\hlkwc{M} \hlstd{= M_boot,} \hlkwc{U} \hlstd{= U_boot,} \hlkwc{u} \hlstd{= u)}
    \hlstd{Env_boot} \hlkwb{<-} \hlkwd{as.numeric}\hlstd{((G_boot} \hlopt{%*%} \hlkwd{t}\hlstd{(G_boot))} \hlopt{%*%} \hlstd{b_boot)}
    \hlstd{Env_boot_fixedu} \hlkwb{<-} \hlkwd{as.numeric}\hlstd{((G_boot_fixedu} \hlopt{%*%}
      \hlkwd{t}\hlstd{(G_boot_fixedu))} \hlopt{%*%} \hlstd{b_boot)}
    \hlstd{Env_wt_boot} \hlkwb{<-} \hlkwd{as.numeric}\hlstd{(}\hlkwd{wtenv}\hlstd{(}\hlkwc{M} \hlstd{= M_boot,} \hlkwc{U} \hlstd{= U_boot,}
      \hlkwc{wt} \hlstd{= wt_bic_boot,} \hlkwc{b} \hlstd{= b_boot))}

    \hlcom{## descrepancy}
    \hlkwd{c}\hlstd{((Env_wt} \hlopt{-} \hlstd{Env_wt_boot),}
      \hlstd{(Env} \hlopt{-} \hlstd{Env_boot),}
      \hlstd{(Env} \hlopt{-} \hlstd{Env_boot_fixedu),}
      \hlstd{(b} \hlopt{-} \hlstd{b_boot), u_boot)}
  \hlstd{\}}
  \hlstd{output}
\hlstd{\}}
\end{alltt}
\end{kframe}
\end{knitrout}

\end{document}